\documentclass[twocolumn,superscriptaddress,preprintnumbers,prb,nofootinbib]{revtex4-2}
\usepackage{eurosym,hyperref}
\usepackage{graphicx,bm,amsmath,amssymb}
\usepackage{orcidlink}
\usepackage{braket}
\usepackage{todonotes}
\usepackage{color}

\hypersetup{colorlinks=true,citecolor=blue,linkcolor=magenta}
\allowdisplaybreaks

\def\gz{\ifmmode{Z\hskip -4.8pt Z}
    \else{\hbox{$Z\hskip -4.8pt Z$}}\fi}

\newcommand{\be}{\begin{equation}}
\newcommand{\ee}{\end{equation}}
\newcommand{\bea}{\begin{eqnarray}}
\newcommand{\eea}{\end{eqnarray}}

\newcommand{\ligand}{\underbar{\textit{L}}}

\usepackage{glossaries}
\newacronym{1D}{1D}{one-dimensional}
\newacronym{2D}{2D}{two-dimensional}
\newacronym{ED}{ED}{exact diagonalization}
\newacronym{CDW}{CDW}{charge-density-wave}
\newacronym{RIXS}{RIXS}{resonant inelastic x-ray scattering}
\newacronym{DMRG}{DMRG}{density matrix renormalization group}
\newacronym{DQMC}{DQMC}{determinant quantum Monte Carlo}
\newacronym{QMC}{QMC}{quantum Monte Carlo}
\newacronym{CTI}{CTI}{charge-transfer insulator}
\newacronym{NCTI}{NCTI}{negative charge-transfer insulator}
\newacronym{ZSA}{ZSA}{Zaanen-Sawatzky-Allen}
\newacronym{ZRS}{ZRS}{Zhang-Rice singlet}
\newacronym{LHB}{LHB}{lower Hubbard band}
\newacronym{UHB}{UHB}{upper Hubbard band}
\newacronym{TMO}{TMO}{Transition metal oxide}
\newacronym{CT}{CT}{charge-transfer}
\newacronym{TM}{TM}{transition metal}
\newacronym{eph}{$e$-ph}{electron-phonon}
\newacronym{XAS}{XAS}{x-ray absorption spectroscopy}
\newacronym{ARPES}{ARPES}{angle-resolved photoemission spectroscopy}
\newacronym{INS}{INS}{inelastic neutron scattering}
\newacronym{tDRMG}{tDRMG}{time-dependent density-matrix renormalization group}

\begin{document}

\title{Band mixing effects in one-dimensional charge transfer insulators}

\author{Samuel Milner}
\affiliation{Physics Department, Northeastern University, Boston, MA 02115, USA}

\author{Steven Johnston\orcidlink{0000-0002-2343-0113}}
\affiliation{Department of Physics and Astronomy, The University of Tennessee, Knoxville, Tennessee 37996, USA}
\affiliation{Institute for Advanced Materials and Manufacturing, University of Tennessee, Knoxville, Tennessee 37996, USA\looseness=-1}

\author{A. E. Feiguin\orcidlink{0000-0001-5509-2561}}
\affiliation{Physics Department, Northeastern University, Boston, MA 02115, USA}

\begin{abstract}
The low-energy properties of transition metal oxides (TMOs) are governed by the electrons occupying strongly correlated $d$-orbitals that are hybridized with surrounding ligand oxygen $p$ orbitals to varying degrees. Their physics is thus established by a complex interplay between the transition-metal (TM)-ligand hopping $t$, charge transfer energy $\Delta_\mathrm{CT}$, and on-site TM Hubbard repulsion $U$. Here, we study the spectral properties of a one-dimensional (1D) analog of such a $pd$ system, with alternating TM $d$ and ligand anion $p$ orbitals situated along a chain. Using the density matrix renormalization group method, we study the model's single-particle spectral function, x-ray absorption spectrum, and dynamical spin structure factor as a function of $\Delta_\mathrm{CT}$ and $U$. In particular, we present results spanning from the Mott insulating ($\Delta_\mathrm{CT} > U$) to negative charge transfer regime $\Delta_\mathrm{CT} < 0$ to better understand the ground and momentum-resolved excited state properties of these different regimes. Our results can guide new studies on TMOs that seek to situate them within the Mott-Hubbard/charge transfer insulator classification scheme. 
\end{abstract}

\maketitle

\section{Introduction}

\glspl*{TMO} are a broad class of quantum materials that have long held the attention of materials scientists and condensed matter physicists~\cite{Rao1989transition, Imada1998, Ngai2014correlated, Chakhalian2014colloquium, Zaanen1985}. These materials host transition metal cations surrounded by oxygen anions and exhibit a rich set of electronic, magnetic, optical, and structural properties that can be tuned through subtle changes in different tuning parameters (e.g., strain, composition, temperature, pressure, etc.). As such, they have the potential to advance many technologies in energy storage, sensing, catalysis, spintronics, and beyond~\cite{Catalano2018, Bibes2011, Hwang2012, Mannhart2010, Chen2017}. 

The low-energy electronic structure of \glspl*{TMO} and, consequently, their functional properties are determined by their partially filled \gls*{TM} $3d$ orbitals, which are often hybridized to the $2p$ orbitals of the surrounding oxygen ligands. A unifying organizational principle behind these materials is the \gls*{ZSA} scheme~\cite{Zaanen1985}, which classifies each material based on the relative values of the \gls*{CT} energy $\Delta_\mathrm{CT}$ and interatomic \gls*{TM} Hubbard repulsion $U$. Defined in the atomic limit, $U$ characterizes the energy associated with $d_i^nd_{i+1}^n\rightarrow d_i^{n-1}d_{i+1}^{n+1}$ charge excitations between the \gls*{TM} cations, while $\Delta_\mathrm{CT}$ is the energy associated with transferring a hole from the \gls*{TM} site to the surrounding ligand orbitals, i.e., $d_i^n\rightarrow d^{n+1}_i\underbar{\textit{L}}$, where $\underbar{\textit{L}}$ denotes a ligand hole. If $U > \Delta_\mathrm{CT}$, charge transfer to the ligand sites is the primary barrier to transport, and the system is a \gls*{CTI}. Conversely, if $U < \Delta_\mathrm{CT}$, Hubbard-like excitations between the \gls*{TM} sites limit transport, and the system is a Mott-Hubbard insulator. If $U\approx \Delta_\mathrm{CT}$, then the system is said to be of a mixed character. There also exists so-called \glspl*{NCTI}, where $\Delta_\mathrm{CT} < 0$ such that holes self-dope onto the ligand orbitals~\cite{Khomskii, Green2024negative}. Correlated materials with cations in high oxidation states often fall into this latter category~\cite{Green2024negative, Mizokawa1991origin, Bisogni2016groundstate, Plumb2016momentum, Akao2003charge}. 

Properly classifying a material has important implications for understanding its functional properties. For example, the quasi-1D and 2D cuprates are \glspl*{CTI}, which helps account for their large superexchange interactions and robust antiferromagnetism~\cite{Anderson1950, Anderson1959, Anderson1963, Zaanen1987, Eskes1990}. Similarly, the recent discovery of superconductivity in the low-valence nickelates~\cite{Li2019superconductivity, Pan2022superconductivity, Sun2023signatures, Wang2024experimental} launched a significant effort~\cite{Hu2019twoband, Nomura2019formation, jiang, Gu2020, Sakakibara2020model, Hepting2020electronic, Adhikary2020orbital, Wang2020synthesis, Botana2020similarities, Lechermann2021doping, Petocchi2020normal, Karp2020manybody, 
Shen2022role} aimed at classifying these materials. Understanding the fundamental differences between nickel and copper oxides can illuminate the basic ingredients for high-$T_\mathrm{c}$ superconductivity. Similarly, several classes of \glspl*{NCTI} have been identified with strong coupling to \gls*{TM}-O bond-stretching phonon modes~\cite{Park2012site, Johnston2014, Khazraie2018oxygen, CohenStead2023hybrid} that arises directly from the strong hybridization between the ligand and \gls*{TM} states. This interaction has been proposed as the mechanism behind charge ordered phases~\cite{Johnston2014, CohenStead2023hybrid, Park2012site}, bipolaronic transport~\cite{Shamblin2018experimental, Tyunina2023small}, and high-$T_\mathrm{c}$ superconductivity~\cite{Pickett2001other, Sleight2015bismuthates, Kim2022superconductivity}.

Given the diverse physics realized by variations in $\Delta_\mathrm{CT}$ and $U$, it is important to develop spectroscopic methods for determining which regime a material occupies. Significant progress has been made in understanding the fundamental properties of \glspl*{TMO} through the use of small cluster  exact diagonalization calculations in the context of core-level spectroscopies~\cite{Sawatzky1984magnitude, Eskes1990, Mizokawa1994, Haverkort2012multiplet, jiang, Bisogni2016groundstate, Mizokawa1991origin, Mizokawa2000spin, Green2024negative, Khomskii, Yamaguchi2024atomic, Haverkort2016quanty, Freeland2016evolution, Chen2013doping}. These include calculations incorporating all five $d$ orbitals and effective three-band models (one $d_{x^2-y^2}$ and two $p$ bands) using \gls*{ED}~\cite{Dobry1994, Lau2011, Lau2011b}. However, scaling these many-body calculations to extended clusters needed for calculating other dynamical properties has been challenging. For example, sophisticated numerical approaches such as the \gls*{DMRG} have only recently been applied to studying the excitation spectra of corner-shared CuO$_3$ chains~\cite{Li2021particle, Nocera2018computing} and ladders \cite{White2015}. The fermion sign problem limits quantum Monte Carlo calculations~\cite{Guerrero1998, Chiciak2018, Huang2017, Mai2024fluctuating}, while cluster sizes can limit quantum embedding techniques~\cite{Weber2012, Mai2021orbital, Cui2020Groundstate}. 

The complexity of the underlying problem is such that one needs to resort to effective low-energy theories with fewer degrees of freedom. While the single band Hubbard or $t$-$J$ models may contain the fundamental ingredients to explain high-$T_\mathrm{c}$ superconductivity~\cite{Dagotto1994, Imada1998, ScalapinoReview, qin2021hubbard, SimonsCollab2015, SimonsCollab2020}, these mappings rely on two main assumptions: (i) that the the fundamental physics takes place in CuO$_2$ (or NiO$_2$) planes, and (ii) the \gls*{ZRS} construction holds for describing the charge carriers in doped \glspl*{CTI}~\cite{ZR, fedro, simon, brt, fei, bel, Belinicher94, ali94}. However, the validity of the \gls*{ZRS} picture is subject to the relative position of the $d$ and $p$ orbitals and remains controversial in systems with large \gls*{CT} energies~\cite{ebra, hamad2, adol, hamad1, jiang, ali20}. Moreover, the mapping onto an effective single band Hamiltonian assumes that higher-energy charge transfer states can be ignored, which is not all obvious when $\Delta_\mathrm{CT}$ is small or negative. Optical conductivity experiments~\cite{Santoso2017} in Zn-LSCO, for example, reveal a mixture of states with singlet and triplet characters that can only be explained using a multi-band model. 

\begin{figure}
    \centering
    \includegraphics[width=\linewidth]{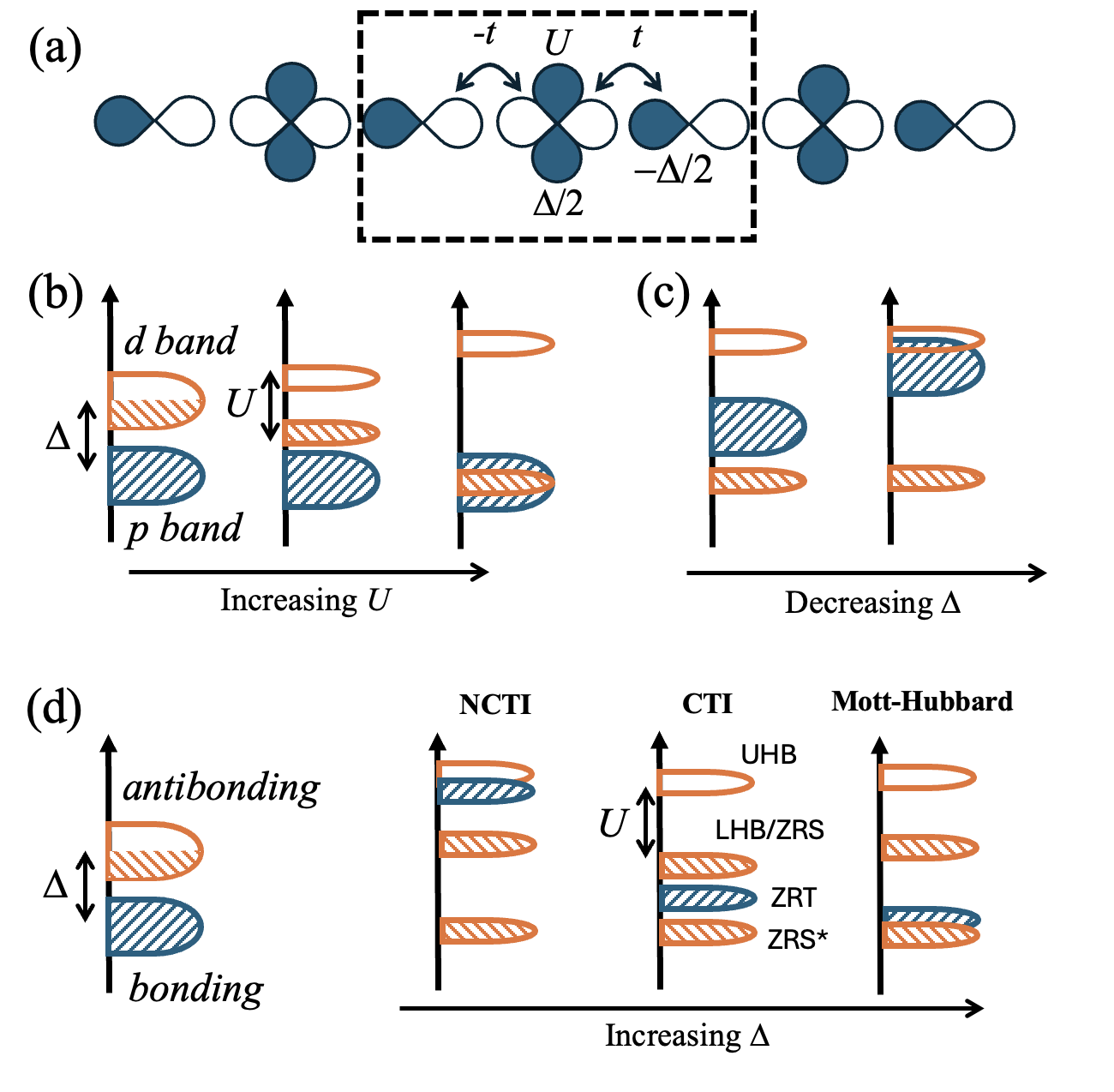}
    \caption{(a) Cartoon of the Cu-O chain studied in this work, described by a two-band Emery Hamiltonian, see Eqs.~\eqref{eq:H}-\eqref{eq:HU}. In our calculation, we use a similar geometry, with $L$ unit cells and an extra oxygen, as shown here for $L=3$. (b), (c) Illustration of the density of states for the atomic picture ignoring hybridization effects between the bands (see text), with $p$ and $d$ states separated by an energy splitting $\Delta$. Shading represents different occupied states.  Panels in (b) show the effect of introducing a Coulomb repulsion $U$ on the $d$ orbitals for positive $\Delta$. This regime corresponds to a Mott-Hubbard insulator. The effects of decreasing $\Delta$ are depicted in (c).
    Finally, panel (d) shows both hybridization and many-body effects, and illustrates the crossover from negative charge transfer insulator to Mott-Hubbard insulator as $\Delta$ varies from negative to positive, shifting the relative center of mass of the bands. }
    \label{fig:cartoon}
\end{figure}

Motivated by these considerations, here we present a systematic study of the excitation spectra of simplified multi-orbital charge-transfer chains using the \gls*{DMRG} method. By focusing on a \gls*{1D} geometry, we can calculate the model's excitation spectra with high momentum resolution and explore a wide regime of parameters. Albeit \gls*{1D}, our model contains the basic ingredients needed to study the crossover between different regimes of the \gls*{ZSA} classification scheme, with carriers that can be defined in terms analogous to the \gls*{ZRS} in higher dimensions~\cite{penc}. By focusing on a \gls*{1D} model we can also make direct connections with a large body of existing literature. For example, undoped \gls*{1D} cuprates such as Sr$_2$CuO$_{3}$ and SrCuO$_2$ have been widely studied using a variety of experimental probes~\cite{neudert, Kim1996observation, Kojima1997reducton, Eisaki1997spin, Zaliznyak2004spinons, Kim2006distinct, Enderle2010twospinon, Drechsler2011comment, Schlappa2012spin, Valentina2014femtosecond, walters, Bounoua2018angle, Schlappa2018probing, Kumar2022unraveling, Adhikary2020orbital}
while \gls*{ARPES}~\cite{chen} and \gls*{RIXS}~\cite{Li2025doping} 
experiments have recently been carried out on the related doped compound Ba$_{2-x}$Sr$_x$CuO$_{3+\delta}$. This family of materials has been analyzed using both multi-orbital models and effective single-band Hamiltonians~\cite{chen, wang, tangb, wang2b, Kumar2018multi, Li2021particle, Feiguin2023, Li2025doping}. For simplicity, here we focus on a two-band model alternating \gls*{TM} and O orbitals along the chain. Despite its apparent simplicity, this model exhibits rich physics as a function of the band splitting, hopping, and interaction strengths.

The manuscript is organized as follows: Section~\ref{sec:model} introduces our model and outlines the details of our \gls*{DMRG} calculations. Section~\ref{sec:results} then presents results, including predictions of the single-particle spectral function (Sec.~\ref{sec:results_arpes}), \gls*{XAS} (Sec.~\ref{sec:results_xas}), and dynamical spin structure factor (Sec.~\ref{sec:results_sqw}). In each case, we are particularly interested in tracking the evolution of various spectral features as a function of the Hubbard interaction $U$ and \gls*{CT} energy $\Delta_\mathrm{CT}$. Finally, Sec.~\ref{sec:conclusions} discusses our results and summarizes our  findings. 

\section{Model and methods}\label{sec:model}
\subsection{Model: the one-dimensional $pd$-chain}
We consider a two-band model with alternating 3$d_{x^2-y^2}$ and $2p$ orbitals along a \gls*{1D} chain~\cite{penc,Zaanen1988}, as shown in Fig.~\ref{fig:cartoon}a. The unit cell contains one $p$ and one $d$ orbital, while the index $i$ indicates the position of the $d$ orbital along the chain. The model's Hamiltonian, written in electron language, is 
\begin{equation}\label{eq:H}
    H = H_{0} + H_U. 
\end{equation}
Here, 
\begin{equation}\label{eq:H0}
H_{0} =-t\sum\limits_{i,\sigma,\delta } P_\delta \left(
d_{i,\sigma}^{\dagger }p^{\phantom\dagger}_{i+\delta,\sigma }
+\text{%
H.c.}\right)+\frac{\Delta}{2}\sum\limits_{i,\sigma,\delta } 
(n_{i}^{d} - n_{i}^{p}), 
\end{equation}
includes the noninteracting terms and
\begin{equation}\label{eq:HU}
H_U = U\sum\limits_{i}
n_{i,\uparrow}^{d}n_{i,\downarrow}^{d}
\end{equation}
is the on-site Hubbard interaction for the $d$ orbitals. 
In these expressions, the fermionic operators $d^\dagger_{i,\sigma}$ ($d_{i,\sigma}$) and $p^\dagger_{i,\sigma}$ ($p_{i,\sigma}$) create (annihilate) a spin $\sigma~(=\uparrow,\downarrow)$ electron on the $3d_{x^2-y^2}$ and $2p$ orbitals in unit cell $i$, respectively, $n^{p/d}$ is the corresponding number operator, $i+\delta = i \pm 1$ indicates the $p$ orbitals situated on either side of the $d$ orbitals, and $P_{\delta=\pm} = \mp 1$ is a phase factor for the $pd$ hopping.\footnote{The sign in the hopping can be removed with a simple gauge transformation~\cite{penc}. However, we keep it in place to avoid artificial phases in the momentum-resolved correlation functions.} The on-site Coulomb interaction for the $d$ orbitals is $U$. The nearest-neighbor $pd$ hopping $t=1$ sets our unit of energy. 

Throughout this work, we follow the \gls*{ZSA}~\cite{Zaanen1985} scheme for describing the system, which starts from the atomic limit ($t = 0$).\footnote{To reconcile our model with \gls*{ZSA}'s, notice that they set $\epsilon_d=0$ and $\epsilon_p = \Delta$ so there is a downward shift of the energy levels by $-\Delta/2$ in our case.} Using our definition of the Hamiltonian, the $d$ levels will split into states centered at $\Delta/2$ and $\Delta/2+U$, while the $p$ levels are centered at $-\Delta/2$. The \gls*{CT} energy $\Delta_\mathrm{CT}$ is equal to the energy difference between the $d^2\underbar{\textit{L}}^1$ and $d^1\underbar{\textit{L}}^0$ configurations in the atomic limit and thus given by $\Delta_\mathrm{CT} = \epsilon_d - \epsilon_p + U = \Delta + U$. Once we reintroduce the electron hopping, these states form bands and acquire a finite bandwidth $W$, as sketched in Figs.~\ref{fig:cartoon}(b)-(f).

Even though the \gls*{ZSA} picture provides some intuition for the electronic structure, it misses a few important aspects about the effects of the lattice and the hybridization between orbitals. To see this, one can instead start from the noninteracting limit, where the orbitals form bonding (-) and antibonding (+) bands with dispersion 
\begin{equation}\label{eq:noninteracting_bands}
\epsilon_\pm(k)=\pm \sqrt{4t^2\sin^2{(ka/2)}+\Delta^2/4},
\end{equation}
where $a$ is the lattice constant.\footnote{Analogs to these bands also appear in the 4-band model studied in Ref.~\onlinecite{Li2021particle} for corner-shared Cu-O chains, which also contains a nonbonding and a flat band that does not contribute to the physics of the system. Therefore, despite the apparent simplicity of our Hamiltonian, we expect it to realize many features relevant to real materials. }
For large $\Delta > t$, the antibonding band has mostly $d$ character and, in the presence of interactions, it will split into a \gls*{LHB} and an \gls*{UHB}~\cite{Varma1993}. The system is classified as a Mott-Hubbard insulator when the chemical potential lies between the two Hubbard bands. When $\Delta \lessapprox t$, the system is in a regime of strong hybridization, and for negative $\Delta$, but small compared to $U$, the system becomes a \gls*{CTI}. Finally, for large negative $\Delta$, the system enters a \gls*{NCTI} regime~\cite{Khomskii} in which the antibonding band has mostly $p$ character, and the system can become metallic (or ``self-doped'') by mixing with the \gls*{UHB}.  

\subsection{Methods}
We calculate the model's spectral functions using the \gls*{tDRMG}~\cite{white2004a,daley2004,vietri,Paeckel2019}. 
Specifically, we compute the single particle two-time correlator 
\begin{equation}
 G_{d/p}(x,t)=\mathrm{i}\langle O^\dagger_\sigma(i,t) O^{\phantom\dagger}_\sigma(L/2,0)\rangle, 
\end{equation}
where the operator $O^\dagger_\sigma(i,t) = d^\dagger_{i,\sigma}(t)$ or 
$p^\dagger_{i,\sigma}(t)$ defines the character of the excitation. This quantity is Fourier transformed to frequency and momentum, yielding the spectral function $A_{d/p}(k,\omega)=- \text{Im} {G_{p/d}(k,\omega)}/\pi$. Similar expressions are used to compute the dynamical spin structure factor $S^z(k,\omega)$, where the spin operator $O=S^z_{p/d}$ is used instead. We can also calculate the \gls*{XAS} spectra using a similar approach but with a modified Hamiltonian that accounts for the core hole's interaction with the valence electrons (see Sec.~\ref{sec:results_xas})~\cite{Zawadzki2023timedependent}.  

We carried out all calculations on a chain with open boundary conditions and an extra oxygen $p$ orbital on the left side of the chain. The total number of orbitals is $2L+1$, where $L$ is the number of unit cells. This cluster geometry preserves the reflection symmetry of the lattice with respect to the $d$ orbital located at the center of the chain $i = L/2$. To reduce artifacts from Fourier transforming in a finite time window, we used a ``Hann window'' technique, convoluting the results with an exponential function that yields a Lorentzian lineshape with a width $\epsilon=0.1$~\cite{oppenheim99}. This practice also helps mitigate errors that tend to accumulate at longer times. We typically use $m=400$ \gls*{DMRG} states (guaranteeing a truncation error below $10^{-6}$), a time step $\delta t=0.02$, and a maximum time $\sigma=t_{\max}=40$. The undoped (i.e., ``half-filled'' in the cuprate literature) stoichiometric case corresponds to three electrons per unit cell, two for each $p$ orbital and one for each $d$, translating into a total of $N_e=3L+2$ electrons. Finally, we performed all simulations using a third-order Suzuki-trotter decomposition of the time evolution operator.

\section{Results}\label{sec:results} 
\subsection{Ground State Properties}\label{sec:results_cluster}
Before discussing our model's spectral properties, it is helpful to examine its ground state character as a function of model parameters. To this end, we carry out small cluster \gls*{ED} calculations, considering only two $p$ orbitals and one $d$ orbital in a $p$-$d$-$p$ geometry, as sketched in the dashed box of Fig.~\ref{fig:cartoon}(a). 

We first solve this cluster for the stoichiometric filling with $N=5$ electrons. The cluster's ground state in this sector can be written as a linear combination $\ket{\Psi_\mathrm{gs}} = \alpha \ket{d^1\ligand^0} + \beta\ket{d^2\ligand^1}$, where 
\begin{equation}
\begin{split}
    \ket{d^1\ligand^0}&= \ket{2,\uparrow,2}, \\
    \ket{d^2\ligand^1}&= \tfrac{1}{\sqrt{2}}\left[\ket{\uparrow,2,2}-\ket{2,2,\uparrow}\right]. 
\end{split}
\end{equation}
Here, each number/spin denotes the occupation of the orbitals, $d^n$ indicates $n$ electrons in the $d$ orbital, and $\ligand^n$ denotes $n$ holes occupying the surrounding ligand oxygen orbitals (see also App.~\ref{sec:Happendix}). 

\begin{figure}[t]
	\centering
    \includegraphics[width=0.48\textwidth]{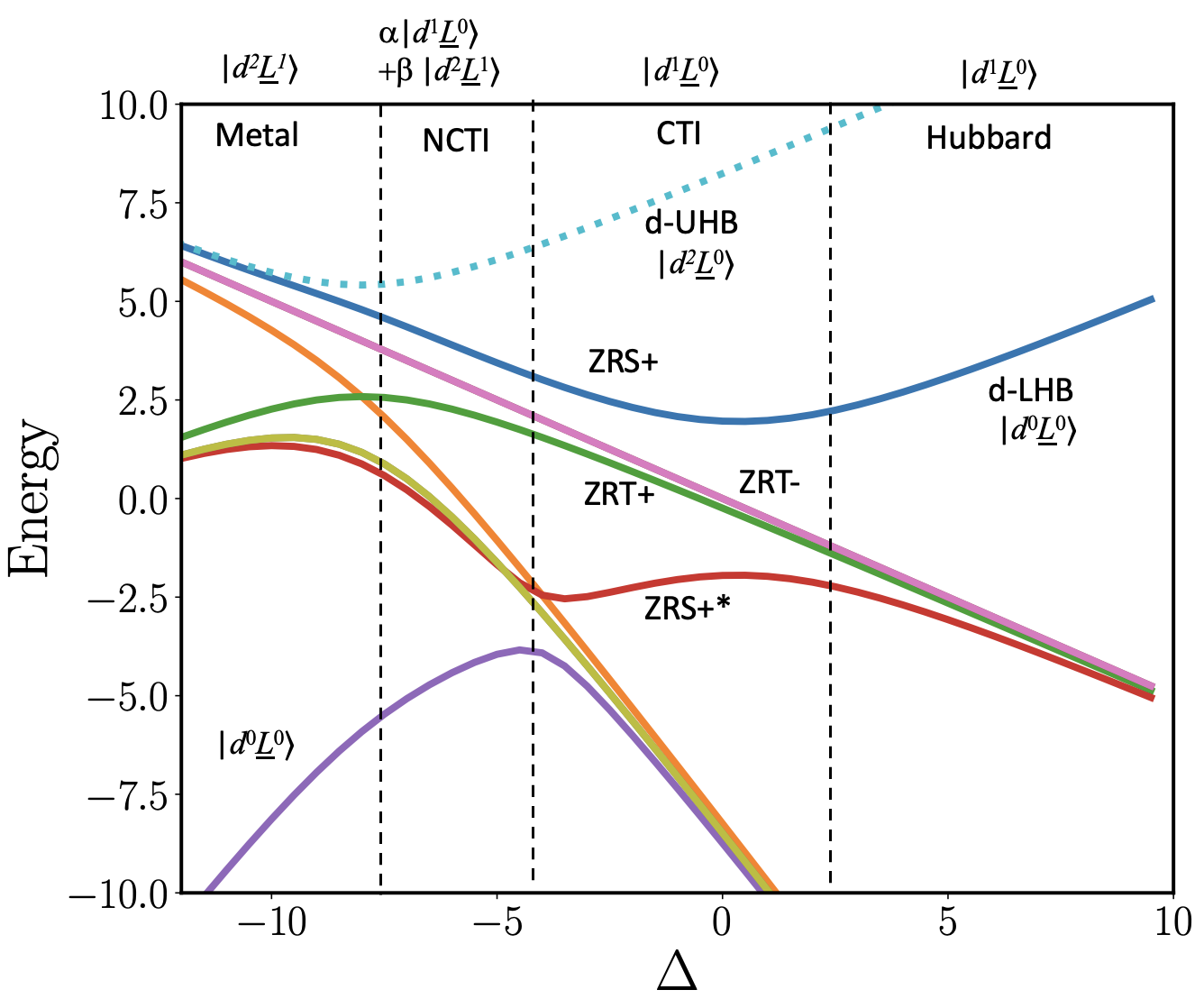}
	\caption{The electron addition and removal spectrum of a three site $p$-$d$-$p$ cluster, as shown in the dashed box of Fig.~\ref{fig:cartoon}. The energies were obtained by exact diagonalization with $U=8t$. The full (dashed) lines correspond to the single-particle removal (addition) states. Vertical dash lines correspond to $\Delta_\mathrm{CT} = -8$ ($\Delta_\mathrm{CT} = 0$) and $\Delta = 0$ ($\Delta_\mathrm{CT}=U$), which correspond to the approximate positions of the crossovers between the different regimes. The labels above each region indicate the dominant character of the one-hole ground state. } \label{fig:cluster}
\end{figure}

\begin{figure}[ht]
	\centering
   \includegraphics[width=0.48\textwidth]{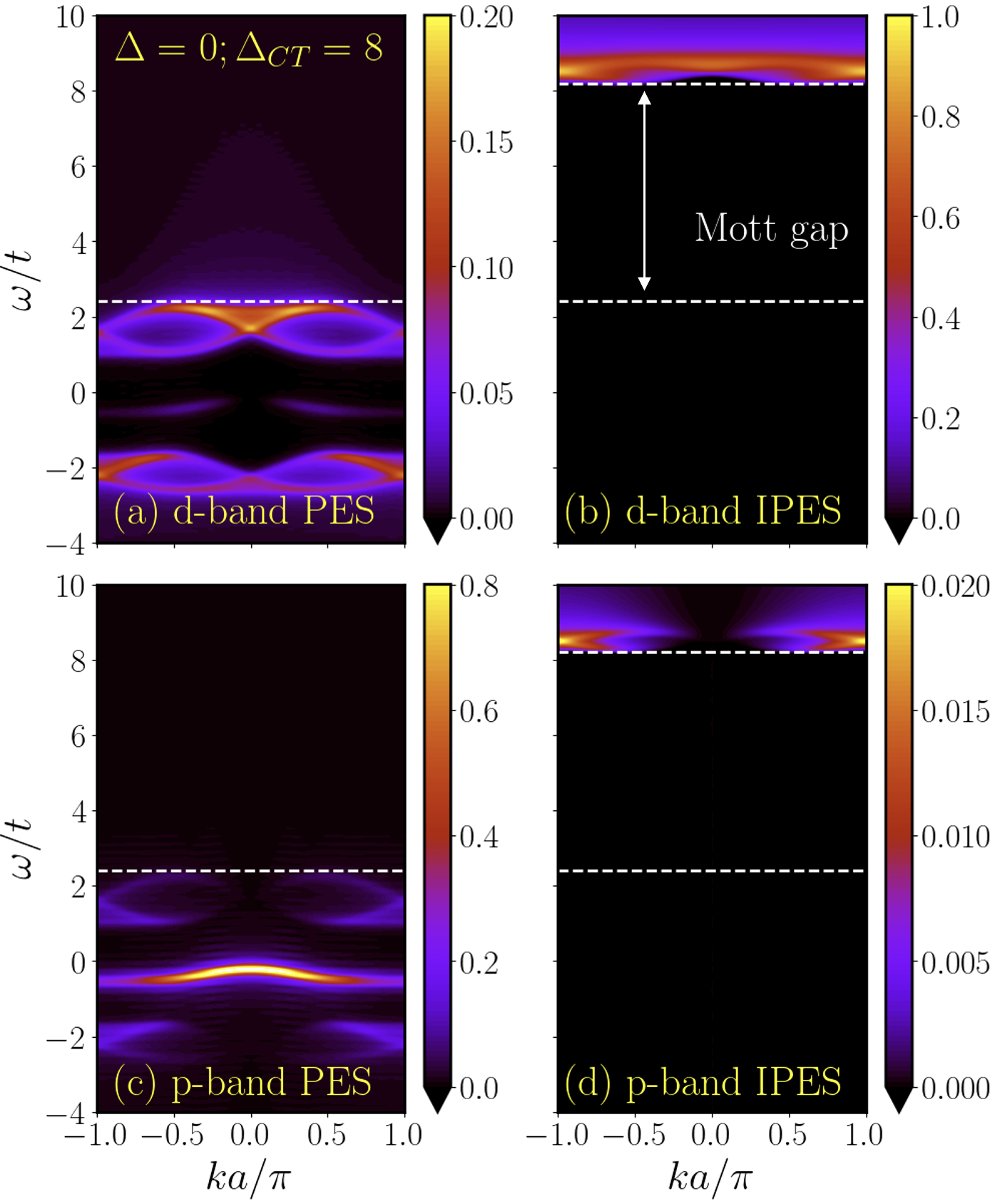}
	\caption{Photoemission (electron removal --PES) and inverse photoemission (electron addition -- IPES) spectra for the ligand (a);(b) $d$ orbitals and (c);(d) $p$ orbitals for fixed $U=8t$. The density is fixed at $N_e=3L+2$, where $L=64$ is the number of unit cells} \label{fig:delta0}
\end{figure}

Adding an electron produces a filled cluster described by the eigenstate $\ket{d^2\ligand^0} = \ket{2,2,2}$. Conversely, the eigenstates of two-hole cluster in the $S^z_\text{tot}=0$ sector are spanned by the basis
\begin{equation}
\begin{split}
    \ket{d^0\underbar{L}^0}&=\ket{2,0,2}, \\
    \ket{\overline{2\uparrow\downarrow}\pm}&=\tfrac{1}{\sqrt{2}}\left[\ket{2,\uparrow,\downarrow} \mp \ket{\downarrow,\uparrow, 2}\right], \\  
    \ket{\overline{2\downarrow\uparrow}\pm}&=\tfrac{1}{\sqrt{2}}\left[\ket{2,\downarrow,\uparrow} \mp \ket{\uparrow,\downarrow, 2}\right], \\
    \ket{\overline{\downarrow 2\uparrow}\pm}&=\tfrac{1}{\sqrt{2}}\left[\ket{\downarrow,2,\uparrow} \pm \ket{\uparrow,2,\downarrow}\right],~\text{and} \\
    \ket{\overline{220}\pm}&=\tfrac{1}{\sqrt{2}}\left[\ket{2,2,0} \pm \ket{0,2,2}\right].
\end{split}
\end{equation}
The states $\ket{\overline{2\uparrow\downarrow}\pm}$ and $\ket{\overline{2\downarrow\uparrow}\pm}$ are related by time reversal symmetry. The Hamiltonians in the symmetric and antisymmetric channels are $5\times 5$ and $4 \times 4$ matrices (their explicit form is given in  App.~\ref{sec:Happendix}), which can be diagonalized numerically. The excited state spectrum is shown in Fig.~\ref{fig:cluster} for $U = 8t$. Here, the electron removal spectrum is obtained from the energy difference $E_0(5)-E_n(4)$ while the electron addition spectrum is obtained from $E_n(6)-E_0(5)$, where $E_n(N_e)$ denotes the eigenenergy of the cluster with $N_e$ electrons. 

\begin{figure*}
	\centering
   \includegraphics[width=1.0\textwidth]{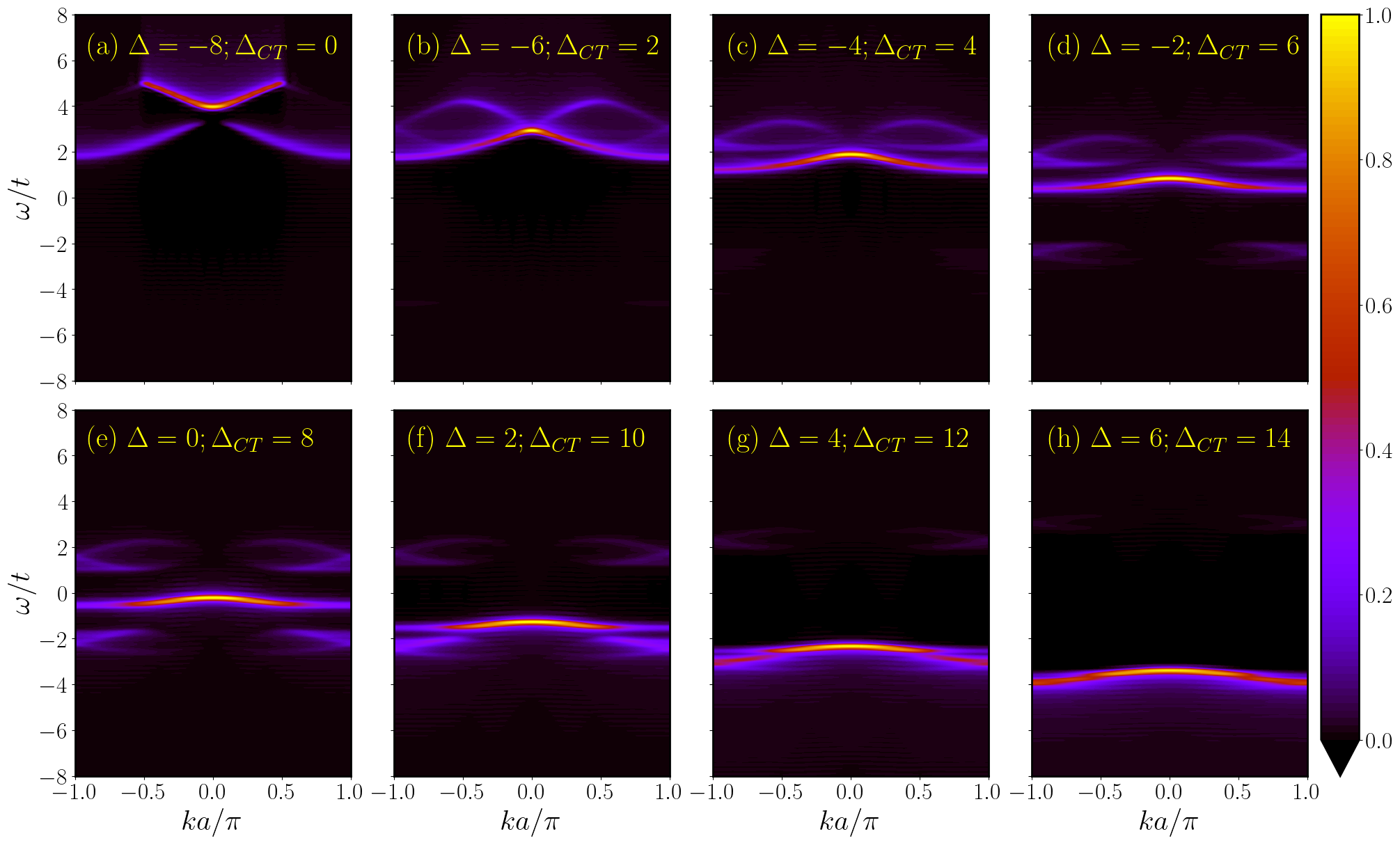}
	\caption{Photoemission (electron removal) spectra  for the ligand $p$ orbitals for $U=8$ as the system crosses from the Mott-Hubbard regime to the \gls*{CTI} and \gls*{NCTI} regimes for decreasing energy splitting $\Delta$. The density is fixed at $N_e=3L+2$, where $L=64$ is the number of unit cells (see text).} \label{fig:arpes_p}
\end{figure*}

\begin{figure*}
	\centering
   \includegraphics[width=1.0\textwidth]{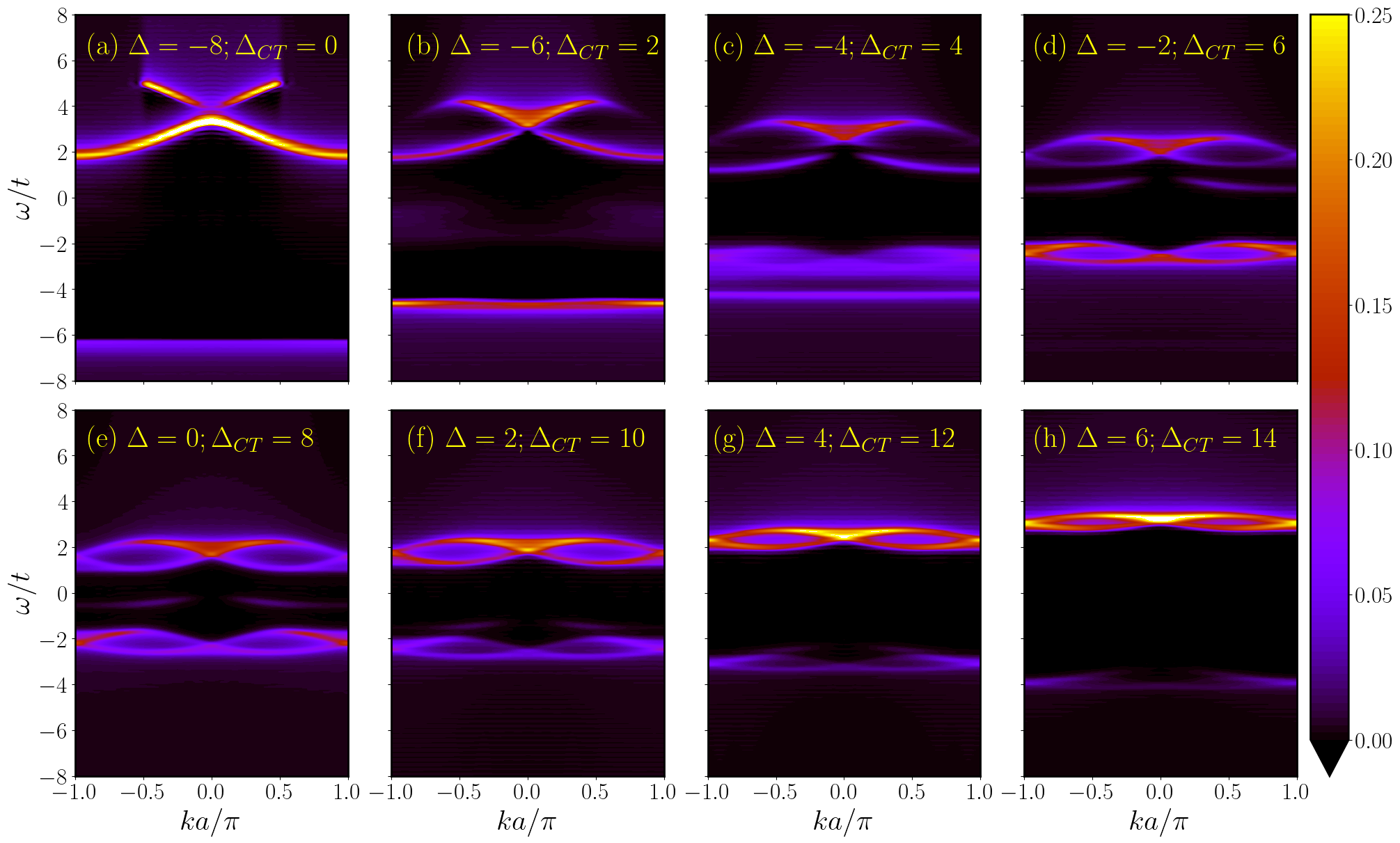}
	\caption{Same as Fig.~\ref{fig:arpes_p} but for the $d$ orbitals. } \label{fig:arpes_d}
\end{figure*}
 
The two-hole eigenstates in the $S^z_\text{tot} = 0$ sector can be written in a general form  
\begin{equation}
\begin{split}
\ket{\Psi}=\alpha\ket{d^0\underbar{L}^0}&+  \tfrac{\beta}{\sqrt{2}}\left[\ket{\overline{2\uparrow\downarrow}\pm}\pm
\ket{\overline{2\downarrow\uparrow}\pm}\right] \\
&+\delta \ket{\overline{\downarrow2\uparrow}\pm}
+\gamma\ket{\overline{220}\pm}.
\end{split}
\end{equation}
The state equivalent to the \gls*{ZRS} in this model is 
\begin{equation}
\ket{\text{ZRS}+}=\alpha\ket{d^0\underbar{L}^0}+\tfrac{\beta}{\sqrt{2}}\left[
\ket{\overline{2\uparrow\downarrow}+}-\ket{\overline{2\downarrow\uparrow}+}\right],
\label{ZRS}
\end{equation}
which is symmetric under reflections about the $d$ orbital. Notice that the additional hole is pure of $p$~($d$) character if $\alpha$~($\beta$) is zero but has mixed character in general. We can also define a state similar to the \gls*{ZRS} but using the antibonding combination of the $p$-states hybridized with the $d$ orbital
\begin{equation}
\ket{\text{ZRS}+^*}=\beta\ket{d^0\underbar{L}^0}-\tfrac{\alpha}{\sqrt{2}}\left[
\ket{\overline{2\uparrow\downarrow}+}-\ket{\overline{2\downarrow\uparrow}+}\right].
\label{ZRS2}
\end{equation}
These two states undergo an avoided level crossing at $\Delta\sim 0$ ($\Delta_\mathrm{CT} = U$). In addition to the \gls*{ZRS} excitations, we can define an antisymmetric counterpart $\ket{\text{ZRS}-}$ that does not include the $\ket{d^0\underbar{L}^0}$ state, as well as Zhang-Rice triplets $\ket{\text{ZRT}\pm }$ with the added hole completely localized in the $p$ orbital.

Figure~\ref{fig:cluster} shows results for the cluster's electron addition and removal eigenstate spectrum for $U = 8t$. 
The vertical dashed lines indicate the approximate transitions between the Mott-Hubbard, \gls*{CTI}, and \gls*{NCTI} regimes. (Note: the locations of these divisions are a bit arbitrary since they denote crossover points rather than phase transitions.) The labels on the top of each region highlight the dominant character of the half-filled cluster's ground state. The dashed curve follows the energy of the $\ket{d^2\ligand^0}$ state, which we associate with the upper Hubbard band. In the Mott-Hubbard regime ($\Delta > 0,~\Delta_\mathrm{CT} > U$), the additional doped hole has predominantly $d$ character. Conversely, 
in the \gls*{CTI} regime ($0 > \Delta > -U,~U>\Delta_\mathrm{CT} > 0$), the doped hole has predominantly $p$ character. In the \gls*{NCTI} regime ($\Delta < -U$, $\Delta_\mathrm{CT} < 0$), the lowest energy states will develop important additional weight in the states involving a double occupied $d$ orbital and two $p$ holes, i.e. $\ket{\overline{220}+}$ and $\ket{\overline{\downarrow 2 \uparrow}+}$~\cite{Green2024negative}. 
The lowest energy hole excitation, which we associate with the highest energy state in Fig.~\ref{fig:cluster}, is $|\text{ZRS}+\rangle$. It is followed by the $|\text{ZRS}-\rangle$ state, the Zhang-Rice triplets $|\text{ZRT}\pm \rangle$, and the antibonding $|\text{ZRS}+*\rangle$ states. The lower band can be associated with excitations with dominant $d$ characters.  

When contrasting with the \gls*{ZSA} picture, we first notice that the lower $d$ band splits into symmetric and antisymmetric bands. One 
usually projects out the antisymmetric band in the Zhang-Rice treatment of the problem since they are high energy states~\cite{ZR}. The symmetric band can be associated with the \gls*{ZRS} band in the multi-orbital problem, and it always appears above the triplet $p$ band. The most relevant parameter to characterize the problem is the splitting between the triplet $p$ states and the \gls*{ZRS} band, which gets smaller as $\Delta$ becomes more and more negative. When $\Delta \lesssim -U$, the system enters the \gls*{NCTI} regime, and the triplet $p$, \gls*{ZRS}, and \gls*{UHB} states all mix, and the system becomes metallic.

\subsection{Electron addition and removal spectra}\label{sec:results_arpes}
Having examined a simplified energy level diagram for our model, we now turn our attention to the electron addition and removal spectra, focusing first on the case with $U=8t$ and $\Delta = 0$ ($\Delta_\mathrm{CT} = 8t)$, as shown in  Fig.~\ref{fig:delta0}. This case corresponds to the strongly covalent or mixed Mott-Hubbard/\gls*{CTI} regime. We remind the reader that we use electron language in the following discussion, so the \gls*{UHB} (\gls*{LHB}) corresponds to states with two (zero) electrons occupying a $d$ orbital. 

In the half-filled, noninteracting case, the system's electronic structure consists of bonding and antibonding $pd$ bands [see Eq.~\eqref{eq:noninteracting_bands}] with the Fermi level located in the center of the antibonding states. Introducing a large $U$ redistributes the spectral weight into Hubbard and Zhang-Rice bands. 
The \gls*{UHB} is apparent in the electron addition spectra, appearing in Figs.~\ref{fig:delta0}(b) and (d) as a band of states centered at $\omega \approx U$. 
In this case, the \gls*{UHB}'s spectral weight has a majority $d$ character with a small admixture of $p$ character. [Note the color bar scale change in panel (d).] 
We can also resolve three additional bands below the Fermi level. The highest and lowest-lying bands are the Zhang-Rice singlet states derived from the bonding and antibonding \gls*{ZRS} states $\ket{\text{ZRS}+}$ and  $\ket{\text{ZRS}+*}$, respectively. These states have majority $d$ character at $k = 0$ but acquire $p$ character at the zone boundary. This weight reflects the fact that one cannot construct long wavelength Bloch states using the molecular $p$ orbitals of the form $L^\dagger = \frac{1}{\sqrt{2}}[p^\dagger_{i+1,\sigma}-p^\dagger_{i-1,\sigma}]$. The narrow band of states appearing just below $\omega = 0$ is derived from the Zhang-Rice triplet states; it has majority $p$ character because the $\ket{\text{ZRT}\pm}$ state places the additional hole entirely on the $p$ orbitals. 

\begin{figure*}
	\centering
   \includegraphics[width=\textwidth]{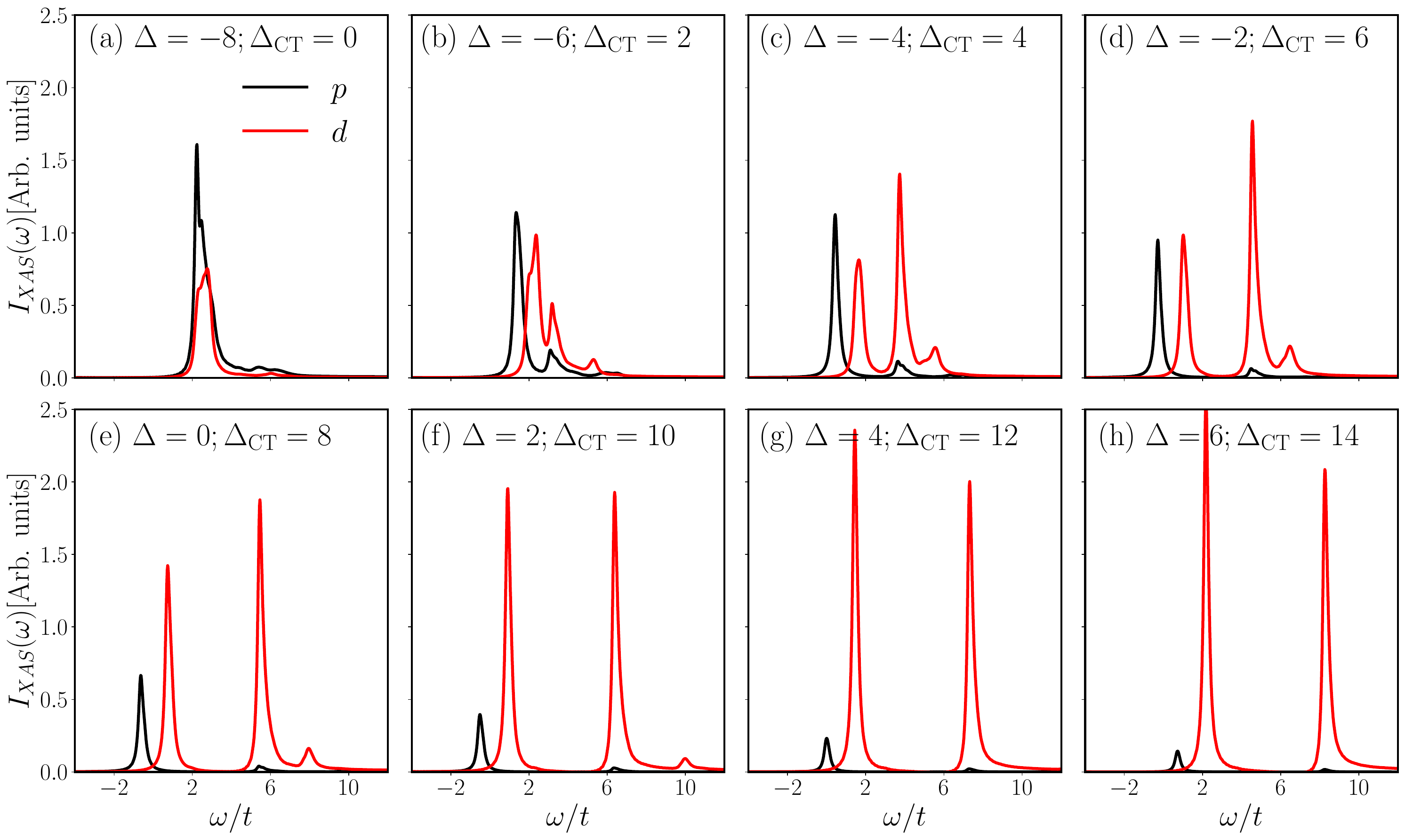}
	\caption{XAS spectra of an Emery chain for different values of $\Delta$. The spectra were calculated for small hole doping ($L=16$; $N_e=44$), fixed $U=8t$, and core-hole potential $U_c=-1.5t$.}\label{fig:xas}
\end{figure*}

Due to the \gls*{1D} geometry, our interacting system cannot be adiabatically connected to a Fermi liquid. Thus, many of the bands display clear signatures of spin-charge separation, with spinon branches in the $d$ bands that form an arc between $\pm k_\mathrm{F}$ and wider holon bands corresponding to the charge excitations~\cite{Ogata1990, Penc1997b, Benthien2007, penc, Feiguin2023}. The bandwidth of the \gls*{UHB} is smaller than the bandwidth of the \gls*{ZRS}-derived states because electrons have to hop over the $p$-$d$-$p$ bridge to propagate, a higher order process involving the virtual double occupation of a $d$ orbital (the effective hopping is of the order $t^2/(U-\Delta)$ \cite{penc}). Notice that the three- or four-band models studied previously include $p$-$p$ hopping processes that allow electrons to bypass the $d$ orbital.

Next, we vary the degree of the hybridization by adjusting $\Delta$. Figs.~\ref{fig:arpes_p} and \ref{fig:arpes_d} plot results for the electron removal (photoemission spectrum) as a function of $\Delta$ while again fixing $U=8t$. With increasing $\Delta > 0$, the splitting between bonding $\ket{\text{ZRS}+}$ and antibonding $\ket{\text{ZRS}+*}$ bands increases while the $\ket{\text{ZRT}\pm}$ band moves downward in energy. For $\Delta = 6t$ ($\Delta_\mathrm{CT} = 14t$), the system is in the Mott insulating regime and two bands strongly overlap, while 
the $d$ spectral weight is mainly concentrated on the antibonding state at high energy (see Fig.~\ref{fig:arpes_d}). 
In this regime, one can project out the $p$ degrees of freedom in the low-energy sector to obtain an effective single-band Hubbard Hamiltonian~\cite{Zaanen1985, Zaanen1988}. 
For $-U < \Delta < 0$ ($0<\Delta_\mathrm{CT}<U$), the system is in the \gls*{CTI} regime. In this case, the $\ket{\text{ZRT}}$ states moves upward in energy with decreasing $\Delta$ and eventually overlap with the antibonding $\ket{\text{ZRS}+*}$ states. Between $\Delta=-6t$ and $\Delta=-8t$, Zhang-Rice and \gls*{UHB} states all mix, and the system becomes metallic~\cite{Mizokawa1994}. 

\subsection{XAS}\label{sec:results_xas}
Next, we examine the \gls*{XAS} spectra for our model to make contact with prior small cluster calculations. \gls*{XAS} is a single photon process in which an electron is excited from a core orbital into an empty or partially filled valence orbital~\cite{deGroot2008, Chantler2024xasprimer}. It thus probes the system's unoccupied density states but with corrections introduced by the interaction between the valence electrons and the core hole created by the absorption process. These corrections can sometimes be strong since the electrons will try to screen the core hole and form a bound state~\cite{deGroot2008, Chantler2024xasprimer, Nocera2023}. 

We calculate the \gls*{XAS} spectra using the time-dependent formalism described in Ref.~\onlinecite{Zawadzki2023timedependent}. In this case, the spectrum is obtained from the imaginary part of the Fourier transform of a single-particle time-dependent correlation function $\langle c_0 \exp{(-\mathrm{i}\mathcal{H}_ct)} c^\dagger_0\rangle$, where the operators act on a reference site $i=0$, which could be either a $p$ or $d$ orbital, depending on the absorption edge that is used in the experiment~\cite{deGroot2008, Chantler2024xasprimer}. The Hamiltonian $\mathcal{H}_c$ in this expression includes the interaction between the core-hole and the valence band electrons. Assuming that the core hole is immobile, we model this interaction using an on-site atomic potential such that 
\begin{equation}
\mathcal{H}_c = H + U_c \sum_\sigma c^\dagger_{0,\sigma}c^{\phantom\dagger}_{0,\sigma},
\end{equation}
where $H$ is the original two-band Hamiltonian given in Eqs.~\eqref{eq:H}-\eqref{eq:HU} and $U_c$ is the attractive potential created by the core hole. The rest of the calculation steps are identical to those used for the photoemission spectrum. We refer the reader to Ref~\cite{Zawadzki2023timedependent} for further details. 

To focus on results that can help intuitively interpret experiments conducted on doped cuprate chains, we consider $L=16$ and $N_e=44$, corresponding to a slightly hole-doped case. The \gls*{XAS} spectra for fixed $U=8t$ are shown in Fig.~\ref{fig:xas} for several values of $\Delta$. Here, we present results for both $K$-edge ($p$ ) and $L$-edge ($d$) experiments. In call cases, we fixed the core-hole potential to $U_c=-1.5t$. 

For a large positive $\Delta=6t$ [$\Delta_\mathrm{CT} = 14t > U$, see Fig.~\ref{fig:xas}(h)], the doped holes predominantly reside in the lower Hubbard $d$ band while the $p$ orbitals are nearly completely filled. This charge distribution Pauli blocks the core electron 
from being excited into the $p$ orbital, and the resulting $K$-edge spectrum is dramatically suppressed as a result. Instead, a core electron can be excited into an unoccupied $d$ orbital at the $L$-edge. In the doped system, these transitions can occur to an empty $d$ orbital or a half-filled one with the opposite spin orientation. The corresponding \gls*{XAS} spectra thus consist of two resonances corresponding to the two possible final state configurations, which appear at $\omega \approx \Delta/2+U_c$ and $\Delta/2+U+2U_c$. 
In this regime, the system behaves qualitatively similar to the atomic limit. 

The \gls*{XAS} spectra preserve these features as $\Delta$ decreases, 
until crossing into the \gls*{CTI} regime for $\Delta \le 0$ ($\Delta_\mathrm{CT} \le U$). The $p$ and $d$ states become strongly hybridized in this regime, and the spectra develop new features. The first is the secondary peak in the $d$ spectrum at energy $\sim \Delta/2+U+U_c$, which corresponds to the ``poor-screened'' resonance in which the core electron is excited into a delocalized state, leaving the $d$ orbital at the site where the core hole is created half-filled. The second is the increased intensity of the $p$ resonance, which eventually merges with the lower $p$ resonance at sufficiently small $\Delta$. These results show that the doped holes have predominant $d$ character in the Mott-Hubbard regime, while they acquire a mixed $d$-$p$ character in the \gls*{CTI} regime. We can thus associate the lower $d$ resonance to the Zhang-Rice state in the latter case.  

\begin{figure*}[ht]
	\centering
   \includegraphics[width=0.8\textwidth]{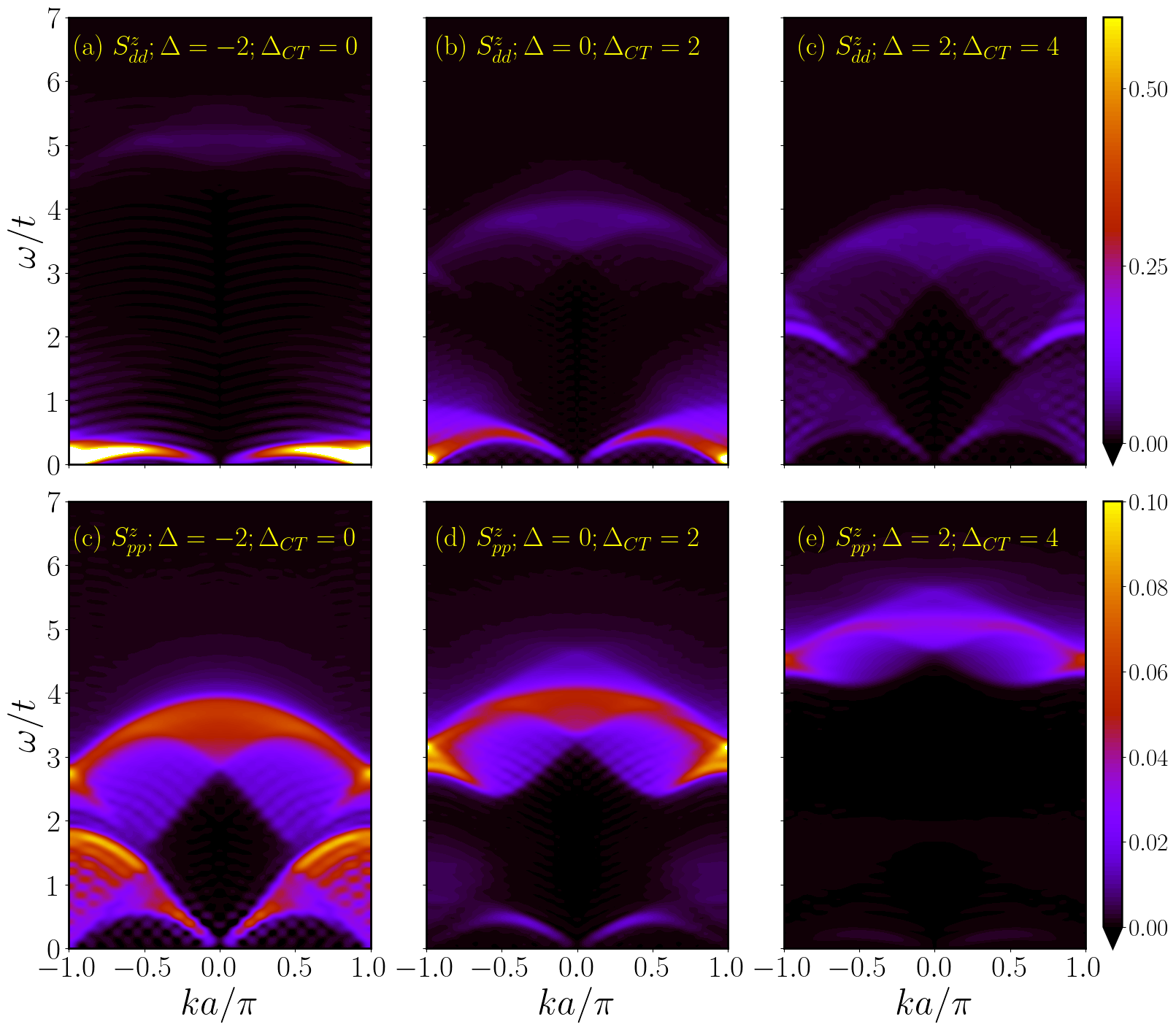}
	\caption{Spin dynamic structure factor $S^z_{\alpha,\alpha}(k,\omega)$ obtained in $L=32$ site half-filed chains with $U = 2t$ and different values of $\Delta$ as indicated in each panel. The top row shows the projection for the $\alpha = d$ orbitals, while the bottom row shows the projection for the $\alpha = p$ orbitals.}\label{fig:sqw}
\end{figure*}

\begin{figure}
	\centering
   \includegraphics[width=\columnwidth]{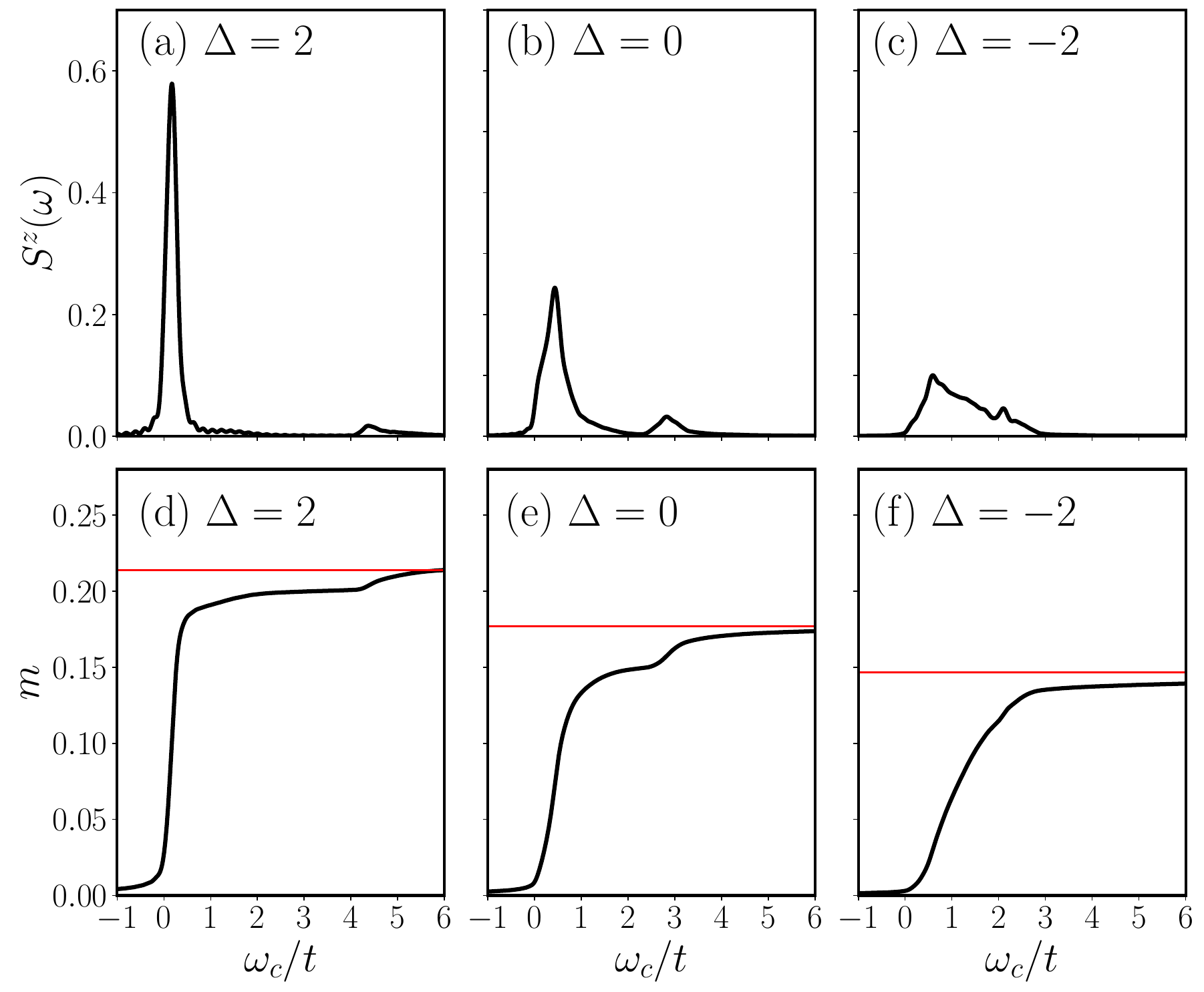}
	\caption{(a)-(c) Momentum integrated spin dynamic structure factor $S^z(\omega)$ for $U=2$ and different values of $\Delta$. Panels (d)-(f) show the corresponding integrated spectral weight as a function of the cut-off frequency $\omega_c$ [see Eq.~\eqref{eq:sum_rule}]. The horizontal lines show the value of the local spin moment calculated directly using Eq.~\eqref{eq:mtot}.}\label{fig:spin}
\end{figure} 

\subsection{Spin excitations}\label{sec:results_sqw}
Finally, we examine the model's spin excitations. This analysis is motivated by efforts to understand the so-called missing spectral weight problem in neutron scattering experiments conducted on low-dimensional cuprates~\cite{Bourges1997superexchange, Coldea2001spin, Lorenzana2005sum, Zaliznyak2004spinons, Walters2009effect, Li2021particle}.  

The dynamical spin structure factor $S(q,\omega)$ obeys a sum rule 
relating its integrated spectral weight to the total magnetic moment in the unit cell. For a \gls*{1D} system, can define the integrated spectral weight up to a cut-off $\omega_c$ 
\begin{equation}\label{eq:sum_rule}
    m(\omega_c) = \frac{3}{N} \sum_{\alpha,\beta,q} \int_0^\infty S^z_{\alpha,\beta}(q,\omega) e^{\mathrm{i}q(r_\alpha-r_\beta)}, 
\end{equation}
where $\alpha$ and $\beta$ are orbital indices in the unit cell, $r_\alpha$ is the basis vector for the orbital, and $S^z_{\alpha,\beta}(q,\omega)$ is the dynamical spin structure factor.  In the $\omega_c \rightarrow \infty$ limit, $m(\omega_c)$ should recover a value set by the total local moment in the unit cell. In particular, for our model 
\begin{equation}
    \lim_{\omega_c \rightarrow \infty} m(\omega_c) = \langle (S_d^z)^2\rangle+\langle (S^z_p)^2 \rangle + 2\langle S^z_pS^z_d \rangle, 
\label{eq:mtot}
\end{equation}
where the expectation values are evaluated using the two-orbital unit cell. (In practice, we use the unit cell at the center of the chain.)

Early experiments on cuprate spin chains~\cite{Zaliznyak2004spinons} and \gls*{2D} high-T$_\mathrm{c}$ superconducting cuprates~\cite{Bourges1997superexchange, Coldea2001spin, Lorenzana2005sum} found that integrating the observed low-energy spectral weight ($\omega_c \lesssim 0.6$ eV) often recovered less than a third of the total expected weight for a Cu$^{2+}$ ion, even after accounting for quantum fluctuations. It was later recognized that Cu-O hybridization effects play a key role here. For example, $\approx 80\%$ of the sum rule is recovered from \gls*{INS} measurements on Sr$_2$CuO$_3$ once the Cu-O form factor is taken into account~\cite{Walters2009effect}. A subsequent \gls*{DQMC}/\gls*{DMRG} study of multi-orbital $pd$-model for the corner-shared cuprates~\cite{Li2021particle} showed that remaining weight transfers to high-energy antibonding oxygen states. Notably, no such transfer occurs for similar simulations for the single-band Hubbard model~\cite{Li2021particle}. 

Motivated by this, we now explore the evolution of the magnetic excitation spectrum and resulting deviations from the expected sum rule for our model as the degree of $p$-$d$ hybridization varies. In general, the elementary magnetic excitations of our \gls*{1D} chains are spinons. While the system may exhibit short-ranged or algebraically decaying spin-spin correlations, it cannot develop true long-range order and coherent magnon excitations. Since spinons are domain wall-like excitations, they are created in pairs when the system is probed with scattering techniques like \gls*{INS} or \gls*{RIXS}. Hence, the resulting magnetic excitation spectrum displays a continuum of two-spinon excitations, characteristic of \gls*{1D} spin chains. 

The present case adds a new ingredient: the charge-transfer energy $\Delta_\mathrm{CT}$. In our model, the spin-spin interactions are mediated by superexchange mechanisms through the ligand $p$ orbital and involve an intermediate process in which a $p$ electron virtually double occupies a $d$ state. This process is significantly suppressed for large $\Delta_\mathrm{CT}$ or $U$, resulting in a weak interaction between spins in this limit. For this reason, here we focus on a parameter regime with smaller $U=2$ and $\Delta$ to study the spin excitations, which produces a reasonably large band width for the spinon excitations. Specifically, we calculate the orbital-resolved dynamic spin structure factors $S^z_{\alpha\beta}(q,\omega)$ for $\alpha=\beta = p,d$ on $L=32$ chains. 

The results for various values of $\Delta$ spanning from the strongly covalent $(\Delta=-2t,~\Delta_\text{CT} = 0)$ to the mixed Mott/\gls*{CTI} $(\Delta=0,~\Delta_\text{CT} = 2t = U)$ to the Mott-Hubbard $(\Delta=2t,~\Delta_\text{CT} = 4t > U)$ regimes are shown in Fig.~\ref{fig:sqw}. In all cases, spin excitations are dominated by contributions from the $d$ orbitals, consistent with most of the magnetic moment residing on those orbitals. We further observe a nonnegligible amount of spectral weight associated with the $p$ orbitals, which is concentrated at higher energies. Varying the value of $\Delta$ affects the spectra in several notable ways. For example, increasing $\Delta$ increases the bandwidth of the excitations derived from the $d$ states. It also transfers spectral weight from the low-energy $d$-derived excitations to the higher energy $p$-derived ones while causing the former to shift to lower energies. These effects are consistent with 
expectations for increased hybridization between the $d$ and $p$ orbitals. 


Figure~\ref{fig:spin} further quantifies the redistribution of spectral weight of the magnetic excitations. Here the top row plots the momentum-integrated dynamical spin structure factor summed over all components $S^z(\omega) = \frac{1}{N}\sum_{\alpha,\beta,q}S_{\alpha\beta}^z(q,\omega)$, while the bottom panels show the corresponding $m(\omega_c)$ as a function of the energy cut-off. The red lines in the bottom panels indicate the average value of the moment squared, as obtained from Eq.~\eqref{eq:mtot}. The results show that the total moment is distributed in energy, with some weight at very high energies. In particular, one must integrate to very high energy for $\Delta_\mathrm{CT} \ge U$ to recover the total expected sum rule. These results are consistent with a prior study that focused on the $\Delta > 0$ case~\cite{Li2021particle}, but derived here for different regimes of the \gls*{ZSA} classification scheme. 

\section{Summary \& Conclusions}\label{sec:conclusions}
We have examined the ground and excited state properties of a \gls*{1D} two-orbital model analogous to the $pd$-models used to describe charge-transfer \glspl*{TMO}. By focusing on a \gls*{1D} system, we could calculate the model's excited state properties with high momentum resolution in different regimes of the \gls*{ZSA} classification scheme. In doing so, we provided predictions for the ground-state character, single-particle spectral function, \gls*{XAS} spectra, and dynamical spin structure factor for parameters spanning from the \gls*{NCTI} to \gls*{CTI} to Mott-Hubbard insulator regimes. Our results for the \gls*{XAS} spectra are broadly consistent with previous results obtained using small cluster methods~\cite{Eskes1990, Eskes1991anomalous, Green2024negative, Bisogni2016groundstate}. However, our results for the other dynamical correlation functions extend well beyond prior work focusing on a single parameter set~\cite{Li2021particle} or previous mean-field treatments of the problem~\cite{Mizokawa2000spin, Johnston2014}. 

One important aspect of our results is the predicted evolution of the magnetic sum rule. As mentioned in the introduction, considerable effort is often expended to situate newly discovered \glspl*{TMO} like the superconducting nickelates within the \gls*{ZSA} scheme. Based on our results shown in Fig.~\ref{fig:spin}, we propose that the amount of missing local moment obtained from integrating the low-energy spin excitations could be used to quantify the degree of TM-O hybridization present in the material and infer the relative values of the charge-transfer energy and Hubbard interaction strength. \gls*{INS} experiments, corrected for the appropriate atomic form factor~\cite{walters}, is ideally suited for such experiments. Analogous experiments with \gls*{RIXS} would be complicated by the fact that the measured spectra are not trivially related to $S(q,w)$~\cite{Jia2016using} with spectral weight that can be modified by other factors like electron-phonon coupling~\cite{Thomas2025theory}. The predicted transfer of spectral weight to high-energy also has important implications for experimental attempts to measure entanglement witnesses like the quantum Fisher information, which often rely on particular sum rules~\cite{Laurell2022magnetic, Scheie2025tutorial}.

Moving forward, it would be interesting to dope our model and study how the canonical picture of anomalous spectral weight transfer in a doping a Mott insulator \cite{Eskes1991anomalous} changes as in the \gls*{CTI} and \gls*{NCTI} regimes. Such calculations can also be compared directly to data taken for the recently synthesized doped cuprate spin chains~\cite{chen} and ladders~\cite{Padma2025beyond, scheie2025cooper}, where anomalous interactions have been inferred from comparing the observed spectra to effective singleband models. 

\section*{Data and Code Availability}
The code and data supporting this study will be deposited in a public online repository once the final version of the paper is accepted for publication. Until that time, both will be made available upon reasonable request. 

\acknowledgements
The authors acknowledge the support of the U.S. Department of Energy, Office of Basic Energy Sciences. This project was initiated under grant No.~DE-SC0014407 and completed under grant No.~DE-SC0022311. We thank A. A. Aligia, A. Nocera, G. A. Sawatzky, and A. de la Torre for illuminating discussions.

\appendix
\section{Cluster Calculations}\label{sec:Happendix}
\noindent{\bf Ground State}. At half-filling, the cluster's ground state can be written as a linear combination of the states 
\begin{equation}
    \ket{d^1\ligand^0}= p^\dagger_{L\uparrow}p^\dagger_{L\downarrow}d^\dagger_{\uparrow} p^\dagger_{R\uparrow}p^\dagger_{R\downarrow}\ket{\text{vac}}=\ket{2,\uparrow,2}
\end{equation}
and 
\begin{align}\nonumber
    \ket{d^2\ligand^1}&=\tfrac{1}{\sqrt{2}} \left[p^\dagger_{L,\uparrow}d^\dagger_\uparrow d^\dagger_\downarrow p^\dagger_{R\uparrow}p^\dagger_{R\downarrow} -p^\dagger_{L,\uparrow}p^\dagger_{L,\downarrow}d^\dagger_\uparrow d^\dagger_\downarrow p^\dagger_{R\uparrow}\right]\ket{\text{vac}}\\
    &=\tfrac{1}{\sqrt{2}}\left[\ket{\uparrow,2,2}-\ket{2,2,\uparrow}\right].
\end{align}
In these expressions, the operators $p^\dagger_{L/R,\sigma}$ create an electron with spin $\sigma$ on the left (L)/right (R) $p$ orbitals of the cluster, while $d^\dagger_\sigma$ does the same on the central $d$ orbital. We only introduce this notation here to define the normal ordering of the operators. In this sector, the Hamiltonian is represented by the $2\times2$ matrix
\begin{equation}
H_0 = \begin{pmatrix}
\Delta & \sqrt{2}t\\
\sqrt{2}t & U
\end{pmatrix},
\end{equation}
yielding the ground state energy
\begin{equation}
E_0=\left(\frac{\Delta+U}{2}\right) -\sqrt{\left(\frac{\Delta-U}{2}\right)^2-2t^2}.
\end{equation}\\

\noindent{\bf Excitations with one hole}. 
The Fock space with one additional hole and $S^z_\text{tot}=0$ is spanned by the states
\begin{equation}
\begin{split}
    \ket{d^0\underbar{L}^0}&=\ket{2,0,2}, \\
     \ket{\overline{220}\pm}&=\tfrac{1}{\sqrt{2}}\left[\ket{2,2,0} \pm \ket{0,2,2}\right], \\  
    \ket{\overline{\downarrow2\uparrow}\pm}&=\tfrac{1}{\sqrt{2}}\left[\ket{\downarrow,2,\uparrow} \pm \ket{\uparrow,2,\downarrow}\right], \\     \ket{\overline{2\uparrow\downarrow}\pm}&=\tfrac{1}{\sqrt{2}}\left[\ket{2,\uparrow,\downarrow} \mp \ket{\downarrow,\uparrow, 2}\right], ~\text{and} \\     \ket{\overline{2\downarrow\uparrow}\pm}&=\tfrac{1}{\sqrt{2}}\left[\ket{2,\downarrow,\uparrow} \mp \ket{\uparrow,\downarrow, 2}\right], 
\end{split}
\end{equation}
where the $+(-)$ sign corresponds to the symmetric (antisymmetric) sector under reflections about the central site. The $\ket{d^0\underbar{L}^0}$ state is symmetric, and the states $ \ket{\overline{220}\pm}$ are triplet (+) or singlet (-) states. In addition, the linear combinations 
$\ket{\overline{2\uparrow\downarrow}\pm} \pm \ket{\overline{2\downarrow\uparrow}\pm}$ yield symmetric and anti-symmetric singlet (-) or triplet (-) wavefunctions.

The Hamiltonian matrices for the two sectors are:
\begin{equation}
H_+ = \begin{pmatrix}
    -2\Delta & 0 &0 &-\sqrt{2}t &\sqrt{2}t \\
                      0& U& 0& -t& t \\
                      0& 0& U& -t& t \\
                      -\sqrt{2}t& -t& -t& -\Delta & 0 \\
                      \sqrt{2}t& t& t& 0& -\Delta
\end{pmatrix}
\end{equation}
and 
\begin{equation}
H_- = \begin{pmatrix}
U& 0& t,& -t \\
0 & U& -t& -t\\
t& -t& -\Delta& 0\\
-t& -t& 0& -\Delta
\end{pmatrix},
\end{equation}
which can be solved numerically to obtain the eigenvalue spectrum plotted in Fig.~\ref{fig:cluster}.\\

\bibliography{references}

\begin{thebibliography}{134}%
\makeatletter
\providecommand \@ifxundefined [1]{%
 \@ifx{#1\undefined}
}%
\providecommand \@ifnum [1]{%
 \ifnum #1\expandafter \@firstoftwo
 \else \expandafter \@secondoftwo
 \fi
}%
\providecommand \@ifx [1]{%
 \ifx #1\expandafter \@firstoftwo
 \else \expandafter \@secondoftwo
 \fi
}%
\providecommand \natexlab [1]{#1}%
\providecommand \enquote  [1]{``#1''}%
\providecommand \bibnamefont  [1]{#1}%
\providecommand \bibfnamefont [1]{#1}%
\providecommand \citenamefont [1]{#1}%
\providecommand \href@noop [0]{\@secondoftwo}%
\providecommand \href [0]{\begingroup \@sanitize@url \@href}%
\providecommand \@href[1]{\@@startlink{#1}\@@href}%
\providecommand \@@href[1]{\endgroup#1\@@endlink}%
\providecommand \@sanitize@url [0]{\catcode `\\12\catcode `\$12\catcode
  `\&12\catcode `\#12\catcode `\^12\catcode `\_12\catcode `\%12\relax}%
\providecommand \@@startlink[1]{}%
\providecommand \@@endlink[0]{}%
\providecommand \url  [0]{\begingroup\@sanitize@url \@url }%
\providecommand \@url [1]{\endgroup\@href {#1}{\urlprefix }}%
\providecommand \urlprefix  [0]{URL }%
\providecommand \Eprint [0]{\href }%
\providecommand \doibase [0]{https://doi.org/}%
\providecommand \selectlanguage [0]{\@gobble}%
\providecommand \bibinfo  [0]{\@secondoftwo}%
\providecommand \bibfield  [0]{\@secondoftwo}%
\providecommand \translation [1]{[#1]}%
\providecommand \BibitemOpen [0]{}%
\providecommand \bibitemStop [0]{}%
\providecommand \bibitemNoStop [0]{.\EOS\space}%
\providecommand \EOS [0]{\spacefactor3000\relax}%
\providecommand \BibitemShut  [1]{\csname bibitem#1\endcsname}%
\let\auto@bib@innerbib\@empty
\bibitem [{\citenamefont {Rao}(1989)}]{Rao1989transition}%
  \BibitemOpen
  \bibfield  {author} {\bibinfo {author} {\bibfnamefont {C.~N.~R.}\
  \bibnamefont {Rao}},\ }\bibfield  {title} {\bibinfo {title} {Transition metal
  oxides},\ }\href
  {https://doi.org/https://doi.org/10.1146/annurev.pc.40.100189.001451}
  {\bibfield  {journal} {\bibinfo  {journal} {Annual Review of Physical
  Chemistry}\ }\textbf {\bibinfo {volume} {40}},\ \bibinfo {pages} {291}
  (\bibinfo {year} {1989})}\BibitemShut {NoStop}%
\bibitem [{\citenamefont {Imada}\ \emph {et~al.}(1998)\citenamefont {Imada},
  \citenamefont {Fujimori},\ and\ \citenamefont {Tokura}}]{Imada1998}%
  \BibitemOpen
  \bibfield  {author} {\bibinfo {author} {\bibfnamefont {M.}~\bibnamefont
  {Imada}}, \bibinfo {author} {\bibfnamefont {A.}~\bibnamefont {Fujimori}},\
  and\ \bibinfo {author} {\bibfnamefont {Y.}~\bibnamefont {Tokura}},\
  }\bibfield  {title} {\bibinfo {title} {Metal-insulator transitions},\ }\href
  {https://doi.org/10.1103/RevModPhys.70.1039} {\bibfield  {journal} {\bibinfo
  {journal} {Rev. Mod. Phys.}\ }\textbf {\bibinfo {volume} {70}},\ \bibinfo
  {pages} {1039} (\bibinfo {year} {1998})}\BibitemShut {NoStop}%
\bibitem [{\citenamefont {Ngai}\ \emph {et~al.}(2014)\citenamefont {Ngai},
  \citenamefont {Walker},\ and\ \citenamefont {Ahn}}]{Ngai2014correlated}%
  \BibitemOpen
  \bibfield  {author} {\bibinfo {author} {\bibfnamefont {J.}~\bibnamefont
  {Ngai}}, \bibinfo {author} {\bibfnamefont {F.}~\bibnamefont {Walker}},\ and\
  \bibinfo {author} {\bibfnamefont {C.}~\bibnamefont {Ahn}},\ }\bibfield
  {title} {\bibinfo {title} {Correlated oxide physics and electronics},\ }\href
  {https://doi.org/https://doi.org/10.1146/annurev-matsci-070813-113248}
  {\bibfield  {journal} {\bibinfo  {journal} {Annual Review of Materials
  Research}\ }\textbf {\bibinfo {volume} {44}},\ \bibinfo {pages} {1} (\bibinfo
  {year} {2014})}\BibitemShut {NoStop}%
\bibitem [{\citenamefont {Chakhalian}\ \emph {et~al.}(2014)\citenamefont
  {Chakhalian}, \citenamefont {Freeland}, \citenamefont {Millis}, \citenamefont
  {Panagopoulos},\ and\ \citenamefont {Rondinelli}}]{Chakhalian2014colloquium}%
  \BibitemOpen
  \bibfield  {author} {\bibinfo {author} {\bibfnamefont {J.}~\bibnamefont
  {Chakhalian}}, \bibinfo {author} {\bibfnamefont {J.~W.}\ \bibnamefont
  {Freeland}}, \bibinfo {author} {\bibfnamefont {A.~J.}\ \bibnamefont
  {Millis}}, \bibinfo {author} {\bibfnamefont {C.}~\bibnamefont
  {Panagopoulos}},\ and\ \bibinfo {author} {\bibfnamefont {J.~M.}\ \bibnamefont
  {Rondinelli}},\ }\bibfield  {title} {\bibinfo {title} {Colloquium: Emergent
  properties in plane view: Strong correlations at oxide interfaces},\ }\href
  {https://doi.org/10.1103/RevModPhys.86.1189} {\bibfield  {journal} {\bibinfo
  {journal} {Rev. Mod. Phys.}\ }\textbf {\bibinfo {volume} {86}},\ \bibinfo
  {pages} {1189} (\bibinfo {year} {2014})}\BibitemShut {NoStop}%
\bibitem [{\citenamefont {Zaanen}\ \emph {et~al.}(1985)\citenamefont {Zaanen},
  \citenamefont {Sawatzky},\ and\ \citenamefont {Allen}}]{Zaanen1985}%
  \BibitemOpen
  \bibfield  {author} {\bibinfo {author} {\bibfnamefont {J.}~\bibnamefont
  {Zaanen}}, \bibinfo {author} {\bibfnamefont {G.~A.}\ \bibnamefont
  {Sawatzky}},\ and\ \bibinfo {author} {\bibfnamefont {J.~W.}\ \bibnamefont
  {Allen}},\ }\bibfield  {title} {\bibinfo {title} {Band gaps and electronic
  structure of transition-metal compounds},\ }\href
  {https://doi.org/10.1103/PhysRevLett.55.418} {\bibfield  {journal} {\bibinfo
  {journal} {Phys. Rev. Lett.}\ }\textbf {\bibinfo {volume} {55}},\ \bibinfo
  {pages} {418} (\bibinfo {year} {1985})}\BibitemShut {NoStop}%
\bibitem [{\citenamefont {Catalano}\ \emph {et~al.}(2018)\citenamefont
  {Catalano}, \citenamefont {Gibert}, \citenamefont {Fowlie}, \citenamefont
  {{\'I}{\~n}iguez}, \citenamefont {Triscone},\ and\ \citenamefont
  {Kreisel}}]{Catalano2018}%
  \BibitemOpen
  \bibfield  {author} {\bibinfo {author} {\bibfnamefont {S.}~\bibnamefont
  {Catalano}}, \bibinfo {author} {\bibfnamefont {M.}~\bibnamefont {Gibert}},
  \bibinfo {author} {\bibfnamefont {J.}~\bibnamefont {Fowlie}}, \bibinfo
  {author} {\bibfnamefont {J.}~\bibnamefont {{\'I}{\~n}iguez}}, \bibinfo
  {author} {\bibfnamefont {J.-M.}\ \bibnamefont {Triscone}},\ and\ \bibinfo
  {author} {\bibfnamefont {J.}~\bibnamefont {Kreisel}},\ }\bibfield  {title}
  {\bibinfo {title} {Rare-earth nickelates {RNiO$_3$}: thin films and
  heterostructures},\ }\href {https://doi.org/10.1088/1361-6633/aaa37a}
  {\bibfield  {journal} {\bibinfo  {journal} {Reports on Progress in Physics}\
  }\textbf {\bibinfo {volume} {81}},\ \bibinfo {pages} {046501} (\bibinfo
  {year} {2018})}\BibitemShut {NoStop}%
\bibitem [{\citenamefont {Bibes}\ \emph {et~al.}(2011)\citenamefont {Bibes},
  \citenamefont {Villegas},\ and\ \citenamefont
  {Barth{\'e}l{\'e}my}}]{Bibes2011}%
  \BibitemOpen
  \bibfield  {author} {\bibinfo {author} {\bibfnamefont {M.}~\bibnamefont
  {Bibes}}, \bibinfo {author} {\bibfnamefont {J.~E.}\ \bibnamefont
  {Villegas}},\ and\ \bibinfo {author} {\bibfnamefont {A.}~\bibnamefont
  {Barth{\'e}l{\'e}my}},\ }\bibfield  {title} {\bibinfo {title} {Ultrathin
  oxide films and interfaces for electronics and spintronics},\ }\href
  {https://doi.org/10.1080/00018732.2010.534865} {\bibfield  {journal}
  {\bibinfo  {journal} {Advances in Physics}\ }\textbf {\bibinfo {volume}
  {60}},\ \bibinfo {pages} {5} (\bibinfo {year} {2011})}\BibitemShut {NoStop}%
\bibitem [{\citenamefont {Hwang}\ \emph {et~al.}(2012)\citenamefont {Hwang},
  \citenamefont {Iwasa}, \citenamefont {Kawasaki}, \citenamefont {Keimer},
  \citenamefont {Nagaosa},\ and\ \citenamefont {Tokura}}]{Hwang2012}%
  \BibitemOpen
  \bibfield  {author} {\bibinfo {author} {\bibfnamefont {H.~Y.}\ \bibnamefont
  {Hwang}}, \bibinfo {author} {\bibfnamefont {Y.}~\bibnamefont {Iwasa}},
  \bibinfo {author} {\bibfnamefont {M.}~\bibnamefont {Kawasaki}}, \bibinfo
  {author} {\bibfnamefont {B.}~\bibnamefont {Keimer}}, \bibinfo {author}
  {\bibfnamefont {N.}~\bibnamefont {Nagaosa}},\ and\ \bibinfo {author}
  {\bibfnamefont {Y.}~\bibnamefont {Tokura}},\ }\bibfield  {title} {\bibinfo
  {title} {Emergent phenomena at oxide interfaces},\ }\href
  {https://doi.org/10.1038/nmat3223} {\bibfield  {journal} {\bibinfo  {journal}
  {Nature Materials}\ }\textbf {\bibinfo {volume} {11}},\ \bibinfo {pages}
  {103} (\bibinfo {year} {2012})}\BibitemShut {NoStop}%
\bibitem [{\citenamefont {Mannhart}\ and\ \citenamefont
  {Schlom}(2010)}]{Mannhart2010}%
  \BibitemOpen
  \bibfield  {author} {\bibinfo {author} {\bibfnamefont {J.}~\bibnamefont
  {Mannhart}}\ and\ \bibinfo {author} {\bibfnamefont {D.~G.}\ \bibnamefont
  {Schlom}},\ }\bibfield  {title} {\bibinfo {title} {Oxide interfaces—an
  opportunity for electronics},\ }\href
  {https://doi.org/10.1126/science.1181862} {\bibfield  {journal} {\bibinfo
  {journal} {Science}\ }\textbf {\bibinfo {volume} {327}},\ \bibinfo {pages}
  {1607} (\bibinfo {year} {2010})}\BibitemShut {NoStop}%
\bibitem [{\citenamefont {Chen}\ and\ \citenamefont {Millis}(2017)}]{Chen2017}%
  \BibitemOpen
  \bibfield  {author} {\bibinfo {author} {\bibfnamefont {H.}~\bibnamefont
  {Chen}}\ and\ \bibinfo {author} {\bibfnamefont {A.}~\bibnamefont {Millis}},\
  }\bibfield  {title} {\bibinfo {title} {Charge transfer driven emergent
  phenomena in oxide heterostructures},\ }\href
  {https://doi.org/10.1088/1361-648X/aa6efe} {\bibfield  {journal} {\bibinfo
  {journal} {Journal of Physics: Condensed Matter}\ }\textbf {\bibinfo {volume}
  {29}},\ \bibinfo {pages} {243001} (\bibinfo {year} {2017})}\BibitemShut
  {NoStop}%
\bibitem [{\citenamefont {Khomskii}(1997)}]{Khomskii}%
  \BibitemOpen
  \bibfield  {author} {\bibinfo {author} {\bibfnamefont {D.~I.}\ \bibnamefont
  {Khomskii}},\ }\bibfield  {title} {\bibinfo {title} {Unusual valence,
  negative charge-transfer gaps and self-doping in transition-metal
  compounds},\ }\href@noop {} {\bibfield  {journal} {\bibinfo  {journal}
  {Lithuanian J. Phys.}\ }\textbf {\bibinfo {volume} {37}},\ \bibinfo {pages}
  {65} (\bibinfo {year} {1997})},\ \bibinfo {note} {{Preprint} available at
  \href{https://arxiv.org/abs/cond-mat/0101164}{https://arxiv.org/abs/cond-mat/0101164}.}\BibitemShut
  {Stop}%
\bibitem [{\citenamefont {Green}\ and\ \citenamefont
  {Sawatzky}(2024)}]{Green2024negative}%
  \BibitemOpen
  \bibfield  {author} {\bibinfo {author} {\bibfnamefont {R.~J.}\ \bibnamefont
  {Green}}\ and\ \bibinfo {author} {\bibfnamefont {G.~A.}\ \bibnamefont
  {Sawatzky}},\ }\bibfield  {title} {\bibinfo {title} {Negative charge transfer
  energy in correlated compounds},\ }\href
  {https://doi.org/10.7566/JPSJ.93.121007} {\bibfield  {journal} {\bibinfo
  {journal} {Journal of the Physical Society of Japan}\ }\textbf {\bibinfo
  {volume} {93}},\ \bibinfo {pages} {121007} (\bibinfo {year}
  {2024})}\BibitemShut {NoStop}%
\bibitem [{\citenamefont {Mizokawa}\ \emph {et~al.}(1991)\citenamefont
  {Mizokawa}, \citenamefont {Namatame}, \citenamefont {Fujimori}, \citenamefont
  {Akeyama}, \citenamefont {Kondoh}, \citenamefont {Kuroda},\ and\
  \citenamefont {Kosugi}}]{Mizokawa1991origin}%
  \BibitemOpen
  \bibfield  {author} {\bibinfo {author} {\bibfnamefont {T.}~\bibnamefont
  {Mizokawa}}, \bibinfo {author} {\bibfnamefont {H.}~\bibnamefont {Namatame}},
  \bibinfo {author} {\bibfnamefont {A.}~\bibnamefont {Fujimori}}, \bibinfo
  {author} {\bibfnamefont {K.}~\bibnamefont {Akeyama}}, \bibinfo {author}
  {\bibfnamefont {H.}~\bibnamefont {Kondoh}}, \bibinfo {author} {\bibfnamefont
  {H.}~\bibnamefont {Kuroda}},\ and\ \bibinfo {author} {\bibfnamefont
  {N.}~\bibnamefont {Kosugi}},\ }\bibfield  {title} {\bibinfo {title} {Origin
  of the band gap in the negative charge-transfer-energy compound
  {$\mathrm{NaCuO}_{2}$}},\ }\href
  {https://doi.org/10.1103/PhysRevLett.67.1638} {\bibfield  {journal} {\bibinfo
   {journal} {Phys. Rev. Lett.}\ }\textbf {\bibinfo {volume} {67}},\ \bibinfo
  {pages} {1638} (\bibinfo {year} {1991})}\BibitemShut {NoStop}%
\bibitem [{\citenamefont {Bisogni}\ \emph {et~al.}(2016)\citenamefont
  {Bisogni}, \citenamefont {Catalano}, \citenamefont {Green}, \citenamefont
  {Gibert}, \citenamefont {Scherwitzl}, \citenamefont {Huang}, \citenamefont
  {Strocov}, \citenamefont {Zubko}, \citenamefont {Balandeh}, \citenamefont
  {Triscone}, \citenamefont {Sawatzky},\ and\ \citenamefont
  {Schmitt}}]{Bisogni2016groundstate}%
  \BibitemOpen
  \bibfield  {author} {\bibinfo {author} {\bibfnamefont {V.}~\bibnamefont
  {Bisogni}}, \bibinfo {author} {\bibfnamefont {S.}~\bibnamefont {Catalano}},
  \bibinfo {author} {\bibfnamefont {R.~J.}\ \bibnamefont {Green}}, \bibinfo
  {author} {\bibfnamefont {M.}~\bibnamefont {Gibert}}, \bibinfo {author}
  {\bibfnamefont {R.}~\bibnamefont {Scherwitzl}}, \bibinfo {author}
  {\bibfnamefont {Y.}~\bibnamefont {Huang}}, \bibinfo {author} {\bibfnamefont
  {V.~N.}\ \bibnamefont {Strocov}}, \bibinfo {author} {\bibfnamefont
  {P.}~\bibnamefont {Zubko}}, \bibinfo {author} {\bibfnamefont
  {S.}~\bibnamefont {Balandeh}}, \bibinfo {author} {\bibfnamefont {J.-M.}\
  \bibnamefont {Triscone}}, \bibinfo {author} {\bibfnamefont {G.}~\bibnamefont
  {Sawatzky}},\ and\ \bibinfo {author} {\bibfnamefont {T.}~\bibnamefont
  {Schmitt}},\ }\bibfield  {title} {\bibinfo {title} {Ground-state oxygen holes
  and the metal--insulator transition in the negative charge-transfer
  rare-earth nickelates},\ }\href {https://doi.org/10.1038/ncomms13017}
  {\bibfield  {journal} {\bibinfo  {journal} {Nature Communications}\ }\textbf
  {\bibinfo {volume} {7}},\ \bibinfo {pages} {13017} (\bibinfo {year}
  {2016})}\BibitemShut {NoStop}%
\bibitem [{\citenamefont {Plumb}\ \emph {et~al.}(2016)\citenamefont {Plumb},
  \citenamefont {Gawryluk}, \citenamefont {Wang}, \citenamefont
  {Risti\ifmmode~\acute{c}\else \'{c}\fi{}}, \citenamefont {Park},
  \citenamefont {Lv}, \citenamefont {Wang}, \citenamefont {Matt}, \citenamefont
  {Xu}, \citenamefont {Shang}, \citenamefont {Conder}, \citenamefont {Mesot},
  \citenamefont {Johnston}, \citenamefont {Shi},\ and\ \citenamefont
  {Radovi\ifmmode~\acute{c}\else \'{c}\fi{}}}]{Plumb2016momentum}%
  \BibitemOpen
  \bibfield  {author} {\bibinfo {author} {\bibfnamefont {N.~C.}\ \bibnamefont
  {Plumb}}, \bibinfo {author} {\bibfnamefont {D.~J.}\ \bibnamefont {Gawryluk}},
  \bibinfo {author} {\bibfnamefont {Y.}~\bibnamefont {Wang}}, \bibinfo {author}
  {\bibfnamefont {Z.}~\bibnamefont {Risti\ifmmode~\acute{c}\else \'{c}\fi{}}},
  \bibinfo {author} {\bibfnamefont {J.}~\bibnamefont {Park}}, \bibinfo {author}
  {\bibfnamefont {B.~Q.}\ \bibnamefont {Lv}}, \bibinfo {author} {\bibfnamefont
  {Z.}~\bibnamefont {Wang}}, \bibinfo {author} {\bibfnamefont {C.~E.}\
  \bibnamefont {Matt}}, \bibinfo {author} {\bibfnamefont {N.}~\bibnamefont
  {Xu}}, \bibinfo {author} {\bibfnamefont {T.}~\bibnamefont {Shang}}, \bibinfo
  {author} {\bibfnamefont {K.}~\bibnamefont {Conder}}, \bibinfo {author}
  {\bibfnamefont {J.}~\bibnamefont {Mesot}}, \bibinfo {author} {\bibfnamefont
  {S.}~\bibnamefont {Johnston}}, \bibinfo {author} {\bibfnamefont
  {M.}~\bibnamefont {Shi}},\ and\ \bibinfo {author} {\bibfnamefont
  {M.}~\bibnamefont {Radovi\ifmmode~\acute{c}\else \'{c}\fi{}}},\ }\bibfield
  {title} {\bibinfo {title} {Momentum-resolved electronic structure of the
  high-{$T_{c}$} superconductor parent compound {$\mathrm{BaBiO}_{3}$}},\
  }\href {https://doi.org/10.1103/PhysRevLett.117.037002} {\bibfield  {journal}
  {\bibinfo  {journal} {Phys. Rev. Lett.}\ }\textbf {\bibinfo {volume} {117}},\
  \bibinfo {pages} {037002} (\bibinfo {year} {2016})}\BibitemShut {NoStop}%
\bibitem [{\citenamefont {Akao}\ \emph {et~al.}(2003)\citenamefont {Akao},
  \citenamefont {Azuma}, \citenamefont {Usuda}, \citenamefont {Nishihata},
  \citenamefont {Mizuki}, \citenamefont {Hamada}, \citenamefont {Hayashi},
  \citenamefont {Terashima},\ and\ \citenamefont {Takano}}]{Akao2003charge}%
  \BibitemOpen
  \bibfield  {author} {\bibinfo {author} {\bibfnamefont {T.}~\bibnamefont
  {Akao}}, \bibinfo {author} {\bibfnamefont {Y.}~\bibnamefont {Azuma}},
  \bibinfo {author} {\bibfnamefont {M.}~\bibnamefont {Usuda}}, \bibinfo
  {author} {\bibfnamefont {Y.}~\bibnamefont {Nishihata}}, \bibinfo {author}
  {\bibfnamefont {J.}~\bibnamefont {Mizuki}}, \bibinfo {author} {\bibfnamefont
  {N.}~\bibnamefont {Hamada}}, \bibinfo {author} {\bibfnamefont
  {N.}~\bibnamefont {Hayashi}}, \bibinfo {author} {\bibfnamefont
  {T.}~\bibnamefont {Terashima}},\ and\ \bibinfo {author} {\bibfnamefont
  {M.}~\bibnamefont {Takano}},\ }\bibfield  {title} {\bibinfo {title}
  {Charge-ordered state in single-crystalline
  {$\mathrm{C}\mathrm{a}\mathrm{F}\mathrm{e}\mathrm{O}_{3}$} thin film studied
  by x-ray anomalous diffraction},\ }\href
  {https://doi.org/10.1103/PhysRevLett.91.156405} {\bibfield  {journal}
  {\bibinfo  {journal} {Phys. Rev. Lett.}\ }\textbf {\bibinfo {volume} {91}},\
  \bibinfo {pages} {156405} (\bibinfo {year} {2003})}\BibitemShut {NoStop}%
\bibitem [{\citenamefont {Anderson}(1950)}]{Anderson1950}%
  \BibitemOpen
  \bibfield  {author} {\bibinfo {author} {\bibfnamefont {P.~W.}\ \bibnamefont
  {Anderson}},\ }\bibfield  {title} {\bibinfo {title} {Antiferromagnetism.
  {T}heory of superexchange interaction},\ }\href
  {https://doi.org/10.1103/PhysRev.79.350} {\bibfield  {journal} {\bibinfo
  {journal} {Phys. Rev.}\ }\textbf {\bibinfo {volume} {79}},\ \bibinfo {pages}
  {350} (\bibinfo {year} {1950})}\BibitemShut {NoStop}%
\bibitem [{\citenamefont {Anderson}(1959)}]{Anderson1959}%
  \BibitemOpen
  \bibfield  {author} {\bibinfo {author} {\bibfnamefont {P.~W.}\ \bibnamefont
  {Anderson}},\ }\bibfield  {title} {\bibinfo {title} {New approach to the
  theory of superexchange interactions},\ }\href
  {https://doi.org/10.1103/PhysRev.115.2} {\bibfield  {journal} {\bibinfo
  {journal} {Phys. Rev.}\ }\textbf {\bibinfo {volume} {115}},\ \bibinfo {pages}
  {2} (\bibinfo {year} {1959})}\BibitemShut {NoStop}%
\bibitem [{\citenamefont {Anderson}(1963)}]{Anderson1963}%
  \BibitemOpen
  \bibfield  {author} {\bibinfo {author} {\bibfnamefont {P.~W.}\ \bibnamefont
  {Anderson}},\ }\bibfield  {title} {\bibinfo {title} {Exchange in insulators:
  Superexchange, direct exchange, and double exchange},\ }in\ \href@noop {}
  {\emph {\bibinfo {booktitle} {Magnetism I}}}\ (\bibinfo  {publisher}
  {Academic Press, New York},\ \bibinfo {year} {1963})\ pp.\ \bibinfo {pages}
  {25--85}\BibitemShut {NoStop}%
\bibitem [{\citenamefont {Zaanen}\ and\ \citenamefont
  {Sawatzky}(1987)}]{Zaanen1987}%
  \BibitemOpen
  \bibfield  {author} {\bibinfo {author} {\bibfnamefont {J.}~\bibnamefont
  {Zaanen}}\ and\ \bibinfo {author} {\bibfnamefont {G.~A.}\ \bibnamefont
  {Sawatzky}},\ }\bibfield  {title} {\bibinfo {title} {The electronic structure
  and superexchange interactions in transition-metal compounds},\ }\href
  {https://doi.org/10.1139/p87-201} {\bibfield  {journal} {\bibinfo  {journal}
  {Canadian Journal of Physics}\ }\textbf {\bibinfo {volume} {65}},\ \bibinfo
  {pages} {1262} (\bibinfo {year} {1987})}\BibitemShut {NoStop}%
\bibitem [{\citenamefont {Eskes}\ \emph {et~al.}(1990)\citenamefont {Eskes},
  \citenamefont {Tjeng},\ and\ \citenamefont {Sawatzky}}]{Eskes1990}%
  \BibitemOpen
  \bibfield  {author} {\bibinfo {author} {\bibfnamefont {H.}~\bibnamefont
  {Eskes}}, \bibinfo {author} {\bibfnamefont {L.~H.}\ \bibnamefont {Tjeng}},\
  and\ \bibinfo {author} {\bibfnamefont {G.~A.}\ \bibnamefont {Sawatzky}},\
  }\bibfield  {title} {\bibinfo {title} {Cluster-model calculation of the
  electronic structure of {CuO}: A model material for the high-${T}_{c}$
  superconductors},\ }\href {https://doi.org/10.1103/PhysRevB.41.288}
  {\bibfield  {journal} {\bibinfo  {journal} {Phys. Rev. B}\ }\textbf {\bibinfo
  {volume} {41}},\ \bibinfo {pages} {288} (\bibinfo {year} {1990})}\BibitemShut
  {NoStop}%
\bibitem [{\citenamefont {Li}\ \emph {et~al.}(2019)\citenamefont {Li},
  \citenamefont {Lee}, \citenamefont {Wang}, \citenamefont {Osada},
  \citenamefont {Crossley}, \citenamefont {Lee}, \citenamefont {Cui},
  \citenamefont {Hikita},\ and\ \citenamefont
  {Hwang}}]{Li2019superconductivity}%
  \BibitemOpen
  \bibfield  {author} {\bibinfo {author} {\bibfnamefont {D.}~\bibnamefont
  {Li}}, \bibinfo {author} {\bibfnamefont {K.}~\bibnamefont {Lee}}, \bibinfo
  {author} {\bibfnamefont {B.~Y.}\ \bibnamefont {Wang}}, \bibinfo {author}
  {\bibfnamefont {M.}~\bibnamefont {Osada}}, \bibinfo {author} {\bibfnamefont
  {S.}~\bibnamefont {Crossley}}, \bibinfo {author} {\bibfnamefont {H.~R.}\
  \bibnamefont {Lee}}, \bibinfo {author} {\bibfnamefont {Y.}~\bibnamefont
  {Cui}}, \bibinfo {author} {\bibfnamefont {Y.}~\bibnamefont {Hikita}},\ and\
  \bibinfo {author} {\bibfnamefont {H.~Y.}\ \bibnamefont {Hwang}},\ }\bibfield
  {title} {\bibinfo {title} {Superconductivity in an infinite-layer
  nickelate},\ }\href {https://doi.org/10.1038/s41586-019-1496-5} {\bibfield
  {journal} {\bibinfo  {journal} {Nature}\ }\textbf {\bibinfo {volume} {572}},\
  \bibinfo {pages} {624} (\bibinfo {year} {2019})}\BibitemShut {NoStop}%
\bibitem [{\citenamefont {Pan}\ \emph {et~al.}(2022)\citenamefont {Pan},
  \citenamefont {Ferenc~Segedin}, \citenamefont {LaBollita}, \citenamefont
  {Song}, \citenamefont {Nica}, \citenamefont {Goodge}, \citenamefont {Pierce},
  \citenamefont {Doyle}, \citenamefont {Novakov}, \citenamefont
  {C{\'o}rdova~Carrizales}, \citenamefont {N'Diaye}, \citenamefont {Shafer},
  \citenamefont {Paik}, \citenamefont {Heron}, \citenamefont {Mason},
  \citenamefont {Yacoby}, \citenamefont {Kourkoutis}, \citenamefont {Erten},
  \citenamefont {Brooks}, \citenamefont {Botana},\ and\ \citenamefont
  {Mundy}}]{Pan2022superconductivity}%
  \BibitemOpen
  \bibfield  {author} {\bibinfo {author} {\bibfnamefont {G.~A.}\ \bibnamefont
  {Pan}}, \bibinfo {author} {\bibfnamefont {D.}~\bibnamefont {Ferenc~Segedin}},
  \bibinfo {author} {\bibfnamefont {H.}~\bibnamefont {LaBollita}}, \bibinfo
  {author} {\bibfnamefont {Q.}~\bibnamefont {Song}}, \bibinfo {author}
  {\bibfnamefont {E.~M.}\ \bibnamefont {Nica}}, \bibinfo {author}
  {\bibfnamefont {B.~H.}\ \bibnamefont {Goodge}}, \bibinfo {author}
  {\bibfnamefont {A.~T.}\ \bibnamefont {Pierce}}, \bibinfo {author}
  {\bibfnamefont {S.}~\bibnamefont {Doyle}}, \bibinfo {author} {\bibfnamefont
  {S.}~\bibnamefont {Novakov}}, \bibinfo {author} {\bibfnamefont
  {D.}~\bibnamefont {C{\'o}rdova~Carrizales}}, \bibinfo {author} {\bibfnamefont
  {A.~T.}\ \bibnamefont {N'Diaye}}, \bibinfo {author} {\bibfnamefont
  {P.}~\bibnamefont {Shafer}}, \bibinfo {author} {\bibfnamefont
  {H.}~\bibnamefont {Paik}}, \bibinfo {author} {\bibfnamefont {J.~T.}\
  \bibnamefont {Heron}}, \bibinfo {author} {\bibfnamefont {J.~A.}\ \bibnamefont
  {Mason}}, \bibinfo {author} {\bibfnamefont {A.}~\bibnamefont {Yacoby}},
  \bibinfo {author} {\bibfnamefont {L.~F.}\ \bibnamefont {Kourkoutis}},
  \bibinfo {author} {\bibfnamefont {O.}~\bibnamefont {Erten}}, \bibinfo
  {author} {\bibfnamefont {C.~M.}\ \bibnamefont {Brooks}}, \bibinfo {author}
  {\bibfnamefont {A.~S.}\ \bibnamefont {Botana}},\ and\ \bibinfo {author}
  {\bibfnamefont {J.~A.}\ \bibnamefont {Mundy}},\ }\bibfield  {title} {\bibinfo
  {title} {Superconductivity in a quintuple-layer square-planar nickelate},\
  }\href {https://doi.org/10.1038/s41563-021-01142-9} {\bibfield  {journal}
  {\bibinfo  {journal} {Nature Materials}\ }\textbf {\bibinfo {volume} {21}},\
  \bibinfo {pages} {160} (\bibinfo {year} {2022})}\BibitemShut {NoStop}%
\bibitem [{\citenamefont {Sun}\ \emph {et~al.}(2023)\citenamefont {Sun},
  \citenamefont {Huo}, \citenamefont {Hu}, \citenamefont {Li}, \citenamefont
  {Liu}, \citenamefont {Han}, \citenamefont {Tang}, \citenamefont {Mao},
  \citenamefont {Yang}, \citenamefont {Wang}, \citenamefont {Cheng},
  \citenamefont {Yao}, \citenamefont {Zhang},\ and\ \citenamefont
  {Wang}}]{Sun2023signatures}%
  \BibitemOpen
  \bibfield  {author} {\bibinfo {author} {\bibfnamefont {H.}~\bibnamefont
  {Sun}}, \bibinfo {author} {\bibfnamefont {M.}~\bibnamefont {Huo}}, \bibinfo
  {author} {\bibfnamefont {X.}~\bibnamefont {Hu}}, \bibinfo {author}
  {\bibfnamefont {J.}~\bibnamefont {Li}}, \bibinfo {author} {\bibfnamefont
  {Z.}~\bibnamefont {Liu}}, \bibinfo {author} {\bibfnamefont {Y.}~\bibnamefont
  {Han}}, \bibinfo {author} {\bibfnamefont {L.}~\bibnamefont {Tang}}, \bibinfo
  {author} {\bibfnamefont {Z.}~\bibnamefont {Mao}}, \bibinfo {author}
  {\bibfnamefont {P.}~\bibnamefont {Yang}}, \bibinfo {author} {\bibfnamefont
  {B.}~\bibnamefont {Wang}}, \bibinfo {author} {\bibfnamefont {J.}~\bibnamefont
  {Cheng}}, \bibinfo {author} {\bibfnamefont {D.-X.}\ \bibnamefont {Yao}},
  \bibinfo {author} {\bibfnamefont {G.-M.}\ \bibnamefont {Zhang}},\ and\
  \bibinfo {author} {\bibfnamefont {M.}~\bibnamefont {Wang}},\ }\bibfield
  {title} {\bibinfo {title} {Signatures of superconductivity near 80 {K} in a
  nickelate under high pressure},\ }\href
  {https://doi.org/10.1038/s41586-023-06408-7} {\bibfield  {journal} {\bibinfo
  {journal} {Nature}\ }\textbf {\bibinfo {volume} {621}},\ \bibinfo {pages}
  {493} (\bibinfo {year} {2023})}\BibitemShut {NoStop}%
\bibitem [{\citenamefont {Wang}\ \emph {et~al.}(2024)\citenamefont {Wang},
  \citenamefont {Lee},\ and\ \citenamefont {Goodge}}]{Wang2024experimental}%
  \BibitemOpen
  \bibfield  {author} {\bibinfo {author} {\bibfnamefont {B.~Y.}\ \bibnamefont
  {Wang}}, \bibinfo {author} {\bibfnamefont {K.}~\bibnamefont {Lee}},\ and\
  \bibinfo {author} {\bibfnamefont {B.~H.}\ \bibnamefont {Goodge}},\ }\bibfield
   {title} {\bibinfo {title} {Experimental progress in superconducting
  nickelates},\ }\href
  {https://doi.org/https://doi.org/10.1146/annurev-conmatphys-032922-093307}
  {\bibfield  {journal} {\bibinfo  {journal} {Annual Review of Condensed Matter
  Physics}\ }\textbf {\bibinfo {volume} {15}},\ \bibinfo {pages} {305}
  (\bibinfo {year} {2024})}\BibitemShut {NoStop}%
\bibitem [{\citenamefont {Hu}\ and\ \citenamefont {Wu}(2019)}]{Hu2019twoband}%
  \BibitemOpen
  \bibfield  {author} {\bibinfo {author} {\bibfnamefont {L.-H.}\ \bibnamefont
  {Hu}}\ and\ \bibinfo {author} {\bibfnamefont {C.}~\bibnamefont {Wu}},\
  }\bibfield  {title} {\bibinfo {title} {Two-band model for magnetism and
  superconductivity in nickelates},\ }\href
  {https://doi.org/10.1103/PhysRevResearch.1.032046} {\bibfield  {journal}
  {\bibinfo  {journal} {Phys. Rev. Res.}\ }\textbf {\bibinfo {volume} {1}},\
  \bibinfo {pages} {032046} (\bibinfo {year} {2019})}\BibitemShut {NoStop}%
\bibitem [{\citenamefont {Nomura}\ \emph {et~al.}(2019)\citenamefont {Nomura},
  \citenamefont {Hirayama}, \citenamefont {Tadano}, \citenamefont {Yoshimoto},
  \citenamefont {Nakamura},\ and\ \citenamefont {Arita}}]{Nomura2019formation}%
  \BibitemOpen
  \bibfield  {author} {\bibinfo {author} {\bibfnamefont {Y.}~\bibnamefont
  {Nomura}}, \bibinfo {author} {\bibfnamefont {M.}~\bibnamefont {Hirayama}},
  \bibinfo {author} {\bibfnamefont {T.}~\bibnamefont {Tadano}}, \bibinfo
  {author} {\bibfnamefont {Y.}~\bibnamefont {Yoshimoto}}, \bibinfo {author}
  {\bibfnamefont {K.}~\bibnamefont {Nakamura}},\ and\ \bibinfo {author}
  {\bibfnamefont {R.}~\bibnamefont {Arita}},\ }\bibfield  {title} {\bibinfo
  {title} {Formation of a two-dimensional single-component correlated electron
  system and band engineering in the nickelate superconductor
  {$\mathrm{NdNiO}_{2}$}},\ }\href
  {https://doi.org/10.1103/PhysRevB.100.205138} {\bibfield  {journal} {\bibinfo
   {journal} {Phys. Rev. B}\ }\textbf {\bibinfo {volume} {100}},\ \bibinfo
  {pages} {205138} (\bibinfo {year} {2019})}\BibitemShut {NoStop}%
\bibitem [{\citenamefont {Jiang}\ \emph {et~al.}(2020)\citenamefont {Jiang},
  \citenamefont {Moeller}, \citenamefont {Berciu},\ and\ \citenamefont
  {Sawatzky}}]{jiang}%
  \BibitemOpen
  \bibfield  {author} {\bibinfo {author} {\bibfnamefont {M.}~\bibnamefont
  {Jiang}}, \bibinfo {author} {\bibfnamefont {M.}~\bibnamefont {Moeller}},
  \bibinfo {author} {\bibfnamefont {M.}~\bibnamefont {Berciu}},\ and\ \bibinfo
  {author} {\bibfnamefont {G.~A.}\ \bibnamefont {Sawatzky}},\ }\bibfield
  {title} {\bibinfo {title} {Relevance of {$\mathrm{Cu}-3d$} multiplet
  structure in models of high-{${T}_{c}$} cuprates},\ }\href
  {https://doi.org/10.1103/PhysRevB.101.035151} {\bibfield  {journal} {\bibinfo
   {journal} {Phys. Rev. B}\ }\textbf {\bibinfo {volume} {101}},\ \bibinfo
  {pages} {035151} (\bibinfo {year} {2020})}\BibitemShut {NoStop}%
\bibitem [{\citenamefont {Gu}\ \emph {et~al.}(2020)\citenamefont {Gu},
  \citenamefont {Zhu}, \citenamefont {Wang}, \citenamefont {Hu},\ and\
  \citenamefont {Chen}}]{Gu2020}%
  \BibitemOpen
  \bibfield  {author} {\bibinfo {author} {\bibfnamefont {Y.}~\bibnamefont
  {Gu}}, \bibinfo {author} {\bibfnamefont {S.}~\bibnamefont {Zhu}}, \bibinfo
  {author} {\bibfnamefont {X.}~\bibnamefont {Wang}}, \bibinfo {author}
  {\bibfnamefont {J.}~\bibnamefont {Hu}},\ and\ \bibinfo {author}
  {\bibfnamefont {H.}~\bibnamefont {Chen}},\ }\bibfield  {title} {\bibinfo
  {title} {A substantial hybridization between correlated {Ni}-d orbital and
  itinerant electrons in infinite-layer nickelates},\ }\href
  {https://doi.org/10.1038/s42005-020-0347-x} {\bibfield  {journal} {\bibinfo
  {journal} {Communications Physics}\ }\textbf {\bibinfo {volume} {3}},\
  \bibinfo {pages} {84} (\bibinfo {year} {2020})}\BibitemShut {NoStop}%
\bibitem [{\citenamefont {Sakakibara}\ \emph {et~al.}(2020)\citenamefont
  {Sakakibara}, \citenamefont {Usui}, \citenamefont {Suzuki}, \citenamefont
  {Kotani}, \citenamefont {Aoki},\ and\ \citenamefont
  {Kuroki}}]{Sakakibara2020model}%
  \BibitemOpen
  \bibfield  {author} {\bibinfo {author} {\bibfnamefont {H.}~\bibnamefont
  {Sakakibara}}, \bibinfo {author} {\bibfnamefont {H.}~\bibnamefont {Usui}},
  \bibinfo {author} {\bibfnamefont {K.}~\bibnamefont {Suzuki}}, \bibinfo
  {author} {\bibfnamefont {T.}~\bibnamefont {Kotani}}, \bibinfo {author}
  {\bibfnamefont {H.}~\bibnamefont {Aoki}},\ and\ \bibinfo {author}
  {\bibfnamefont {K.}~\bibnamefont {Kuroki}},\ }\bibfield  {title} {\bibinfo
  {title} {Model construction and a possibility of cupratelike pairing in a new
  ${d}^{9}$ nickelate superconductor
  {$(\mathrm{Nd},\mathrm{Sr}){\mathrm{NiO}}_{2}$}},\ }\href
  {https://doi.org/10.1103/PhysRevLett.125.077003} {\bibfield  {journal}
  {\bibinfo  {journal} {Phys. Rev. Lett.}\ }\textbf {\bibinfo {volume} {125}},\
  \bibinfo {pages} {077003} (\bibinfo {year} {2020})}\BibitemShut {NoStop}%
\bibitem [{\citenamefont {Hepting}\ \emph {et~al.}(2020)\citenamefont
  {Hepting}, \citenamefont {Li}, \citenamefont {Jia}, \citenamefont {Lu},
  \citenamefont {Paris}, \citenamefont {Tseng}, \citenamefont {Feng},
  \citenamefont {Osada}, \citenamefont {Been}, \citenamefont {Hikita},
  \citenamefont {Chuang}, \citenamefont {Hussain}, \citenamefont {Zhou},
  \citenamefont {Nag}, \citenamefont {Garcia-Fernandez}, \citenamefont {Rossi},
  \citenamefont {Huang}, \citenamefont {Huang}, \citenamefont {Shen},
  \citenamefont {Schmitt}, \citenamefont {Hwang}, \citenamefont {Moritz},
  \citenamefont {Zaanen}, \citenamefont {Devereaux},\ and\ \citenamefont
  {Lee}}]{Hepting2020electronic}%
  \BibitemOpen
  \bibfield  {author} {\bibinfo {author} {\bibfnamefont {M.}~\bibnamefont
  {Hepting}}, \bibinfo {author} {\bibfnamefont {D.}~\bibnamefont {Li}},
  \bibinfo {author} {\bibfnamefont {C.~J.}\ \bibnamefont {Jia}}, \bibinfo
  {author} {\bibfnamefont {H.}~\bibnamefont {Lu}}, \bibinfo {author}
  {\bibfnamefont {E.}~\bibnamefont {Paris}}, \bibinfo {author} {\bibfnamefont
  {Y.}~\bibnamefont {Tseng}}, \bibinfo {author} {\bibfnamefont
  {X.}~\bibnamefont {Feng}}, \bibinfo {author} {\bibfnamefont {M.}~\bibnamefont
  {Osada}}, \bibinfo {author} {\bibfnamefont {E.}~\bibnamefont {Been}},
  \bibinfo {author} {\bibfnamefont {Y.}~\bibnamefont {Hikita}}, \bibinfo
  {author} {\bibfnamefont {Y.~D.}\ \bibnamefont {Chuang}}, \bibinfo {author}
  {\bibfnamefont {Z.}~\bibnamefont {Hussain}}, \bibinfo {author} {\bibfnamefont
  {K.~J.}\ \bibnamefont {Zhou}}, \bibinfo {author} {\bibfnamefont
  {A.}~\bibnamefont {Nag}}, \bibinfo {author} {\bibfnamefont {M.}~\bibnamefont
  {Garcia-Fernandez}}, \bibinfo {author} {\bibfnamefont {M.}~\bibnamefont
  {Rossi}}, \bibinfo {author} {\bibfnamefont {H.~Y.}\ \bibnamefont {Huang}},
  \bibinfo {author} {\bibfnamefont {D.~J.}\ \bibnamefont {Huang}}, \bibinfo
  {author} {\bibfnamefont {Z.~X.}\ \bibnamefont {Shen}}, \bibinfo {author}
  {\bibfnamefont {T.}~\bibnamefont {Schmitt}}, \bibinfo {author} {\bibfnamefont
  {H.~Y.}\ \bibnamefont {Hwang}}, \bibinfo {author} {\bibfnamefont
  {B.}~\bibnamefont {Moritz}}, \bibinfo {author} {\bibfnamefont
  {J.}~\bibnamefont {Zaanen}}, \bibinfo {author} {\bibfnamefont {T.~P.}\
  \bibnamefont {Devereaux}},\ and\ \bibinfo {author} {\bibfnamefont {W.~S.}\
  \bibnamefont {Lee}},\ }\bibfield  {title} {\bibinfo {title} {Electronic
  structure of the parent compound of superconducting infinite-layer
  nickelates},\ }\href {https://doi.org/10.1038/s41563-019-0585-z} {\bibfield
  {journal} {\bibinfo  {journal} {Nature Materials}\ }\textbf {\bibinfo
  {volume} {19}},\ \bibinfo {pages} {381} (\bibinfo {year} {2020})}\BibitemShut
  {NoStop}%
\bibitem [{\citenamefont {Adhikary}\ \emph {et~al.}(2020)\citenamefont
  {Adhikary}, \citenamefont {Bandyopadhyay}, \citenamefont {Das}, \citenamefont
  {Dasgupta},\ and\ \citenamefont {Saha-Dasgupta}}]{Adhikary2020orbital}%
  \BibitemOpen
  \bibfield  {author} {\bibinfo {author} {\bibfnamefont {P.}~\bibnamefont
  {Adhikary}}, \bibinfo {author} {\bibfnamefont {S.}~\bibnamefont
  {Bandyopadhyay}}, \bibinfo {author} {\bibfnamefont {T.}~\bibnamefont {Das}},
  \bibinfo {author} {\bibfnamefont {I.}~\bibnamefont {Dasgupta}},\ and\
  \bibinfo {author} {\bibfnamefont {T.}~\bibnamefont {Saha-Dasgupta}},\
  }\bibfield  {title} {\bibinfo {title} {Orbital-selective superconductivity in
  a two-band model of infinite-layer nickelates},\ }\href
  {https://doi.org/10.1103/PhysRevB.102.100501} {\bibfield  {journal} {\bibinfo
   {journal} {Phys. Rev. B}\ }\textbf {\bibinfo {volume} {102}},\ \bibinfo
  {pages} {100501} (\bibinfo {year} {2020})}\BibitemShut {NoStop}%
\bibitem [{\citenamefont {Wang}\ \emph {et~al.}(2020)\citenamefont {Wang},
  \citenamefont {Zheng}, \citenamefont {Krivyakina}, \citenamefont {Chmaissem},
  \citenamefont {Lopes}, \citenamefont {Lynn}, \citenamefont {Gallington},
  \citenamefont {Ren}, \citenamefont {Rosenkranz}, \citenamefont {Mitchell},\
  and\ \citenamefont {Phelan}}]{Wang2020synthesis}%
  \BibitemOpen
  \bibfield  {author} {\bibinfo {author} {\bibfnamefont {B.-X.}\ \bibnamefont
  {Wang}}, \bibinfo {author} {\bibfnamefont {H.}~\bibnamefont {Zheng}},
  \bibinfo {author} {\bibfnamefont {E.}~\bibnamefont {Krivyakina}}, \bibinfo
  {author} {\bibfnamefont {O.}~\bibnamefont {Chmaissem}}, \bibinfo {author}
  {\bibfnamefont {P.~P.}\ \bibnamefont {Lopes}}, \bibinfo {author}
  {\bibfnamefont {J.~W.}\ \bibnamefont {Lynn}}, \bibinfo {author}
  {\bibfnamefont {L.~C.}\ \bibnamefont {Gallington}}, \bibinfo {author}
  {\bibfnamefont {Y.}~\bibnamefont {Ren}}, \bibinfo {author} {\bibfnamefont
  {S.}~\bibnamefont {Rosenkranz}}, \bibinfo {author} {\bibfnamefont {J.~F.}\
  \bibnamefont {Mitchell}},\ and\ \bibinfo {author} {\bibfnamefont
  {D.}~\bibnamefont {Phelan}},\ }\bibfield  {title} {\bibinfo {title}
  {Synthesis and characterization of bulk
  {${\mathrm{Nd}}_{1\ensuremath{-}x}{\mathrm{Sr}}_{x}\mathrm{Ni}{\mathrm{O}}_{2}$}
  and
  {${\mathrm{Nd}}_{1\ensuremath{-}x}{\mathrm{Sr}}_{x}\mathrm{Ni}{\mathrm{O}}_{3}$}},\
  }\href {https://doi.org/10.1103/PhysRevMaterials.4.084409} {\bibfield
  {journal} {\bibinfo  {journal} {Phys. Rev. Mater.}\ }\textbf {\bibinfo
  {volume} {4}},\ \bibinfo {pages} {084409} (\bibinfo {year}
  {2020})}\BibitemShut {NoStop}%
\bibitem [{\citenamefont {Botana}\ and\ \citenamefont
  {Norman}(2020)}]{Botana2020similarities}%
  \BibitemOpen
  \bibfield  {author} {\bibinfo {author} {\bibfnamefont {A.~S.}\ \bibnamefont
  {Botana}}\ and\ \bibinfo {author} {\bibfnamefont {M.~R.}\ \bibnamefont
  {Norman}},\ }\bibfield  {title} {\bibinfo {title} {Similarities and
  differences between {$\mathrm{LaNiO}_{2}$} and {${\mathrm{CaCuO}}_{2}$} and
  implications for superconductivity},\ }\href
  {https://doi.org/10.1103/PhysRevX.10.011024} {\bibfield  {journal} {\bibinfo
  {journal} {Phys. Rev. X}\ }\textbf {\bibinfo {volume} {10}},\ \bibinfo
  {pages} {011024} (\bibinfo {year} {2020})}\BibitemShut {NoStop}%
\bibitem [{\citenamefont {Lechermann}(2021)}]{Lechermann2021doping}%
  \BibitemOpen
  \bibfield  {author} {\bibinfo {author} {\bibfnamefont {F.}~\bibnamefont
  {Lechermann}},\ }\bibfield  {title} {\bibinfo {title} {Doping-dependent
  character and possible magnetic ordering of {${\mathrm{NdNiO}}_{2}$}},\
  }\href {https://doi.org/10.1103/PhysRevMaterials.5.044803} {\bibfield
  {journal} {\bibinfo  {journal} {Phys. Rev. Mater.}\ }\textbf {\bibinfo
  {volume} {5}},\ \bibinfo {pages} {044803} (\bibinfo {year}
  {2021})}\BibitemShut {NoStop}%
\bibitem [{\citenamefont {Petocchi}\ \emph {et~al.}(2020)\citenamefont
  {Petocchi}, \citenamefont {Christiansson}, \citenamefont {Nilsson},
  \citenamefont {Aryasetiawan},\ and\ \citenamefont
  {Werner}}]{Petocchi2020normal}%
  \BibitemOpen
  \bibfield  {author} {\bibinfo {author} {\bibfnamefont {F.}~\bibnamefont
  {Petocchi}}, \bibinfo {author} {\bibfnamefont {V.}~\bibnamefont
  {Christiansson}}, \bibinfo {author} {\bibfnamefont {F.}~\bibnamefont
  {Nilsson}}, \bibinfo {author} {\bibfnamefont {F.}~\bibnamefont
  {Aryasetiawan}},\ and\ \bibinfo {author} {\bibfnamefont {P.}~\bibnamefont
  {Werner}},\ }\bibfield  {title} {\bibinfo {title} {Normal state of
  {${\mathrm{Nd}}_{1\ensuremath{-}x}{\mathrm{Sr}}_{x}{\mathrm{NiO}}_{2}$} from
  self-consistent ${GW}+\mathrm{EDMFT}$},\ }\href
  {https://doi.org/10.1103/PhysRevX.10.041047} {\bibfield  {journal} {\bibinfo
  {journal} {Phys. Rev. X}\ }\textbf {\bibinfo {volume} {10}},\ \bibinfo
  {pages} {041047} (\bibinfo {year} {2020})}\BibitemShut {NoStop}%
\bibitem [{\citenamefont {Karp}\ \emph {et~al.}(2020)\citenamefont {Karp},
  \citenamefont {Botana}, \citenamefont {Norman}, \citenamefont {Park},
  \citenamefont {Zingl},\ and\ \citenamefont {Millis}}]{Karp2020manybody}%
  \BibitemOpen
  \bibfield  {author} {\bibinfo {author} {\bibfnamefont {J.}~\bibnamefont
  {Karp}}, \bibinfo {author} {\bibfnamefont {A.~S.}\ \bibnamefont {Botana}},
  \bibinfo {author} {\bibfnamefont {M.~R.}\ \bibnamefont {Norman}}, \bibinfo
  {author} {\bibfnamefont {H.}~\bibnamefont {Park}}, \bibinfo {author}
  {\bibfnamefont {M.}~\bibnamefont {Zingl}},\ and\ \bibinfo {author}
  {\bibfnamefont {A.}~\bibnamefont {Millis}},\ }\bibfield  {title} {\bibinfo
  {title} {Many-body electronic structure of {$\mathrm{NdNiO}_{2}$} and
  {$\mathrm{CaCuO}_{2}$}},\ }\href {https://doi.org/10.1103/PhysRevX.10.021061}
  {\bibfield  {journal} {\bibinfo  {journal} {Phys. Rev. X}\ }\textbf {\bibinfo
  {volume} {10}},\ \bibinfo {pages} {021061} (\bibinfo {year}
  {2020})}\BibitemShut {NoStop}%
\bibitem [{\citenamefont {Shen}\ \emph {et~al.}(2022)\citenamefont {Shen},
  \citenamefont {Sears}, \citenamefont {Fabbris}, \citenamefont {Li},
  \citenamefont {Pelliciari}, \citenamefont {Jarrige}, \citenamefont {He},
  \citenamefont {Bo\ifmmode \check{z}\else \v{z}\fi{}ovi\ifmmode~\acute{c}\else
  \'{c}\fi{}}, \citenamefont {Mitrano}, \citenamefont {Zhang}, \citenamefont
  {Mitchell}, \citenamefont {Botana}, \citenamefont {Bisogni}, \citenamefont
  {Norman}, \citenamefont {Johnston},\ and\ \citenamefont
  {Dean}}]{Shen2022role}%
  \BibitemOpen
  \bibfield  {author} {\bibinfo {author} {\bibfnamefont {Y.}~\bibnamefont
  {Shen}}, \bibinfo {author} {\bibfnamefont {J.}~\bibnamefont {Sears}},
  \bibinfo {author} {\bibfnamefont {G.}~\bibnamefont {Fabbris}}, \bibinfo
  {author} {\bibfnamefont {J.}~\bibnamefont {Li}}, \bibinfo {author}
  {\bibfnamefont {J.}~\bibnamefont {Pelliciari}}, \bibinfo {author}
  {\bibfnamefont {I.}~\bibnamefont {Jarrige}}, \bibinfo {author} {\bibfnamefont
  {X.}~\bibnamefont {He}}, \bibinfo {author} {\bibfnamefont {I.}~\bibnamefont
  {Bo\ifmmode \check{z}\else \v{z}\fi{}ovi\ifmmode~\acute{c}\else \'{c}\fi{}}},
  \bibinfo {author} {\bibfnamefont {M.}~\bibnamefont {Mitrano}}, \bibinfo
  {author} {\bibfnamefont {J.}~\bibnamefont {Zhang}}, \bibinfo {author}
  {\bibfnamefont {J.~F.}\ \bibnamefont {Mitchell}}, \bibinfo {author}
  {\bibfnamefont {A.~S.}\ \bibnamefont {Botana}}, \bibinfo {author}
  {\bibfnamefont {V.}~\bibnamefont {Bisogni}}, \bibinfo {author} {\bibfnamefont
  {M.~R.}\ \bibnamefont {Norman}}, \bibinfo {author} {\bibfnamefont
  {S.}~\bibnamefont {Johnston}},\ and\ \bibinfo {author} {\bibfnamefont
  {M.~P.~M.}\ \bibnamefont {Dean}},\ }\bibfield  {title} {\bibinfo {title}
  {Role of oxygen states in the low valence nickelate
  {${\mathrm{La}}_{4}{\mathrm{Ni}}_{3}{\mathrm{O}}_{8}$}},\ }\href
  {https://doi.org/10.1103/PhysRevX.12.011055} {\bibfield  {journal} {\bibinfo
  {journal} {Phys. Rev. X}\ }\textbf {\bibinfo {volume} {12}},\ \bibinfo
  {pages} {011055} (\bibinfo {year} {2022})}\BibitemShut {NoStop}%
\bibitem [{\citenamefont {Park}\ \emph {et~al.}(2012)\citenamefont {Park},
  \citenamefont {Millis},\ and\ \citenamefont {Marianetti}}]{Park2012site}%
  \BibitemOpen
  \bibfield  {author} {\bibinfo {author} {\bibfnamefont {H.}~\bibnamefont
  {Park}}, \bibinfo {author} {\bibfnamefont {A.~J.}\ \bibnamefont {Millis}},\
  and\ \bibinfo {author} {\bibfnamefont {C.~A.}\ \bibnamefont {Marianetti}},\
  }\bibfield  {title} {\bibinfo {title} {Site-selective {M}ott transition in
  rare-earth-element nickelates},\ }\href
  {https://doi.org/10.1103/PhysRevLett.109.156402} {\bibfield  {journal}
  {\bibinfo  {journal} {Phys. Rev. Lett.}\ }\textbf {\bibinfo {volume} {109}},\
  \bibinfo {pages} {156402} (\bibinfo {year} {2012})}\BibitemShut {NoStop}%
\bibitem [{\citenamefont {Johnston}\ \emph {et~al.}(2014)\citenamefont
  {Johnston}, \citenamefont {Mukherjee}, \citenamefont {Elfimov}, \citenamefont
  {Berciu},\ and\ \citenamefont {Sawatzky}}]{Johnston2014}%
  \BibitemOpen
  \bibfield  {author} {\bibinfo {author} {\bibfnamefont {S.}~\bibnamefont
  {Johnston}}, \bibinfo {author} {\bibfnamefont {A.}~\bibnamefont {Mukherjee}},
  \bibinfo {author} {\bibfnamefont {I.}~\bibnamefont {Elfimov}}, \bibinfo
  {author} {\bibfnamefont {M.}~\bibnamefont {Berciu}},\ and\ \bibinfo {author}
  {\bibfnamefont {G.~A.}\ \bibnamefont {Sawatzky}},\ }\bibfield  {title}
  {\bibinfo {title} {Charge disproportionation without charge transfer in the
  rare-earth-element nickelates as a possible mechanism for the metal-insulator
  transition},\ }\href {https://doi.org/10.1103/PhysRevLett.112.106404}
  {\bibfield  {journal} {\bibinfo  {journal} {Phys. Rev. Lett.}\ }\textbf
  {\bibinfo {volume} {112}},\ \bibinfo {pages} {106404} (\bibinfo {year}
  {2014})}\BibitemShut {NoStop}%
\bibitem [{\citenamefont {Khazraie}\ \emph {et~al.}(2018)\citenamefont
  {Khazraie}, \citenamefont {Foyevtsova}, \citenamefont {Elfimov},\ and\
  \citenamefont {Sawatzky}}]{Khazraie2018oxygen}%
  \BibitemOpen
  \bibfield  {author} {\bibinfo {author} {\bibfnamefont {A.}~\bibnamefont
  {Khazraie}}, \bibinfo {author} {\bibfnamefont {K.}~\bibnamefont
  {Foyevtsova}}, \bibinfo {author} {\bibfnamefont {I.}~\bibnamefont
  {Elfimov}},\ and\ \bibinfo {author} {\bibfnamefont {G.~A.}\ \bibnamefont
  {Sawatzky}},\ }\bibfield  {title} {\bibinfo {title} {Oxygen holes and
  hybridization in the bismuthates},\ }\href
  {https://doi.org/10.1103/PhysRevB.97.075103} {\bibfield  {journal} {\bibinfo
  {journal} {Phys. Rev. B}\ }\textbf {\bibinfo {volume} {97}},\ \bibinfo
  {pages} {075103} (\bibinfo {year} {2018})}\BibitemShut {NoStop}%
\bibitem [{\citenamefont {Cohen-Stead}\ \emph {et~al.}(2023)\citenamefont
  {Cohen-Stead}, \citenamefont {Barros}, \citenamefont {Scalettar},\ and\
  \citenamefont {Johnston}}]{CohenStead2023hybrid}%
  \BibitemOpen
  \bibfield  {author} {\bibinfo {author} {\bibfnamefont {B.}~\bibnamefont
  {Cohen-Stead}}, \bibinfo {author} {\bibfnamefont {K.}~\bibnamefont {Barros}},
  \bibinfo {author} {\bibfnamefont {R.}~\bibnamefont {Scalettar}},\ and\
  \bibinfo {author} {\bibfnamefont {S.}~\bibnamefont {Johnston}},\ }\bibfield
  {title} {\bibinfo {title} {A hybrid {M}onte {C}arlo study of bond-stretching
  electron--phonon interactions and charge order in {BaBiO$_3$}},\ }\href
  {https://doi.org/10.1038/s41524-023-00998-6} {\bibfield  {journal} {\bibinfo
  {journal} {npj Computational Materials}\ }\textbf {\bibinfo {volume} {9}},\
  \bibinfo {pages} {40} (\bibinfo {year} {2023})}\BibitemShut {NoStop}%
\bibitem [{\citenamefont {Shamblin}\ \emph {et~al.}(2018)\citenamefont
  {Shamblin}, \citenamefont {Heres}, \citenamefont {Zhou}, \citenamefont
  {Sangoro}, \citenamefont {Lang}, \citenamefont {Neuefeind}, \citenamefont
  {Alonso},\ and\ \citenamefont {Johnston}}]{Shamblin2018experimental}%
  \BibitemOpen
  \bibfield  {author} {\bibinfo {author} {\bibfnamefont {J.}~\bibnamefont
  {Shamblin}}, \bibinfo {author} {\bibfnamefont {M.}~\bibnamefont {Heres}},
  \bibinfo {author} {\bibfnamefont {H.}~\bibnamefont {Zhou}}, \bibinfo {author}
  {\bibfnamefont {J.}~\bibnamefont {Sangoro}}, \bibinfo {author} {\bibfnamefont
  {M.}~\bibnamefont {Lang}}, \bibinfo {author} {\bibfnamefont {J.}~\bibnamefont
  {Neuefeind}}, \bibinfo {author} {\bibfnamefont {J.~A.}\ \bibnamefont
  {Alonso}},\ and\ \bibinfo {author} {\bibfnamefont {S.}~\bibnamefont
  {Johnston}},\ }\bibfield  {title} {\bibinfo {title} {Experimental evidence
  for bipolaron condensation as a mechanism for the metal-insulator transition
  in rare-earth nickelates},\ }\href
  {https://doi.org/10.1038/s41467-017-02561-6} {\bibfield  {journal} {\bibinfo
  {journal} {Nature Communications}\ }\textbf {\bibinfo {volume} {9}},\
  \bibinfo {pages} {86} (\bibinfo {year} {2018})}\BibitemShut {NoStop}%
\bibitem [{\citenamefont {Tyunina}\ \emph {et~al.}(2023)\citenamefont
  {Tyunina}, \citenamefont {Savinov}, \citenamefont {Pacherova},\ and\
  \citenamefont {Dejneka}}]{Tyunina2023small}%
  \BibitemOpen
  \bibfield  {author} {\bibinfo {author} {\bibfnamefont {M.}~\bibnamefont
  {Tyunina}}, \bibinfo {author} {\bibfnamefont {M.}~\bibnamefont {Savinov}},
  \bibinfo {author} {\bibfnamefont {O.}~\bibnamefont {Pacherova}},\ and\
  \bibinfo {author} {\bibfnamefont {A.}~\bibnamefont {Dejneka}},\ }\bibfield
  {title} {\bibinfo {title} {Small-polaron transport in perovskite
  nickelates},\ }\href {https://doi.org/10.1038/s41598-023-39821-z} {\bibfield
  {journal} {\bibinfo  {journal} {Scientific Reports}\ }\textbf {\bibinfo
  {volume} {13}},\ \bibinfo {pages} {12493} (\bibinfo {year}
  {2023})}\BibitemShut {NoStop}%
\bibitem [{\citenamefont {Pickett}(2001)}]{Pickett2001other}%
  \BibitemOpen
  \bibfield  {author} {\bibinfo {author} {\bibfnamefont {W.}~\bibnamefont
  {Pickett}},\ }\bibfield  {title} {\bibinfo {title} {The other
  high-temperature superconductors},\ }\href
  {https://doi.org/https://doi.org/10.1016/S0921-4526(00)00787-0} {\bibfield
  {journal} {\bibinfo  {journal} {Physica B: Condensed Matter}\ }\textbf
  {\bibinfo {volume} {296}},\ \bibinfo {pages} {112} (\bibinfo {year}
  {2001})}\BibitemShut {NoStop}%
\bibitem [{\citenamefont {Sleight}(2015)}]{Sleight2015bismuthates}%
  \BibitemOpen
  \bibfield  {author} {\bibinfo {author} {\bibfnamefont {A.~W.}\ \bibnamefont
  {Sleight}},\ }\bibfield  {title} {\bibinfo {title} {Bismuthates: {BaBiO$_3$}
  and related superconducting phases},\ }\href
  {https://doi.org/https://doi.org/10.1016/j.physc.2015.02.012} {\bibfield
  {journal} {\bibinfo  {journal} {Physica C: Superconductivity and its
  Applications}\ }\textbf {\bibinfo {volume} {514}},\ \bibinfo {pages} {152}
  (\bibinfo {year} {2015})}\BibitemShut {NoStop}%
\bibitem [{\citenamefont {Kim}\ \emph {et~al.}(2022)\citenamefont {Kim},
  \citenamefont {McNally}, \citenamefont {Kim}, \citenamefont {Oudah},
  \citenamefont {Gibbs}, \citenamefont {Manuel}, \citenamefont {Green},
  \citenamefont {Sutarto}, \citenamefont {Takayama}, \citenamefont {Yaresko},
  \citenamefont {Wedig}, \citenamefont {Isobe}, \citenamefont {Kremer},
  \citenamefont {Bonn}, \citenamefont {Keimer},\ and\ \citenamefont
  {Takagi}}]{Kim2022superconductivity}%
  \BibitemOpen
  \bibfield  {author} {\bibinfo {author} {\bibfnamefont {M.}~\bibnamefont
  {Kim}}, \bibinfo {author} {\bibfnamefont {G.~M.}\ \bibnamefont {McNally}},
  \bibinfo {author} {\bibfnamefont {H.-H.}\ \bibnamefont {Kim}}, \bibinfo
  {author} {\bibfnamefont {M.}~\bibnamefont {Oudah}}, \bibinfo {author}
  {\bibfnamefont {A.~S.}\ \bibnamefont {Gibbs}}, \bibinfo {author}
  {\bibfnamefont {P.}~\bibnamefont {Manuel}}, \bibinfo {author} {\bibfnamefont
  {R.~J.}\ \bibnamefont {Green}}, \bibinfo {author} {\bibfnamefont
  {R.}~\bibnamefont {Sutarto}}, \bibinfo {author} {\bibfnamefont
  {T.}~\bibnamefont {Takayama}}, \bibinfo {author} {\bibfnamefont
  {A.}~\bibnamefont {Yaresko}}, \bibinfo {author} {\bibfnamefont
  {U.}~\bibnamefont {Wedig}}, \bibinfo {author} {\bibfnamefont
  {M.}~\bibnamefont {Isobe}}, \bibinfo {author} {\bibfnamefont {R.~K.}\
  \bibnamefont {Kremer}}, \bibinfo {author} {\bibfnamefont {D.~A.}\
  \bibnamefont {Bonn}}, \bibinfo {author} {\bibfnamefont {B.}~\bibnamefont
  {Keimer}},\ and\ \bibinfo {author} {\bibfnamefont {H.}~\bibnamefont
  {Takagi}},\ }\bibfield  {title} {\bibinfo {title} {Superconductivity in
  {(Ba,K)SbO$_3$}},\ }\href {https://doi.org/10.1038/s41563-022-01203-7}
  {\bibfield  {journal} {\bibinfo  {journal} {Nature Materials}\ }\textbf
  {\bibinfo {volume} {21}},\ \bibinfo {pages} {627} (\bibinfo {year}
  {2022})}\BibitemShut {NoStop}%
\bibitem [{\citenamefont {Sawatzky}\ and\ \citenamefont
  {Allen}(1984)}]{Sawatzky1984magnitude}%
  \BibitemOpen
  \bibfield  {author} {\bibinfo {author} {\bibfnamefont {G.~A.}\ \bibnamefont
  {Sawatzky}}\ and\ \bibinfo {author} {\bibfnamefont {J.~W.}\ \bibnamefont
  {Allen}},\ }\bibfield  {title} {\bibinfo {title} {Magnitude and origin of the
  band gap in {NiO}},\ }\href {https://doi.org/10.1103/PhysRevLett.53.2339}
  {\bibfield  {journal} {\bibinfo  {journal} {Phys. Rev. Lett.}\ }\textbf
  {\bibinfo {volume} {53}},\ \bibinfo {pages} {2339} (\bibinfo {year}
  {1984})}\BibitemShut {NoStop}%
\bibitem [{\citenamefont {Mizokawa}\ \emph {et~al.}(1994)\citenamefont
  {Mizokawa}, \citenamefont {Fujimori}, \citenamefont {Namatame}, \citenamefont
  {Akeyama},\ and\ \citenamefont {Kosugi}}]{Mizokawa1994}%
  \BibitemOpen
  \bibfield  {author} {\bibinfo {author} {\bibfnamefont {T.}~\bibnamefont
  {Mizokawa}}, \bibinfo {author} {\bibfnamefont {A.}~\bibnamefont {Fujimori}},
  \bibinfo {author} {\bibfnamefont {H.}~\bibnamefont {Namatame}}, \bibinfo
  {author} {\bibfnamefont {K.}~\bibnamefont {Akeyama}},\ and\ \bibinfo {author}
  {\bibfnamefont {N.}~\bibnamefont {Kosugi}},\ }\bibfield  {title} {\bibinfo
  {title} {Electronic structure of the local-singlet insulator
  {$\mathrm{NaCuO}_{2}$}},\ }\href {https://doi.org/10.1103/PhysRevB.49.7193}
  {\bibfield  {journal} {\bibinfo  {journal} {Phys. Rev. B}\ }\textbf {\bibinfo
  {volume} {49}},\ \bibinfo {pages} {7193} (\bibinfo {year}
  {1994})}\BibitemShut {NoStop}%
\bibitem [{\citenamefont {Haverkort}\ \emph {et~al.}(2012)\citenamefont
  {Haverkort}, \citenamefont {Zwierzycki},\ and\ \citenamefont
  {Andersen}}]{Haverkort2012multiplet}%
  \BibitemOpen
  \bibfield  {author} {\bibinfo {author} {\bibfnamefont {M.~W.}\ \bibnamefont
  {Haverkort}}, \bibinfo {author} {\bibfnamefont {M.}~\bibnamefont
  {Zwierzycki}},\ and\ \bibinfo {author} {\bibfnamefont {O.~K.}\ \bibnamefont
  {Andersen}},\ }\bibfield  {title} {\bibinfo {title} {Multiplet ligand-field
  theory using {W}annier orbitals},\ }\href
  {https://doi.org/10.1103/PhysRevB.85.165113} {\bibfield  {journal} {\bibinfo
  {journal} {Phys. Rev. B}\ }\textbf {\bibinfo {volume} {85}},\ \bibinfo
  {pages} {165113} (\bibinfo {year} {2012})}\BibitemShut {NoStop}%
\bibitem [{\citenamefont {Mizokawa}\ \emph {et~al.}(2000)\citenamefont
  {Mizokawa}, \citenamefont {Khomskii},\ and\ \citenamefont
  {Sawatzky}}]{Mizokawa2000spin}%
  \BibitemOpen
  \bibfield  {author} {\bibinfo {author} {\bibfnamefont {T.}~\bibnamefont
  {Mizokawa}}, \bibinfo {author} {\bibfnamefont {D.~I.}\ \bibnamefont
  {Khomskii}},\ and\ \bibinfo {author} {\bibfnamefont {G.~A.}\ \bibnamefont
  {Sawatzky}},\ }\bibfield  {title} {\bibinfo {title} {Spin and charge ordering
  in self-doped {M}ott insulators},\ }\href
  {https://doi.org/10.1103/PhysRevB.61.11263} {\bibfield  {journal} {\bibinfo
  {journal} {Phys. Rev. B}\ }\textbf {\bibinfo {volume} {61}},\ \bibinfo
  {pages} {11263} (\bibinfo {year} {2000})}\BibitemShut {NoStop}%
\bibitem [{\citenamefont {Yamaguchi}\ \emph {et~al.}(2024)\citenamefont
  {Yamaguchi}, \citenamefont {Higashi}, \citenamefont {Regoutz}, \citenamefont
  {Takahashi}, \citenamefont {Lazemi}, \citenamefont {Che}, \citenamefont
  {de~Groot},\ and\ \citenamefont {Hariki}}]{Yamaguchi2024atomic}%
  \BibitemOpen
  \bibfield  {author} {\bibinfo {author} {\bibfnamefont {T.}~\bibnamefont
  {Yamaguchi}}, \bibinfo {author} {\bibfnamefont {K.}~\bibnamefont {Higashi}},
  \bibinfo {author} {\bibfnamefont {A.}~\bibnamefont {Regoutz}}, \bibinfo
  {author} {\bibfnamefont {Y.}~\bibnamefont {Takahashi}}, \bibinfo {author}
  {\bibfnamefont {M.}~\bibnamefont {Lazemi}}, \bibinfo {author} {\bibfnamefont
  {Q.}~\bibnamefont {Che}}, \bibinfo {author} {\bibfnamefont {F.~M.~F.}\
  \bibnamefont {de~Groot}},\ and\ \bibinfo {author} {\bibfnamefont
  {A.}~\bibnamefont {Hariki}},\ }\bibfield  {title} {\bibinfo {title} {Atomic
  multiplet and charge transfer screening effects in $1s$ and $2p$ core-level
  x-ray photoelectron spectra of early $3d$ transition-metal oxides},\ }\href
  {https://doi.org/10.1103/PhysRevB.109.205143} {\bibfield  {journal} {\bibinfo
   {journal} {Phys. Rev. B}\ }\textbf {\bibinfo {volume} {109}},\ \bibinfo
  {pages} {205143} (\bibinfo {year} {2024})}\BibitemShut {NoStop}%
\bibitem [{\citenamefont {Haverkort}(2016)}]{Haverkort2016quanty}%
  \BibitemOpen
  \bibfield  {author} {\bibinfo {author} {\bibfnamefont {M.~W.}\ \bibnamefont
  {Haverkort}},\ }\bibfield  {title} {\bibinfo {title} {Quanty for core level
  spectroscopy - excitons, resonances and band excitations in time and
  frequency domain},\ }\href {https://doi.org/10.1088/1742-6596/712/1/012001}
  {\bibfield  {journal} {\bibinfo  {journal} {Journal of Physics: Conference
  Series}\ }\textbf {\bibinfo {volume} {712}},\ \bibinfo {pages} {012001}
  (\bibinfo {year} {2016})}\BibitemShut {NoStop}%
\bibitem [{\citenamefont {Freeland}\ \emph {et~al.}(2016)\citenamefont
  {Freeland}, \citenamefont {{van Veenendaal}},\ and\ \citenamefont
  {Chakhalian}}]{Freeland2016evolution}%
  \BibitemOpen
  \bibfield  {author} {\bibinfo {author} {\bibfnamefont {J.~W.}\ \bibnamefont
  {Freeland}}, \bibinfo {author} {\bibfnamefont {M.}~\bibnamefont {{van
  Veenendaal}}},\ and\ \bibinfo {author} {\bibfnamefont {J.}~\bibnamefont
  {Chakhalian}},\ }\bibfield  {title} {\bibinfo {title} {Evolution of
  electronic structure across the rare-earth {RNiO$_3$} series},\ }\href
  {https://doi.org/https://doi.org/10.1016/j.elspec.2015.07.006} {\bibfield
  {journal} {\bibinfo  {journal} {Journal of Electron Spectroscopy and Related
  Phenomena}\ }\textbf {\bibinfo {volume} {208}},\ \bibinfo {pages} {56}
  (\bibinfo {year} {2016})}\BibitemShut {NoStop}%
\bibitem [{\citenamefont {Chen}\ \emph {et~al.}(2013)\citenamefont {Chen},
  \citenamefont {Sentef}, \citenamefont {Kung}, \citenamefont {Jia},
  \citenamefont {Thomale}, \citenamefont {Moritz}, \citenamefont {Kampf},\ and\
  \citenamefont {Devereaux}}]{Chen2013doping}%
  \BibitemOpen
  \bibfield  {author} {\bibinfo {author} {\bibfnamefont {C.-C.}\ \bibnamefont
  {Chen}}, \bibinfo {author} {\bibfnamefont {M.}~\bibnamefont {Sentef}},
  \bibinfo {author} {\bibfnamefont {Y.~F.}\ \bibnamefont {Kung}}, \bibinfo
  {author} {\bibfnamefont {C.~J.}\ \bibnamefont {Jia}}, \bibinfo {author}
  {\bibfnamefont {R.}~\bibnamefont {Thomale}}, \bibinfo {author} {\bibfnamefont
  {B.}~\bibnamefont {Moritz}}, \bibinfo {author} {\bibfnamefont {A.~P.}\
  \bibnamefont {Kampf}},\ and\ \bibinfo {author} {\bibfnamefont {T.~P.}\
  \bibnamefont {Devereaux}},\ }\bibfield  {title} {\bibinfo {title} {Doping
  evolution of the oxygen {$K$}-edge x-ray absorption spectra of cuprate
  superconductors using a three-orbital {H}ubbard model},\ }\href
  {https://doi.org/10.1103/PhysRevB.87.165144} {\bibfield  {journal} {\bibinfo
  {journal} {Phys. Rev. B}\ }\textbf {\bibinfo {volume} {87}},\ \bibinfo
  {pages} {165144} (\bibinfo {year} {2013})}\BibitemShut {NoStop}%
\bibitem [{\citenamefont {Dobry}\ \emph {et~al.}(1994)\citenamefont {Dobry},
  \citenamefont {Greco}, \citenamefont {Lorenzana},\ and\ \citenamefont
  {Riera}}]{Dobry1994}%
  \BibitemOpen
  \bibfield  {author} {\bibinfo {author} {\bibfnamefont {A.}~\bibnamefont
  {Dobry}}, \bibinfo {author} {\bibfnamefont {A.}~\bibnamefont {Greco}},
  \bibinfo {author} {\bibfnamefont {J.}~\bibnamefont {Lorenzana}},\ and\
  \bibinfo {author} {\bibfnamefont {J.}~\bibnamefont {Riera}},\ }\bibfield
  {title} {\bibinfo {title} {Polarons in the three-band {P}eierls-{H}ubbard
  model: An exact diagonalization study},\ }\href
  {https://doi.org/10.1103/PhysRevB.49.505} {\bibfield  {journal} {\bibinfo
  {journal} {Phys. Rev. B}\ }\textbf {\bibinfo {volume} {49}},\ \bibinfo
  {pages} {505} (\bibinfo {year} {1994})}\BibitemShut {NoStop}%
\bibitem [{\citenamefont {Lau}\ \emph {et~al.}(2011{\natexlab{a}})\citenamefont
  {Lau}, \citenamefont {Berciu},\ and\ \citenamefont {Sawatzky}}]{Lau2011}%
  \BibitemOpen
  \bibfield  {author} {\bibinfo {author} {\bibfnamefont {B.}~\bibnamefont
  {Lau}}, \bibinfo {author} {\bibfnamefont {M.}~\bibnamefont {Berciu}},\ and\
  \bibinfo {author} {\bibfnamefont {G.~A.}\ \bibnamefont {Sawatzky}},\
  }\bibfield  {title} {\bibinfo {title} {High-spin polaron in lightly doped
  {$\mathrm{CuO}_{2}$} planes},\ }\href
  {https://doi.org/10.1103/PhysRevLett.106.036401} {\bibfield  {journal}
  {\bibinfo  {journal} {Phys. Rev. Lett.}\ }\textbf {\bibinfo {volume} {106}},\
  \bibinfo {pages} {036401} (\bibinfo {year} {2011}{\natexlab{a}})}\BibitemShut
  {NoStop}%
\bibitem [{\citenamefont {Lau}\ \emph {et~al.}(2011{\natexlab{b}})\citenamefont
  {Lau}, \citenamefont {Berciu},\ and\ \citenamefont {Sawatzky}}]{Lau2011b}%
  \BibitemOpen
  \bibfield  {author} {\bibinfo {author} {\bibfnamefont {B.}~\bibnamefont
  {Lau}}, \bibinfo {author} {\bibfnamefont {M.}~\bibnamefont {Berciu}},\ and\
  \bibinfo {author} {\bibfnamefont {G.~A.}\ \bibnamefont {Sawatzky}},\
  }\bibfield  {title} {\bibinfo {title} {Computational approach to a doped
  antiferromagnet: Correlations between two spin polarons in the lightly doped
  {CuO$_{2}$} plane},\ }\href {https://doi.org/10.1103/PhysRevB.84.165102}
  {\bibfield  {journal} {\bibinfo  {journal} {Phys. Rev. B}\ }\textbf {\bibinfo
  {volume} {84}},\ \bibinfo {pages} {165102} (\bibinfo {year}
  {2011}{\natexlab{b}})}\BibitemShut {NoStop}%
\bibitem [{\citenamefont {Li}\ \emph {et~al.}(2021)\citenamefont {Li},
  \citenamefont {Nocera}, \citenamefont {Kumar},\ and\ \citenamefont
  {Johnston}}]{Li2021particle}%
  \BibitemOpen
  \bibfield  {author} {\bibinfo {author} {\bibfnamefont {S.}~\bibnamefont
  {Li}}, \bibinfo {author} {\bibfnamefont {A.}~\bibnamefont {Nocera}}, \bibinfo
  {author} {\bibfnamefont {U.}~\bibnamefont {Kumar}},\ and\ \bibinfo {author}
  {\bibfnamefont {S.}~\bibnamefont {Johnston}},\ }\bibfield  {title} {\bibinfo
  {title} {Particle-hole asymmetry in the dynamical spin and charge responses
  of corner-shared {1D} cuprates},\ }\href
  {https://doi.org/10.1038/s42005-021-00718-w} {\bibfield  {journal} {\bibinfo
  {journal} {Communications Physics}\ }\textbf {\bibinfo {volume} {4}},\
  \bibinfo {pages} {217} (\bibinfo {year} {2021})}\BibitemShut {NoStop}%
\bibitem [{\citenamefont {Nocera}\ \emph {et~al.}(2018)\citenamefont {Nocera},
  \citenamefont {Kumar}, \citenamefont {Kaushal}, \citenamefont {Alvarez},
  \citenamefont {Dagotto},\ and\ \citenamefont
  {Johnston}}]{Nocera2018computing}%
  \BibitemOpen
  \bibfield  {author} {\bibinfo {author} {\bibfnamefont {A.}~\bibnamefont
  {Nocera}}, \bibinfo {author} {\bibfnamefont {U.}~\bibnamefont {Kumar}},
  \bibinfo {author} {\bibfnamefont {N.}~\bibnamefont {Kaushal}}, \bibinfo
  {author} {\bibfnamefont {G.}~\bibnamefont {Alvarez}}, \bibinfo {author}
  {\bibfnamefont {E.}~\bibnamefont {Dagotto}},\ and\ \bibinfo {author}
  {\bibfnamefont {S.}~\bibnamefont {Johnston}},\ }\bibfield  {title} {\bibinfo
  {title} {Computing resonant inelastic x-ray scattering spectra using the
  density matrix renormalization group method},\ }\href
  {https://doi.org/10.1038/s41598-018-29218-8} {\bibfield  {journal} {\bibinfo
  {journal} {Scientific Reports}\ }\textbf {\bibinfo {volume} {8}},\ \bibinfo
  {pages} {11080} (\bibinfo {year} {2018})}\BibitemShut {NoStop}%
\bibitem [{\citenamefont {White}\ and\ \citenamefont
  {Scalapino}(2015)}]{White2015}%
  \BibitemOpen
  \bibfield  {author} {\bibinfo {author} {\bibfnamefont {S.~R.}\ \bibnamefont
  {White}}\ and\ \bibinfo {author} {\bibfnamefont {D.~J.}\ \bibnamefont
  {Scalapino}},\ }\bibfield  {title} {\bibinfo {title} {Doping asymmetry and
  striping in a three-orbital {$\mathrm{CuO}_{2}$} {H}ubbard model},\ }\href
  {https://doi.org/10.1103/PhysRevB.92.205112} {\bibfield  {journal} {\bibinfo
  {journal} {Phys. Rev. B}\ }\textbf {\bibinfo {volume} {92}},\ \bibinfo
  {pages} {205112} (\bibinfo {year} {2015})}\BibitemShut {NoStop}%
\bibitem [{\citenamefont {Guerrero}\ \emph {et~al.}(1998)\citenamefont
  {Guerrero}, \citenamefont {Gubernatis},\ and\ \citenamefont
  {Zhang}}]{Guerrero1998}%
  \BibitemOpen
  \bibfield  {author} {\bibinfo {author} {\bibfnamefont {M.}~\bibnamefont
  {Guerrero}}, \bibinfo {author} {\bibfnamefont {J.~E.}\ \bibnamefont
  {Gubernatis}},\ and\ \bibinfo {author} {\bibfnamefont {S.}~\bibnamefont
  {Zhang}},\ }\bibfield  {title} {\bibinfo {title} {Quantum {M}onte {C}arlo
  study of hole binding and pairing correlations in the three-band {H}ubbard
  model},\ }\href {https://doi.org/10.1103/PhysRevB.57.11980} {\bibfield
  {journal} {\bibinfo  {journal} {Phys. Rev. B}\ }\textbf {\bibinfo {volume}
  {57}},\ \bibinfo {pages} {11980} (\bibinfo {year} {1998})}\BibitemShut
  {NoStop}%
\bibitem [{\citenamefont {Chiciak}\ \emph {et~al.}(2018)\citenamefont
  {Chiciak}, \citenamefont {Vitali}, \citenamefont {Shi},\ and\ \citenamefont
  {Zhang}}]{Chiciak2018}%
  \BibitemOpen
  \bibfield  {author} {\bibinfo {author} {\bibfnamefont {A.}~\bibnamefont
  {Chiciak}}, \bibinfo {author} {\bibfnamefont {E.}~\bibnamefont {Vitali}},
  \bibinfo {author} {\bibfnamefont {H.}~\bibnamefont {Shi}},\ and\ \bibinfo
  {author} {\bibfnamefont {S.}~\bibnamefont {Zhang}},\ }\bibfield  {title}
  {\bibinfo {title} {Magnetic orders in the hole-doped three-band {H}ubbard
  model: Spin spirals, nematicity, and ferromagnetic domain walls},\ }\href
  {https://doi.org/10.1103/PhysRevB.97.235127} {\bibfield  {journal} {\bibinfo
  {journal} {Phys. Rev. B}\ }\textbf {\bibinfo {volume} {97}},\ \bibinfo
  {pages} {235127} (\bibinfo {year} {2018})}\BibitemShut {NoStop}%
\bibitem [{\citenamefont {Huang}\ \emph {et~al.}(2017)\citenamefont {Huang},
  \citenamefont {Mendl}, \citenamefont {Liu}, \citenamefont {Johnston},
  \citenamefont {Jiang}, \citenamefont {Moritz},\ and\ \citenamefont
  {Devereaux}}]{Huang2017}%
  \BibitemOpen
  \bibfield  {author} {\bibinfo {author} {\bibfnamefont {E.~W.}\ \bibnamefont
  {Huang}}, \bibinfo {author} {\bibfnamefont {C.~B.}\ \bibnamefont {Mendl}},
  \bibinfo {author} {\bibfnamefont {S.}~\bibnamefont {Liu}}, \bibinfo {author}
  {\bibfnamefont {S.}~\bibnamefont {Johnston}}, \bibinfo {author}
  {\bibfnamefont {H.~C.}\ \bibnamefont {Jiang}}, \bibinfo {author}
  {\bibfnamefont {B.}~\bibnamefont {Moritz}},\ and\ \bibinfo {author}
  {\bibfnamefont {T.~P.}\ \bibnamefont {Devereaux}},\ }\bibfield  {title}
  {\bibinfo {title} {Numerical evidence of fluctuating stripes in the normal
  state of high-{T$_c$} cuprate superconductors},\ }\href
  {https://doi.org/10.1126/science.aak9546} {\bibfield  {journal} {\bibinfo
  {journal} {Science}\ }\textbf {\bibinfo {volume} {358}},\ \bibinfo {pages}
  {1161–1164} (\bibinfo {year} {2017})}\BibitemShut {NoStop}%
\bibitem [{\citenamefont {Mai}\ \emph {et~al.}(2024)\citenamefont {Mai},
  \citenamefont {Cohen-Stead}, \citenamefont {Maier},\ and\ \citenamefont
  {Johnston}}]{Mai2024fluctuating}%
  \BibitemOpen
  \bibfield  {author} {\bibinfo {author} {\bibfnamefont {P.}~\bibnamefont
  {Mai}}, \bibinfo {author} {\bibfnamefont {B.}~\bibnamefont {Cohen-Stead}},
  \bibinfo {author} {\bibfnamefont {T.~A.}\ \bibnamefont {Maier}},\ and\
  \bibinfo {author} {\bibfnamefont {S.}~\bibnamefont {Johnston}},\ }\bibfield
  {title} {\bibinfo {title} {Fluctuating charge-density-wave correlations in
  the three-band {H}ubbard model},\ }\href
  {https://doi.org/10.1073/pnas.2408717121} {\bibfield  {journal} {\bibinfo
  {journal} {Proceedings of the National Academy of Sciences}\ }\textbf
  {\bibinfo {volume} {121}},\ \bibinfo {pages} {e2408717121} (\bibinfo {year}
  {2024})}\BibitemShut {NoStop}%
\bibitem [{\citenamefont {Weber}\ \emph {et~al.}(2012)\citenamefont {Weber},
  \citenamefont {Yee}, \citenamefont {Haule},\ and\ \citenamefont
  {Kotliar}}]{Weber2012}%
  \BibitemOpen
  \bibfield  {author} {\bibinfo {author} {\bibfnamefont {C.}~\bibnamefont
  {Weber}}, \bibinfo {author} {\bibfnamefont {C.}~\bibnamefont {Yee}}, \bibinfo
  {author} {\bibfnamefont {K.}~\bibnamefont {Haule}},\ and\ \bibinfo {author}
  {\bibfnamefont {G.}~\bibnamefont {Kotliar}},\ }\bibfield  {title} {\bibinfo
  {title} {Scaling of the transition temperature of hole-doped cuprate
  superconductors with the charge-transfer energy},\ }\href
  {https://doi.org/10.1209/0295-5075/100/37001} {\bibfield  {journal} {\bibinfo
   {journal} {Europhysics Letters}\ }\textbf {\bibinfo {volume} {100}},\
  \bibinfo {pages} {37001} (\bibinfo {year} {2012})}\BibitemShut {NoStop}%
\bibitem [{\citenamefont {Mai}\ \emph {et~al.}(2021)\citenamefont {Mai},
  \citenamefont {Balduzzi}, \citenamefont {Johnston},\ and\ \citenamefont
  {Maier}}]{Mai2021orbital}%
  \BibitemOpen
  \bibfield  {author} {\bibinfo {author} {\bibfnamefont {P.}~\bibnamefont
  {Mai}}, \bibinfo {author} {\bibfnamefont {G.}~\bibnamefont {Balduzzi}},
  \bibinfo {author} {\bibfnamefont {S.}~\bibnamefont {Johnston}},\ and\
  \bibinfo {author} {\bibfnamefont {T.~A.}\ \bibnamefont {Maier}},\ }\bibfield
  {title} {\bibinfo {title} {Orbital structure of the effective pairing
  interaction in the high-temperature superconducting cuprates},\ }\href
  {https://doi.org/10.1038/s41535-021-00326-5} {\bibfield  {journal} {\bibinfo
  {journal} {npj Quantum Materials}\ }\textbf {\bibinfo {volume} {6}},\
  \bibinfo {pages} {26} (\bibinfo {year} {2021})}\BibitemShut {NoStop}%
\bibitem [{\citenamefont {Cui}\ \emph {et~al.}(2020)\citenamefont {Cui},
  \citenamefont {Sun}, \citenamefont {Ray}, \citenamefont {Zheng},
  \citenamefont {Sun},\ and\ \citenamefont {Chan}}]{Cui2020Groundstate}%
  \BibitemOpen
  \bibfield  {author} {\bibinfo {author} {\bibfnamefont {Z.-H.}\ \bibnamefont
  {Cui}}, \bibinfo {author} {\bibfnamefont {C.}~\bibnamefont {Sun}}, \bibinfo
  {author} {\bibfnamefont {U.}~\bibnamefont {Ray}}, \bibinfo {author}
  {\bibfnamefont {B.-X.}\ \bibnamefont {Zheng}}, \bibinfo {author}
  {\bibfnamefont {Q.}~\bibnamefont {Sun}},\ and\ \bibinfo {author}
  {\bibfnamefont {G.~K.-L.}\ \bibnamefont {Chan}},\ }\bibfield  {title}
  {\bibinfo {title} {Ground-state phase diagram of the three-band {H}ubbard
  model from density matrix embedding theory},\ }\href
  {https://doi.org/10.1103/PhysRevResearch.2.043259} {\bibfield  {journal}
  {\bibinfo  {journal} {Phys. Rev. Res.}\ }\textbf {\bibinfo {volume} {2}},\
  \bibinfo {pages} {043259} (\bibinfo {year} {2020})}\BibitemShut {NoStop}%
\bibitem [{\citenamefont {Dagotto}(1994)}]{Dagotto1994}%
  \BibitemOpen
  \bibfield  {author} {\bibinfo {author} {\bibfnamefont {E.}~\bibnamefont
  {Dagotto}},\ }\bibfield  {title} {\bibinfo {title} {Correlated electrons in
  high-temperature superconductors},\ }\href
  {https://doi.org/10.1103/RevModPhys.66.763} {\bibfield  {journal} {\bibinfo
  {journal} {Rev. Mod. Phys.}\ }\textbf {\bibinfo {volume} {66}},\ \bibinfo
  {pages} {763} (\bibinfo {year} {1994})}\BibitemShut {NoStop}%
\bibitem [{\citenamefont {Scalapino}(2007)}]{ScalapinoReview}%
  \BibitemOpen
  \bibfield  {author} {\bibinfo {author} {\bibfnamefont {D.}~\bibnamefont
  {Scalapino}},\ }\bibfield  {title} {\bibinfo {title} {Numerical studies of
  the {2D} {H}ubbard model},\ }in\ \href
  {https://doi.org/10.1007/978-0-387-68734-6_13} {\emph {\bibinfo {booktitle}
  {Handbook of High-Temperature Superconductivity}}},\ \bibinfo {editor}
  {edited by\ \bibinfo {editor} {\bibfnamefont {J.}~\bibnamefont {Schrieffer}}\
  and\ \bibinfo {editor} {\bibfnamefont {J.}~\bibnamefont {Brooks}}}\ (\bibinfo
   {publisher} {Springer New York},\ \bibinfo {year} {2007})\ pp.\ \bibinfo
  {pages} {495--526}\BibitemShut {NoStop}%
\bibitem [{\citenamefont {Qin}\ \emph {et~al.}(2022)\citenamefont {Qin},
  \citenamefont {Sch\"{a}fer}, \citenamefont {Andergassen}, \citenamefont
  {Corboz},\ and\ \citenamefont {Gull}}]{qin2021hubbard}%
  \BibitemOpen
  \bibfield  {author} {\bibinfo {author} {\bibfnamefont {M.}~\bibnamefont
  {Qin}}, \bibinfo {author} {\bibfnamefont {T.}~\bibnamefont {Sch\"{a}fer}},
  \bibinfo {author} {\bibfnamefont {S.}~\bibnamefont {Andergassen}}, \bibinfo
  {author} {\bibfnamefont {P.}~\bibnamefont {Corboz}},\ and\ \bibinfo {author}
  {\bibfnamefont {E.}~\bibnamefont {Gull}},\ }\bibfield  {title} {\bibinfo
  {title} {The {H}ubbard model: A computational perspective},\ }\href
  {https://doi.org/10.1146/annurev-conmatphys-090921-033948} {\bibfield
  {journal} {\bibinfo  {journal} {Annual Review of Condensed Matter Physics}\
  }\textbf {\bibinfo {volume} {13}},\ \bibinfo {pages} {275} (\bibinfo {year}
  {2022})}\BibitemShut {NoStop}%
\bibitem [{\citenamefont {LeBlanc}\ \emph {et~al.}(2015)\citenamefont
  {LeBlanc}, \citenamefont {Antipov}, \citenamefont {Becca}, \citenamefont
  {Bulik}, \citenamefont {Chan}, \citenamefont {Chung}, \citenamefont {Deng},
  \citenamefont {Ferrero}, \citenamefont {Henderson}, \citenamefont
  {Jim\'enez-Hoyos}, \citenamefont {Kozik}, \citenamefont {Liu}, \citenamefont
  {Millis}, \citenamefont {Prokof'ev}, \citenamefont {Qin}, \citenamefont
  {Scuseria}, \citenamefont {Shi}, \citenamefont {Svistunov}, \citenamefont
  {Tocchio}, \citenamefont {Tupitsyn}, \citenamefont {White}, \citenamefont
  {Zhang}, \citenamefont {Zheng}, \citenamefont {Zhu},\ and\ \citenamefont
  {Gull}}]{SimonsCollab2015}%
  \BibitemOpen
  \bibfield  {author} {\bibinfo {author} {\bibfnamefont {J.~P.~F.}\
  \bibnamefont {LeBlanc}}, \bibinfo {author} {\bibfnamefont {A.~E.}\
  \bibnamefont {Antipov}}, \bibinfo {author} {\bibfnamefont {F.}~\bibnamefont
  {Becca}}, \bibinfo {author} {\bibfnamefont {I.~W.}\ \bibnamefont {Bulik}},
  \bibinfo {author} {\bibfnamefont {G.~K.-L.}\ \bibnamefont {Chan}}, \bibinfo
  {author} {\bibfnamefont {C.-M.}\ \bibnamefont {Chung}}, \bibinfo {author}
  {\bibfnamefont {Y.}~\bibnamefont {Deng}}, \bibinfo {author} {\bibfnamefont
  {M.}~\bibnamefont {Ferrero}}, \bibinfo {author} {\bibfnamefont {T.~M.}\
  \bibnamefont {Henderson}}, \bibinfo {author} {\bibfnamefont {C.~A.}\
  \bibnamefont {Jim\'enez-Hoyos}}, \bibinfo {author} {\bibfnamefont
  {E.}~\bibnamefont {Kozik}}, \bibinfo {author} {\bibfnamefont {X.-W.}\
  \bibnamefont {Liu}}, \bibinfo {author} {\bibfnamefont {A.~J.}\ \bibnamefont
  {Millis}}, \bibinfo {author} {\bibfnamefont {N.~V.}\ \bibnamefont
  {Prokof'ev}}, \bibinfo {author} {\bibfnamefont {M.}~\bibnamefont {Qin}},
  \bibinfo {author} {\bibfnamefont {G.~E.}\ \bibnamefont {Scuseria}}, \bibinfo
  {author} {\bibfnamefont {H.}~\bibnamefont {Shi}}, \bibinfo {author}
  {\bibfnamefont {B.~V.}\ \bibnamefont {Svistunov}}, \bibinfo {author}
  {\bibfnamefont {L.~F.}\ \bibnamefont {Tocchio}}, \bibinfo {author}
  {\bibfnamefont {I.~S.}\ \bibnamefont {Tupitsyn}}, \bibinfo {author}
  {\bibfnamefont {S.~R.}\ \bibnamefont {White}}, \bibinfo {author}
  {\bibfnamefont {S.}~\bibnamefont {Zhang}}, \bibinfo {author} {\bibfnamefont
  {B.-X.}\ \bibnamefont {Zheng}}, \bibinfo {author} {\bibfnamefont
  {Z.}~\bibnamefont {Zhu}},\ and\ \bibinfo {author} {\bibfnamefont
  {E.}~\bibnamefont {Gull}} (\bibinfo {collaboration} {Simons Collaboration on
  the Many-Electron Problem}),\ }\bibfield  {title} {\bibinfo {title}
  {Solutions of the two-dimensional {H}ubbard model: Benchmarks and results
  from a wide range of numerical algorithms},\ }\href
  {https://doi.org/10.1103/PhysRevX.5.041041} {\bibfield  {journal} {\bibinfo
  {journal} {Phys. Rev. X}\ }\textbf {\bibinfo {volume} {5}},\ \bibinfo {pages}
  {041041} (\bibinfo {year} {2015})}\BibitemShut {NoStop}%
\bibitem [{\citenamefont {Qin}\ \emph {et~al.}(2020)\citenamefont {Qin},
  \citenamefont {Chung}, \citenamefont {Shi}, \citenamefont {Vitali},
  \citenamefont {Hubig}, \citenamefont {Schollw\"ock}, \citenamefont {White},\
  and\ \citenamefont {Zhang}}]{SimonsCollab2020}%
  \BibitemOpen
  \bibfield  {author} {\bibinfo {author} {\bibfnamefont {M.}~\bibnamefont
  {Qin}}, \bibinfo {author} {\bibfnamefont {C.-M.}\ \bibnamefont {Chung}},
  \bibinfo {author} {\bibfnamefont {H.}~\bibnamefont {Shi}}, \bibinfo {author}
  {\bibfnamefont {E.}~\bibnamefont {Vitali}}, \bibinfo {author} {\bibfnamefont
  {C.}~\bibnamefont {Hubig}}, \bibinfo {author} {\bibfnamefont
  {U.}~\bibnamefont {Schollw\"ock}}, \bibinfo {author} {\bibfnamefont {S.~R.}\
  \bibnamefont {White}},\ and\ \bibinfo {author} {\bibfnamefont
  {S.}~\bibnamefont {Zhang}} (\bibinfo {collaboration} {Simons Collaboration on
  the Many-Electron Problem}),\ }\bibfield  {title} {\bibinfo {title} {Absence
  of superconductivity in the pure two-dimensional {H}ubbard model},\ }\href
  {https://doi.org/10.1103/PhysRevX.10.031016} {\bibfield  {journal} {\bibinfo
  {journal} {Phys. Rev. X}\ }\textbf {\bibinfo {volume} {10}},\ \bibinfo
  {pages} {031016} (\bibinfo {year} {2020})}\BibitemShut {NoStop}%
\bibitem [{\citenamefont {Zhang}\ and\ \citenamefont {Rice}(1988)}]{ZR}%
  \BibitemOpen
  \bibfield  {author} {\bibinfo {author} {\bibfnamefont {F.~C.}\ \bibnamefont
  {Zhang}}\ and\ \bibinfo {author} {\bibfnamefont {T.~M.}\ \bibnamefont
  {Rice}},\ }\bibfield  {title} {\bibinfo {title} {Effective {H}amiltonian for
  the superconducting {Cu} oxides},\ }\href
  {https://doi.org/10.1103/PhysRevB.37.3759} {\bibfield  {journal} {\bibinfo
  {journal} {Phys. Rev. B}\ }\textbf {\bibinfo {volume} {37}},\ \bibinfo
  {pages} {3759} (\bibinfo {year} {1988})}\BibitemShut {NoStop}%
\bibitem [{\citenamefont {Sch\"uttler}\ and\ \citenamefont
  {Fedro}(1992)}]{fedro}%
  \BibitemOpen
  \bibfield  {author} {\bibinfo {author} {\bibfnamefont {H.-B.}\ \bibnamefont
  {Sch\"uttler}}\ and\ \bibinfo {author} {\bibfnamefont {A.~J.}\ \bibnamefont
  {Fedro}},\ }\bibfield  {title} {\bibinfo {title} {Copper-oxygen charge
  excitations and the effective-single-band theory of cuprate
  superconductors},\ }\href {https://doi.org/10.1103/PhysRevB.45.7588}
  {\bibfield  {journal} {\bibinfo  {journal} {Phys. Rev. B}\ }\textbf {\bibinfo
  {volume} {45}},\ \bibinfo {pages} {7588} (\bibinfo {year}
  {1992})}\BibitemShut {NoStop}%
\bibitem [{\citenamefont {Simon}\ \emph {et~al.}(1993)\citenamefont {Simon},
  \citenamefont {Balina},\ and\ \citenamefont {Aligia}}]{simon}%
  \BibitemOpen
  \bibfield  {author} {\bibinfo {author} {\bibfnamefont {M.}~\bibnamefont
  {Simon}}, \bibinfo {author} {\bibfnamefont {M.}~\bibnamefont {Balina}},\ and\
  \bibinfo {author} {\bibfnamefont {A.}~\bibnamefont {Aligia}},\ }\bibfield
  {title} {\bibinfo {title} {Effective one-band {H}amiltonian for cuprate
  superconductor metal-insulator transition},\ }\href
  {https://doi.org/https://doi.org/10.1016/0921-4534(93)90529-Y} {\bibfield
  {journal} {\bibinfo  {journal} {Physica C: Superconductivity}\ }\textbf
  {\bibinfo {volume} {206}},\ \bibinfo {pages} {297} (\bibinfo {year}
  {1993})}\BibitemShut {NoStop}%
\bibitem [{\citenamefont {Sim\'on}\ and\ \citenamefont {Aligia}(1993)}]{brt}%
  \BibitemOpen
  \bibfield  {author} {\bibinfo {author} {\bibfnamefont {M.~E.}\ \bibnamefont
  {Sim\'on}}\ and\ \bibinfo {author} {\bibfnamefont {A.~A.}\ \bibnamefont
  {Aligia}},\ }\bibfield  {title} {\bibinfo {title} {{B}rinkman-{R}ice
  transition in layered perovskites},\ }\href
  {https://doi.org/10.1103/PhysRevB.48.7471} {\bibfield  {journal} {\bibinfo
  {journal} {Phys. Rev. B}\ }\textbf {\bibinfo {volume} {48}},\ \bibinfo
  {pages} {7471} (\bibinfo {year} {1993})}\BibitemShut {NoStop}%
\bibitem [{\citenamefont {Feiner}\ \emph {et~al.}(1996)\citenamefont {Feiner},
  \citenamefont {Jefferson},\ and\ \citenamefont {Raimondi}}]{fei}%
  \BibitemOpen
  \bibfield  {author} {\bibinfo {author} {\bibfnamefont {L.~F.}\ \bibnamefont
  {Feiner}}, \bibinfo {author} {\bibfnamefont {J.~H.}\ \bibnamefont
  {Jefferson}},\ and\ \bibinfo {author} {\bibfnamefont {R.}~\bibnamefont
  {Raimondi}},\ }\bibfield  {title} {\bibinfo {title} {Effective single-band
  models for the high-{$\mathit{T}_{\mathit{c}}$} cuprates. {I}. {C}oulomb
  interactions},\ }\href {https://doi.org/10.1103/PhysRevB.53.8751} {\bibfield
  {journal} {\bibinfo  {journal} {Phys. Rev. B}\ }\textbf {\bibinfo {volume}
  {53}},\ \bibinfo {pages} {8751} (\bibinfo {year} {1996})}\BibitemShut
  {NoStop}%
\bibitem [{\citenamefont {Belinicher}\ \emph {et~al.}(1994)\citenamefont
  {Belinicher}, \citenamefont {Chernyshev},\ and\ \citenamefont
  {Popovich}}]{bel}%
  \BibitemOpen
  \bibfield  {author} {\bibinfo {author} {\bibfnamefont {V.~I.}\ \bibnamefont
  {Belinicher}}, \bibinfo {author} {\bibfnamefont {A.~L.}\ \bibnamefont
  {Chernyshev}},\ and\ \bibinfo {author} {\bibfnamefont {L.~V.}\ \bibnamefont
  {Popovich}},\ }\bibfield  {title} {\bibinfo {title} {Range of the {$t$-$J$}
  model parameters for {$\mathrm{CuO}_{2}$} planes: Experimental data
  constraints},\ }\href {https://doi.org/10.1103/PhysRevB.50.13768} {\bibfield
  {journal} {\bibinfo  {journal} {Phys. Rev. B}\ }\textbf {\bibinfo {volume}
  {50}},\ \bibinfo {pages} {13768} (\bibinfo {year} {1994})}\BibitemShut
  {NoStop}%
\bibitem [{\citenamefont {Belinicher}\ and\ \citenamefont
  {Chernyshev}(1994)}]{Belinicher94}%
  \BibitemOpen
  \bibfield  {author} {\bibinfo {author} {\bibfnamefont {V.~I.}\ \bibnamefont
  {Belinicher}}\ and\ \bibinfo {author} {\bibfnamefont {A.~L.}\ \bibnamefont
  {Chernyshev}},\ }\bibfield  {title} {\bibinfo {title} {Consistent low-energy
  reduction of the three-band model for copper oxides with {O-O} hopping to the
  effective {$t$-$J$} model},\ }\href
  {https://doi.org/10.1103/PhysRevB.49.9746} {\bibfield  {journal} {\bibinfo
  {journal} {Phys. Rev. B}\ }\textbf {\bibinfo {volume} {49}},\ \bibinfo
  {pages} {9746} (\bibinfo {year} {1994})}\BibitemShut {NoStop}%
\bibitem [{\citenamefont {Aligia}\ \emph {et~al.}(1994)\citenamefont {Aligia},
  \citenamefont {Sim\'on},\ and\ \citenamefont {Batista}}]{ali94}%
  \BibitemOpen
  \bibfield  {author} {\bibinfo {author} {\bibfnamefont {A.~A.}\ \bibnamefont
  {Aligia}}, \bibinfo {author} {\bibfnamefont {M.~E.}\ \bibnamefont
  {Sim\'on}},\ and\ \bibinfo {author} {\bibfnamefont {C.~D.}\ \bibnamefont
  {Batista}},\ }\bibfield  {title} {\bibinfo {title} {Systematic derivation of
  a generalized {$t$-$J$} model},\ }\href
  {https://doi.org/10.1103/PhysRevB.49.13061} {\bibfield  {journal} {\bibinfo
  {journal} {Phys. Rev. B}\ }\textbf {\bibinfo {volume} {49}},\ \bibinfo
  {pages} {13061} (\bibinfo {year} {1994})}\BibitemShut {NoStop}%
\bibitem [{\citenamefont {Ebrahimnejad}\ \emph {et~al.}(2014)\citenamefont
  {Ebrahimnejad}, \citenamefont {Sawatzky},\ and\ \citenamefont
  {Berciu}}]{ebra}%
  \BibitemOpen
  \bibfield  {author} {\bibinfo {author} {\bibfnamefont {H.}~\bibnamefont
  {Ebrahimnejad}}, \bibinfo {author} {\bibfnamefont {G.~A.}\ \bibnamefont
  {Sawatzky}},\ and\ \bibinfo {author} {\bibfnamefont {M.}~\bibnamefont
  {Berciu}},\ }\bibfield  {title} {\bibinfo {title} {The dynamics of a doped
  hole in a cuprate is not controlled by spin fluctuations},\ }\href
  {https://doi.org/10.1038/nphys3130} {\bibfield  {journal} {\bibinfo
  {journal} {Nature Physics}\ }\textbf {\bibinfo {volume} {10}},\ \bibinfo
  {pages} {951} (\bibinfo {year} {2014})}\BibitemShut {NoStop}%
\bibitem [{\citenamefont {Hamad}\ \emph {et~al.}(2021)\citenamefont {Hamad},
  \citenamefont {Manuel},\ and\ \citenamefont {Aligia}}]{hamad2}%
  \BibitemOpen
  \bibfield  {author} {\bibinfo {author} {\bibfnamefont {I.~J.}\ \bibnamefont
  {Hamad}}, \bibinfo {author} {\bibfnamefont {L.~O.}\ \bibnamefont {Manuel}},\
  and\ \bibinfo {author} {\bibfnamefont {A.~A.}\ \bibnamefont {Aligia}},\
  }\bibfield  {title} {\bibinfo {title} {Magnon-assisted dynamics of a hole
  doped in a cuprate superconductor},\ }\href
  {https://doi.org/10.1103/PhysRevB.103.144510} {\bibfield  {journal} {\bibinfo
   {journal} {Phys. Rev. B}\ }\textbf {\bibinfo {volume} {103}},\ \bibinfo
  {pages} {144510} (\bibinfo {year} {2021})}\BibitemShut {NoStop}%
\bibitem [{\citenamefont {Adolphs}\ \emph {et~al.}(2016)\citenamefont
  {Adolphs}, \citenamefont {Moser}, \citenamefont {Sawatzky},\ and\
  \citenamefont {Berciu}}]{adol}%
  \BibitemOpen
  \bibfield  {author} {\bibinfo {author} {\bibfnamefont {C.~P.~J.}\
  \bibnamefont {Adolphs}}, \bibinfo {author} {\bibfnamefont {S.}~\bibnamefont
  {Moser}}, \bibinfo {author} {\bibfnamefont {G.~A.}\ \bibnamefont
  {Sawatzky}},\ and\ \bibinfo {author} {\bibfnamefont {M.}~\bibnamefont
  {Berciu}},\ }\bibfield  {title} {\bibinfo {title} {Non-{Z}hang-{R}ice singlet
  character of the first ionization state of {T}-{CuO}},\ }\href
  {https://doi.org/10.1103/PhysRevLett.116.087002} {\bibfield  {journal}
  {\bibinfo  {journal} {Phys. Rev. Lett.}\ }\textbf {\bibinfo {volume} {116}},\
  \bibinfo {pages} {087002} (\bibinfo {year} {2016})}\BibitemShut {NoStop}%
\bibitem [{\citenamefont {Hamad}\ \emph {et~al.}(2018)\citenamefont {Hamad},
  \citenamefont {Manuel},\ and\ \citenamefont {Aligia}}]{hamad1}%
  \BibitemOpen
  \bibfield  {author} {\bibinfo {author} {\bibfnamefont {I.~J.}\ \bibnamefont
  {Hamad}}, \bibinfo {author} {\bibfnamefont {L.~O.}\ \bibnamefont {Manuel}},\
  and\ \bibinfo {author} {\bibfnamefont {A.~A.}\ \bibnamefont {Aligia}},\
  }\bibfield  {title} {\bibinfo {title} {Generalized one-band model based on
  {Z}hang-{R}ice singlets for tetragonal {CuO}},\ }\href
  {https://doi.org/10.1103/PhysRevLett.120.177001} {\bibfield  {journal}
  {\bibinfo  {journal} {Phys. Rev. Lett.}\ }\textbf {\bibinfo {volume} {120}},\
  \bibinfo {pages} {177001} (\bibinfo {year} {2018})}\BibitemShut {NoStop}%
\bibitem [{\citenamefont {Aligia}(2020)}]{ali20}%
  \BibitemOpen
  \bibfield  {author} {\bibinfo {author} {\bibfnamefont {A.~A.}\ \bibnamefont
  {Aligia}},\ }\bibfield  {title} {\bibinfo {title} {Comment on ``{R}elevance
  of {C}u-$3d$ multiplet structure in models of high-${T}_{c}$ cuprates''},\
  }\href {https://doi.org/10.1103/PhysRevB.102.117101} {\bibfield  {journal}
  {\bibinfo  {journal} {Phys. Rev. B}\ }\textbf {\bibinfo {volume} {102}},\
  \bibinfo {pages} {117101} (\bibinfo {year} {2020})}\BibitemShut {NoStop}%
\bibitem [{\citenamefont {Santoso}\ \emph {et~al.}(2017)\citenamefont
  {Santoso}, \citenamefont {Ku}, \citenamefont {Shirakawa}, \citenamefont
  {Neuber}, \citenamefont {Yin}, \citenamefont {Enoki}, \citenamefont {Fujita},
  \citenamefont {Liang}, \citenamefont {Venkatesan}, \citenamefont {Sawatzky},
  \citenamefont {Kotlov}, \citenamefont {Yunoki}, \citenamefont {R\"ubhausen},\
  and\ \citenamefont {Rusydi}}]{Santoso2017}%
  \BibitemOpen
  \bibfield  {author} {\bibinfo {author} {\bibfnamefont {I.}~\bibnamefont
  {Santoso}}, \bibinfo {author} {\bibfnamefont {W.}~\bibnamefont {Ku}},
  \bibinfo {author} {\bibfnamefont {T.}~\bibnamefont {Shirakawa}}, \bibinfo
  {author} {\bibfnamefont {G.}~\bibnamefont {Neuber}}, \bibinfo {author}
  {\bibfnamefont {X.}~\bibnamefont {Yin}}, \bibinfo {author} {\bibfnamefont
  {M.}~\bibnamefont {Enoki}}, \bibinfo {author} {\bibfnamefont
  {M.}~\bibnamefont {Fujita}}, \bibinfo {author} {\bibfnamefont
  {R.}~\bibnamefont {Liang}}, \bibinfo {author} {\bibfnamefont
  {T.}~\bibnamefont {Venkatesan}}, \bibinfo {author} {\bibfnamefont {G.~A.}\
  \bibnamefont {Sawatzky}}, \bibinfo {author} {\bibfnamefont {A.}~\bibnamefont
  {Kotlov}}, \bibinfo {author} {\bibfnamefont {S.}~\bibnamefont {Yunoki}},
  \bibinfo {author} {\bibfnamefont {M.}~\bibnamefont {R\"ubhausen}},\ and\
  \bibinfo {author} {\bibfnamefont {A.}~\bibnamefont {Rusydi}},\ }\bibfield
  {title} {\bibinfo {title} {Unraveling local spin polarization of
  {Z}hang-{R}ice singlet in lightly hole-doped cuprates using high-energy
  optical conductivity},\ }\href {https://doi.org/10.1103/PhysRevB.95.165108}
  {\bibfield  {journal} {\bibinfo  {journal} {Phys. Rev. B}\ }\textbf {\bibinfo
  {volume} {95}},\ \bibinfo {pages} {165108} (\bibinfo {year}
  {2017})}\BibitemShut {NoStop}%
\bibitem [{\citenamefont {Penc}\ and\ \citenamefont {Stephan}(2000)}]{penc}%
  \BibitemOpen
  \bibfield  {author} {\bibinfo {author} {\bibfnamefont {K.}~\bibnamefont
  {Penc}}\ and\ \bibinfo {author} {\bibfnamefont {W.}~\bibnamefont {Stephan}},\
  }\bibfield  {title} {\bibinfo {title} {Dynamical correlations in
  one-dimensional charge-transfer insulators},\ }\href
  {https://doi.org/10.1103/physrevb.62.12707} {\bibfield  {journal} {\bibinfo
  {journal} {Physical Review B}\ }\textbf {\bibinfo {volume} {62}},\ \bibinfo
  {pages} {12707} (\bibinfo {year} {2000})}\BibitemShut {NoStop}%
\bibitem [{\citenamefont {Neudert}\ \emph {et~al.}(2000)\citenamefont
  {Neudert}, \citenamefont {Drechsler}, \citenamefont {M\'alek}, \citenamefont
  {Rosner}, \citenamefont {Kielwein}, \citenamefont {Hu}, \citenamefont
  {Knupfer}, \citenamefont {Golden}, \citenamefont {Fink}, \citenamefont
  {N\"ucker}, \citenamefont {Merz}, \citenamefont {Schuppler}, \citenamefont
  {Motoyama}, \citenamefont {Eisaki}, \citenamefont {Uchida}, \citenamefont
  {Domke},\ and\ \citenamefont {Kaindl}}]{neudert}%
  \BibitemOpen
  \bibfield  {author} {\bibinfo {author} {\bibfnamefont {R.}~\bibnamefont
  {Neudert}}, \bibinfo {author} {\bibfnamefont {S.-L.}\ \bibnamefont
  {Drechsler}}, \bibinfo {author} {\bibfnamefont {J.}~\bibnamefont {M\'alek}},
  \bibinfo {author} {\bibfnamefont {H.}~\bibnamefont {Rosner}}, \bibinfo
  {author} {\bibfnamefont {M.}~\bibnamefont {Kielwein}}, \bibinfo {author}
  {\bibfnamefont {Z.}~\bibnamefont {Hu}}, \bibinfo {author} {\bibfnamefont
  {M.}~\bibnamefont {Knupfer}}, \bibinfo {author} {\bibfnamefont {M.~S.}\
  \bibnamefont {Golden}}, \bibinfo {author} {\bibfnamefont {J.}~\bibnamefont
  {Fink}}, \bibinfo {author} {\bibfnamefont {N.}~\bibnamefont {N\"ucker}},
  \bibinfo {author} {\bibfnamefont {M.}~\bibnamefont {Merz}}, \bibinfo {author}
  {\bibfnamefont {S.}~\bibnamefont {Schuppler}}, \bibinfo {author}
  {\bibfnamefont {N.}~\bibnamefont {Motoyama}}, \bibinfo {author}
  {\bibfnamefont {H.}~\bibnamefont {Eisaki}}, \bibinfo {author} {\bibfnamefont
  {S.}~\bibnamefont {Uchida}}, \bibinfo {author} {\bibfnamefont
  {M.}~\bibnamefont {Domke}},\ and\ \bibinfo {author} {\bibfnamefont
  {G.}~\bibnamefont {Kaindl}},\ }\bibfield  {title} {\bibinfo {title}
  {Four-band extended {H}ubbard {H}amiltonian for the one-dimensional cuprate
  {$\mathrm{Sr}_{2}\mathrm{CuO}_{3}:$} distribution of oxygen holes and its
  relation to strong intersite {C}oulomb interaction},\ }\href
  {https://doi.org/10.1103/PhysRevB.62.10752} {\bibfield  {journal} {\bibinfo
  {journal} {Phys. Rev. B}\ }\textbf {\bibinfo {volume} {62}},\ \bibinfo
  {pages} {10752} (\bibinfo {year} {2000})}\BibitemShut {NoStop}%
\bibitem [{\citenamefont {Kim}\ \emph {et~al.}(1996)\citenamefont {Kim},
  \citenamefont {Matsuura}, \citenamefont {Shen}, \citenamefont {Motoyama},
  \citenamefont {Eisaki}, \citenamefont {Uchida}, \citenamefont {Tohyama},\
  and\ \citenamefont {Maekawa}}]{Kim1996observation}%
  \BibitemOpen
  \bibfield  {author} {\bibinfo {author} {\bibfnamefont {C.}~\bibnamefont
  {Kim}}, \bibinfo {author} {\bibfnamefont {A.~Y.}\ \bibnamefont {Matsuura}},
  \bibinfo {author} {\bibfnamefont {Z.-X.}\ \bibnamefont {Shen}}, \bibinfo
  {author} {\bibfnamefont {N.}~\bibnamefont {Motoyama}}, \bibinfo {author}
  {\bibfnamefont {H.}~\bibnamefont {Eisaki}}, \bibinfo {author} {\bibfnamefont
  {S.}~\bibnamefont {Uchida}}, \bibinfo {author} {\bibfnamefont
  {T.}~\bibnamefont {Tohyama}},\ and\ \bibinfo {author} {\bibfnamefont
  {S.}~\bibnamefont {Maekawa}},\ }\bibfield  {title} {\bibinfo {title}
  {Observation of spin-charge separation in one-dimensional
  {SrCu${\mathrm{O}}_{2}$}},\ }\href
  {https://doi.org/10.1103/PhysRevLett.77.4054} {\bibfield  {journal} {\bibinfo
   {journal} {Phys. Rev. Lett.}\ }\textbf {\bibinfo {volume} {77}},\ \bibinfo
  {pages} {4054} (\bibinfo {year} {1996})}\BibitemShut {NoStop}%
\bibitem [{\citenamefont {Kojima}\ \emph {et~al.}(1997)\citenamefont {Kojima},
  \citenamefont {Fudamoto}, \citenamefont {Larkin}, \citenamefont {Luke},
  \citenamefont {Merrin}, \citenamefont {Nachumi}, \citenamefont {Uemura},
  \citenamefont {Motoyama}, \citenamefont {Eisaki}, \citenamefont {Uchida},
  \citenamefont {Yamada}, \citenamefont {Endoh}, \citenamefont {Hosoya},
  \citenamefont {Sternlieb},\ and\ \citenamefont
  {Shirane}}]{Kojima1997reducton}%
  \BibitemOpen
  \bibfield  {author} {\bibinfo {author} {\bibfnamefont {K.~M.}\ \bibnamefont
  {Kojima}}, \bibinfo {author} {\bibfnamefont {Y.}~\bibnamefont {Fudamoto}},
  \bibinfo {author} {\bibfnamefont {M.}~\bibnamefont {Larkin}}, \bibinfo
  {author} {\bibfnamefont {G.~M.}\ \bibnamefont {Luke}}, \bibinfo {author}
  {\bibfnamefont {J.}~\bibnamefont {Merrin}}, \bibinfo {author} {\bibfnamefont
  {B.}~\bibnamefont {Nachumi}}, \bibinfo {author} {\bibfnamefont {Y.~J.}\
  \bibnamefont {Uemura}}, \bibinfo {author} {\bibfnamefont {N.}~\bibnamefont
  {Motoyama}}, \bibinfo {author} {\bibfnamefont {H.}~\bibnamefont {Eisaki}},
  \bibinfo {author} {\bibfnamefont {S.}~\bibnamefont {Uchida}}, \bibinfo
  {author} {\bibfnamefont {K.}~\bibnamefont {Yamada}}, \bibinfo {author}
  {\bibfnamefont {Y.}~\bibnamefont {Endoh}}, \bibinfo {author} {\bibfnamefont
  {S.}~\bibnamefont {Hosoya}}, \bibinfo {author} {\bibfnamefont {B.~J.}\
  \bibnamefont {Sternlieb}},\ and\ \bibinfo {author} {\bibfnamefont
  {G.}~\bibnamefont {Shirane}},\ }\bibfield  {title} {\bibinfo {title}
  {Reduction of ordered moment and {N}{\'e}el temperature of
  quasi-one-dimensional antiferromagnets {Sr}$_2${CuO}$_3$ and
  {Ca}$_2${CuO}$_3$},\ }\href {https://doi.org/10.1103/PhysRevLett.78.1787}
  {\bibfield  {journal} {\bibinfo  {journal} {Phys. Rev. Lett.}\ }\textbf
  {\bibinfo {volume} {78}},\ \bibinfo {pages} {1787} (\bibinfo {year}
  {1997})}\BibitemShut {NoStop}%
\bibitem [{\citenamefont {Eisaki}\ \emph {et~al.}(1997)\citenamefont {Eisaki},
  \citenamefont {Motoyama},\ and\ \citenamefont {Uchida}}]{Eisaki1997spin}%
  \BibitemOpen
  \bibfield  {author} {\bibinfo {author} {\bibfnamefont {H.}~\bibnamefont
  {Eisaki}}, \bibinfo {author} {\bibfnamefont {N.}~\bibnamefont {Motoyama}},\
  and\ \bibinfo {author} {\bibfnamefont {S.}~\bibnamefont {Uchida}},\
  }\bibfield  {title} {\bibinfo {title} {Spin magnetic susceptibility of {Cu-O}
  chains in {Sr$_2$CuO$_3$} and {SrCuO$_2$}},\ }\href
  {https://doi.org/https://doi.org/10.1016/S0921-4534(97)00730-2} {\bibfield
  {journal} {\bibinfo  {journal} {Physica C: Superconductivity}\ }\textbf
  {\bibinfo {volume} {282-287}},\ \bibinfo {pages} {1323} (\bibinfo {year}
  {1997})}\BibitemShut {NoStop}%
\bibitem [{\citenamefont {Zaliznyak}\ \emph {et~al.}(2004)\citenamefont
  {Zaliznyak}, \citenamefont {Woo}, \citenamefont {Perring}, \citenamefont
  {Broholm}, \citenamefont {Frost},\ and\ \citenamefont
  {Takagi}}]{Zaliznyak2004spinons}%
  \BibitemOpen
  \bibfield  {author} {\bibinfo {author} {\bibfnamefont {I.~A.}\ \bibnamefont
  {Zaliznyak}}, \bibinfo {author} {\bibfnamefont {H.}~\bibnamefont {Woo}},
  \bibinfo {author} {\bibfnamefont {T.~G.}\ \bibnamefont {Perring}}, \bibinfo
  {author} {\bibfnamefont {C.~L.}\ \bibnamefont {Broholm}}, \bibinfo {author}
  {\bibfnamefont {C.~D.}\ \bibnamefont {Frost}},\ and\ \bibinfo {author}
  {\bibfnamefont {H.}~\bibnamefont {Takagi}},\ }\bibfield  {title} {\bibinfo
  {title} {Spinons in the strongly correlated copper oxide chains in
  {$\mathrm{S}\mathrm{r}\mathrm{C}\mathrm{u}\mathrm{O}_{2}$}},\ }\href
  {https://doi.org/10.1103/PhysRevLett.93.087202} {\bibfield  {journal}
  {\bibinfo  {journal} {Phys. Rev. Lett.}\ }\textbf {\bibinfo {volume} {93}},\
  \bibinfo {pages} {087202} (\bibinfo {year} {2004})}\BibitemShut {NoStop}%
\bibitem [{\citenamefont {Kim}\ \emph {et~al.}(2006)\citenamefont {Kim},
  \citenamefont {Koh}, \citenamefont {Rotenberg}, \citenamefont {Oh},
  \citenamefont {Eisaki}, \citenamefont {Motoyama}, \citenamefont {Uchida},
  \citenamefont {Tohyama}, \citenamefont {Maekawa}, \citenamefont {Shen},\ and\
  \citenamefont {Kim}}]{Kim2006distinct}%
  \BibitemOpen
  \bibfield  {author} {\bibinfo {author} {\bibfnamefont {B.~J.}\ \bibnamefont
  {Kim}}, \bibinfo {author} {\bibfnamefont {H.}~\bibnamefont {Koh}}, \bibinfo
  {author} {\bibfnamefont {E.}~\bibnamefont {Rotenberg}}, \bibinfo {author}
  {\bibfnamefont {S.~J.}\ \bibnamefont {Oh}}, \bibinfo {author} {\bibfnamefont
  {H.}~\bibnamefont {Eisaki}}, \bibinfo {author} {\bibfnamefont
  {N.}~\bibnamefont {Motoyama}}, \bibinfo {author} {\bibfnamefont
  {S.}~\bibnamefont {Uchida}}, \bibinfo {author} {\bibfnamefont
  {T.}~\bibnamefont {Tohyama}}, \bibinfo {author} {\bibfnamefont
  {S.}~\bibnamefont {Maekawa}}, \bibinfo {author} {\bibfnamefont {Z.~X.}\
  \bibnamefont {Shen}},\ and\ \bibinfo {author} {\bibfnamefont
  {C.}~\bibnamefont {Kim}},\ }\bibfield  {title} {\bibinfo {title} {Distinct
  spinon and holon dispersions in photoemission spectral functions from
  one-dimensional {SrCuO$_2$}},\ }\href {https://doi.org/10.1038/nphys316}
  {\bibfield  {journal} {\bibinfo  {journal} {Nature Physics}\ }\textbf
  {\bibinfo {volume} {2}},\ \bibinfo {pages} {397} (\bibinfo {year}
  {2006})}\BibitemShut {NoStop}%
\bibitem [{\citenamefont {Enderle}\ \emph {et~al.}(2010)\citenamefont
  {Enderle}, \citenamefont {F\aa{}k}, \citenamefont {Mikeska}, \citenamefont
  {Kremer}, \citenamefont {Prokofiev},\ and\ \citenamefont
  {Assmus}}]{Enderle2010twospinon}%
  \BibitemOpen
  \bibfield  {author} {\bibinfo {author} {\bibfnamefont {M.}~\bibnamefont
  {Enderle}}, \bibinfo {author} {\bibfnamefont {B.}~\bibnamefont {F\aa{}k}},
  \bibinfo {author} {\bibfnamefont {H.-J.}\ \bibnamefont {Mikeska}}, \bibinfo
  {author} {\bibfnamefont {R.~K.}\ \bibnamefont {Kremer}}, \bibinfo {author}
  {\bibfnamefont {A.}~\bibnamefont {Prokofiev}},\ and\ \bibinfo {author}
  {\bibfnamefont {W.}~\bibnamefont {Assmus}},\ }\bibfield  {title} {\bibinfo
  {title} {Two-spinon and four-spinon continuum in a frustrated ferromagnetic
  spin-$1/2$ chain},\ }\href {https://doi.org/10.1103/PhysRevLett.104.237207}
  {\bibfield  {journal} {\bibinfo  {journal} {Phys. Rev. Lett.}\ }\textbf
  {\bibinfo {volume} {104}},\ \bibinfo {pages} {237207} (\bibinfo {year}
  {2010})}\BibitemShut {NoStop}%
\bibitem [{\citenamefont {Drechsler}\ \emph {et~al.}(2011)\citenamefont
  {Drechsler}, \citenamefont {Nishimoto}, \citenamefont {Kuzian}, \citenamefont
  {M\'alek}, \citenamefont {Lorenz}, \citenamefont {Richter}, \citenamefont
  {van~den Brink}, \citenamefont {Schmitt},\ and\ \citenamefont
  {Rosner}}]{Drechsler2011comment}%
  \BibitemOpen
  \bibfield  {author} {\bibinfo {author} {\bibfnamefont {S.-L.}\ \bibnamefont
  {Drechsler}}, \bibinfo {author} {\bibfnamefont {S.}~\bibnamefont
  {Nishimoto}}, \bibinfo {author} {\bibfnamefont {R.~O.}\ \bibnamefont
  {Kuzian}}, \bibinfo {author} {\bibfnamefont {J.}~\bibnamefont {M\'alek}},
  \bibinfo {author} {\bibfnamefont {W.~E.~A.}\ \bibnamefont {Lorenz}}, \bibinfo
  {author} {\bibfnamefont {J.}~\bibnamefont {Richter}}, \bibinfo {author}
  {\bibfnamefont {J.}~\bibnamefont {van~den Brink}}, \bibinfo {author}
  {\bibfnamefont {M.}~\bibnamefont {Schmitt}},\ and\ \bibinfo {author}
  {\bibfnamefont {H.}~\bibnamefont {Rosner}},\ }\bibfield  {title} {\bibinfo
  {title} {Comment on ``two-spinon and four-spinon continuum in a frustrated
  ferromagnetic spin-$1/2$ chain''},\ }\href
  {https://doi.org/10.1103/PhysRevLett.106.219701} {\bibfield  {journal}
  {\bibinfo  {journal} {Phys. Rev. Lett.}\ }\textbf {\bibinfo {volume} {106}},\
  \bibinfo {pages} {219701} (\bibinfo {year} {2011})}\BibitemShut {NoStop}%
\bibitem [{\citenamefont {Schlappa}\ \emph {et~al.}(2012)\citenamefont
  {Schlappa}, \citenamefont {Wohlfeld}, \citenamefont {Zhou}, \citenamefont
  {Mourigal}, \citenamefont {Haverkort}, \citenamefont {Strocov}, \citenamefont
  {Hozoi}, \citenamefont {Monney}, \citenamefont {Nishimoto}, \citenamefont
  {Singh}, \citenamefont {Revcolevschi}, \citenamefont {Caux}, \citenamefont
  {Patthey}, \citenamefont {R{\o}nnow}, \citenamefont {van~den Brink},\ and\
  \citenamefont {Schmitt}}]{Schlappa2012spin}%
  \BibitemOpen
  \bibfield  {author} {\bibinfo {author} {\bibfnamefont {J.}~\bibnamefont
  {Schlappa}}, \bibinfo {author} {\bibfnamefont {K.}~\bibnamefont {Wohlfeld}},
  \bibinfo {author} {\bibfnamefont {K.~J.}\ \bibnamefont {Zhou}}, \bibinfo
  {author} {\bibfnamefont {M.}~\bibnamefont {Mourigal}}, \bibinfo {author}
  {\bibfnamefont {M.~W.}\ \bibnamefont {Haverkort}}, \bibinfo {author}
  {\bibfnamefont {V.~N.}\ \bibnamefont {Strocov}}, \bibinfo {author}
  {\bibfnamefont {L.}~\bibnamefont {Hozoi}}, \bibinfo {author} {\bibfnamefont
  {C.}~\bibnamefont {Monney}}, \bibinfo {author} {\bibfnamefont
  {S.}~\bibnamefont {Nishimoto}}, \bibinfo {author} {\bibfnamefont
  {S.}~\bibnamefont {Singh}}, \bibinfo {author} {\bibfnamefont
  {A.}~\bibnamefont {Revcolevschi}}, \bibinfo {author} {\bibfnamefont {J.~S.}\
  \bibnamefont {Caux}}, \bibinfo {author} {\bibfnamefont {L.}~\bibnamefont
  {Patthey}}, \bibinfo {author} {\bibfnamefont {H.~M.}\ \bibnamefont
  {R{\o}nnow}}, \bibinfo {author} {\bibfnamefont {J.}~\bibnamefont {van~den
  Brink}},\ and\ \bibinfo {author} {\bibfnamefont {T.}~\bibnamefont
  {Schmitt}},\ }\bibfield  {title} {\bibinfo {title} {Spin--orbital separation
  in the quasi-one-dimensional {M}ott insulator {Sr$_2$CuO$_3$}},\ }\href
  {https://doi.org/10.1038/nature10974} {\bibfield  {journal} {\bibinfo
  {journal} {Nature}\ }\textbf {\bibinfo {volume} {485}},\ \bibinfo {pages}
  {82} (\bibinfo {year} {2012})}\BibitemShut {NoStop}%
\bibitem [{\citenamefont {Bisogni}\ \emph {et~al.}(2014)\citenamefont
  {Bisogni}, \citenamefont {Kourtis}, \citenamefont {Monney}, \citenamefont
  {Zhou}, \citenamefont {Kraus}, \citenamefont {Sekar}, \citenamefont
  {Strocov}, \citenamefont {B\"uchner}, \citenamefont {van~den Brink},
  \citenamefont {Braicovich}, \citenamefont {Schmitt}, \citenamefont
  {Daghofer},\ and\ \citenamefont {Geck}}]{Valentina2014femtosecond}%
  \BibitemOpen
  \bibfield  {author} {\bibinfo {author} {\bibfnamefont {V.}~\bibnamefont
  {Bisogni}}, \bibinfo {author} {\bibfnamefont {S.}~\bibnamefont {Kourtis}},
  \bibinfo {author} {\bibfnamefont {C.}~\bibnamefont {Monney}}, \bibinfo
  {author} {\bibfnamefont {K.}~\bibnamefont {Zhou}}, \bibinfo {author}
  {\bibfnamefont {R.}~\bibnamefont {Kraus}}, \bibinfo {author} {\bibfnamefont
  {C.}~\bibnamefont {Sekar}}, \bibinfo {author} {\bibfnamefont
  {V.}~\bibnamefont {Strocov}}, \bibinfo {author} {\bibfnamefont
  {B.}~\bibnamefont {B\"uchner}}, \bibinfo {author} {\bibfnamefont
  {J.}~\bibnamefont {van~den Brink}}, \bibinfo {author} {\bibfnamefont
  {L.}~\bibnamefont {Braicovich}}, \bibinfo {author} {\bibfnamefont
  {T.}~\bibnamefont {Schmitt}}, \bibinfo {author} {\bibfnamefont
  {M.}~\bibnamefont {Daghofer}},\ and\ \bibinfo {author} {\bibfnamefont
  {J.}~\bibnamefont {Geck}},\ }\bibfield  {title} {\bibinfo {title}
  {Femtosecond dynamics of momentum-dependent magnetic excitations from
  resonant inelastic x-ray scattering in {CaCu}$_2${O}$_3$},\ }\href
  {https://doi.org/10.1103/PhysRevLett.112.147401} {\bibfield  {journal}
  {\bibinfo  {journal} {Phys. Rev. Lett.}\ }\textbf {\bibinfo {volume} {112}},\
  \bibinfo {pages} {147401} (\bibinfo {year} {2014})}\BibitemShut {NoStop}%
\bibitem [{\citenamefont {Walters}\ \emph
  {et~al.}(2009{\natexlab{a}})\citenamefont {Walters}, \citenamefont {Perring},
  \citenamefont {Caux}, \citenamefont {Savici}, \citenamefont {Gu},
  \citenamefont {Lee}, \citenamefont {Ku},\ and\ \citenamefont
  {Zaliznyak}}]{walters}%
  \BibitemOpen
  \bibfield  {author} {\bibinfo {author} {\bibfnamefont {A.~C.}\ \bibnamefont
  {Walters}}, \bibinfo {author} {\bibfnamefont {T.~G.}\ \bibnamefont
  {Perring}}, \bibinfo {author} {\bibfnamefont {J.-S.}\ \bibnamefont {Caux}},
  \bibinfo {author} {\bibfnamefont {A.~T.}\ \bibnamefont {Savici}}, \bibinfo
  {author} {\bibfnamefont {G.~D.}\ \bibnamefont {Gu}}, \bibinfo {author}
  {\bibfnamefont {C.-C.}\ \bibnamefont {Lee}}, \bibinfo {author} {\bibfnamefont
  {W.}~\bibnamefont {Ku}},\ and\ \bibinfo {author} {\bibfnamefont {I.~A.}\
  \bibnamefont {Zaliznyak}},\ }\bibfield  {title} {\bibinfo {title} {Effect of
  covalent bonding on magnetism and the missing neutron intensity in copper
  oxide compounds},\ }\href {https://doi.org/10.1038/nphys1405} {\bibfield
  {journal} {\bibinfo  {journal} {Nature Physics}\ }\textbf {\bibinfo {volume}
  {5}},\ \bibinfo {pages} {867} (\bibinfo {year}
  {2009}{\natexlab{a}})}\BibitemShut {NoStop}%
\bibitem [{\citenamefont {Bounoua}\ \emph {et~al.}(2018)\citenamefont
  {Bounoua}, \citenamefont {Saint-Martin}, \citenamefont {Dai}, \citenamefont
  {R{\"o}del}, \citenamefont {Sengupta}, \citenamefont {Frantzeskakis},
  \citenamefont {Bertran}, \citenamefont {Lefevre}, \citenamefont {Fortuna},
  \citenamefont {Santander-Syro},\ and\ \citenamefont
  {Pinsard-Gaudart}}]{Bounoua2018angle}%
  \BibitemOpen
  \bibfield  {author} {\bibinfo {author} {\bibfnamefont {D.}~\bibnamefont
  {Bounoua}}, \bibinfo {author} {\bibfnamefont {R.}~\bibnamefont
  {Saint-Martin}}, \bibinfo {author} {\bibfnamefont {J.}~\bibnamefont {Dai}},
  \bibinfo {author} {\bibfnamefont {T.}~\bibnamefont {R{\"o}del}}, \bibinfo
  {author} {\bibfnamefont {S.}~\bibnamefont {Sengupta}}, \bibinfo {author}
  {\bibfnamefont {E.}~\bibnamefont {Frantzeskakis}}, \bibinfo {author}
  {\bibfnamefont {F.}~\bibnamefont {Bertran}}, \bibinfo {author} {\bibfnamefont
  {P.}~\bibnamefont {Lefevre}}, \bibinfo {author} {\bibfnamefont
  {F.}~\bibnamefont {Fortuna}}, \bibinfo {author} {\bibfnamefont {A.~F.}\
  \bibnamefont {Santander-Syro}},\ and\ \bibinfo {author} {\bibfnamefont
  {L.}~\bibnamefont {Pinsard-Gaudart}},\ }\bibfield  {title} {\bibinfo {title}
  {Angle resolved photoemission spectroscopy study of the spin-charge
  separation in the strongly correlated cuprates {SrCuO$_2$} and
  {Sr$_2$CuO$_3$} with {$S =0$} impurities},\ }\href
  {https://doi.org/https://doi.org/10.1016/j.elspec.2018.03.011} {\bibfield
  {journal} {\bibinfo  {journal} {Journal of Electron Spectroscopy and Related
  Phenomena}\ }\textbf {\bibinfo {volume} {225}},\ \bibinfo {pages} {49}
  (\bibinfo {year} {2018})}\BibitemShut {NoStop}%
\bibitem [{\citenamefont {Schlappa}\ \emph {et~al.}(2018)\citenamefont
  {Schlappa}, \citenamefont {Kumar}, \citenamefont {Zhou}, \citenamefont
  {Singh}, \citenamefont {Mourigal}, \citenamefont {Strocov}, \citenamefont
  {Revcolevschi}, \citenamefont {Patthey}, \citenamefont {R{\o}nnow},
  \citenamefont {Johnston},\ and\ \citenamefont
  {Schmitt}}]{Schlappa2018probing}%
  \BibitemOpen
  \bibfield  {author} {\bibinfo {author} {\bibfnamefont {J.}~\bibnamefont
  {Schlappa}}, \bibinfo {author} {\bibfnamefont {U.}~\bibnamefont {Kumar}},
  \bibinfo {author} {\bibfnamefont {K.~J.}\ \bibnamefont {Zhou}}, \bibinfo
  {author} {\bibfnamefont {S.}~\bibnamefont {Singh}}, \bibinfo {author}
  {\bibfnamefont {M.}~\bibnamefont {Mourigal}}, \bibinfo {author}
  {\bibfnamefont {V.~N.}\ \bibnamefont {Strocov}}, \bibinfo {author}
  {\bibfnamefont {A.}~\bibnamefont {Revcolevschi}}, \bibinfo {author}
  {\bibfnamefont {L.}~\bibnamefont {Patthey}}, \bibinfo {author} {\bibfnamefont
  {H.~M.}\ \bibnamefont {R{\o}nnow}}, \bibinfo {author} {\bibfnamefont
  {S.}~\bibnamefont {Johnston}},\ and\ \bibinfo {author} {\bibfnamefont
  {T.}~\bibnamefont {Schmitt}},\ }\bibfield  {title} {\bibinfo {title} {Probing
  multi-spinon excitations outside of the two-spinon continuum in the
  antiferromagnetic spin chain cuprate {Sr}$_2${CuO}$_3$},\ }\href
  {https://doi.org/10.1038/s41467-018-07838-y} {\bibfield  {journal} {\bibinfo
  {journal} {Nature Communications}\ }\textbf {\bibinfo {volume} {9}},\
  \bibinfo {pages} {5394} (\bibinfo {year} {2018})}\BibitemShut {NoStop}%
\bibitem [{\citenamefont {Kumar}\ \emph {et~al.}(2022)\citenamefont {Kumar},
  \citenamefont {Nag}, \citenamefont {Li}, \citenamefont {Robarts},
  \citenamefont {Walters}, \citenamefont {Garc\'{\i}a-Fern\'andez},
  \citenamefont {Saint-Martin}, \citenamefont {Revcolevschi}, \citenamefont
  {Schlappa}, \citenamefont {Schmitt}, \citenamefont {Johnston},\ and\
  \citenamefont {Zhou}}]{Kumar2022unraveling}%
  \BibitemOpen
  \bibfield  {author} {\bibinfo {author} {\bibfnamefont {U.}~\bibnamefont
  {Kumar}}, \bibinfo {author} {\bibfnamefont {A.}~\bibnamefont {Nag}}, \bibinfo
  {author} {\bibfnamefont {J.}~\bibnamefont {Li}}, \bibinfo {author}
  {\bibfnamefont {H.~C.}\ \bibnamefont {Robarts}}, \bibinfo {author}
  {\bibfnamefont {A.~C.}\ \bibnamefont {Walters}}, \bibinfo {author}
  {\bibfnamefont {M.}~\bibnamefont {Garc\'{\i}a-Fern\'andez}}, \bibinfo
  {author} {\bibfnamefont {R.}~\bibnamefont {Saint-Martin}}, \bibinfo {author}
  {\bibfnamefont {A.}~\bibnamefont {Revcolevschi}}, \bibinfo {author}
  {\bibfnamefont {J.}~\bibnamefont {Schlappa}}, \bibinfo {author}
  {\bibfnamefont {T.}~\bibnamefont {Schmitt}}, \bibinfo {author} {\bibfnamefont
  {S.}~\bibnamefont {Johnston}},\ and\ \bibinfo {author} {\bibfnamefont
  {K.-J.}\ \bibnamefont {Zhou}},\ }\bibfield  {title} {\bibinfo {title}
  {Unraveling higher-order contributions to spin excitations probed using
  resonant inelastic x-ray scattering},\ }\href
  {https://doi.org/10.1103/PhysRevB.106.L060406} {\bibfield  {journal}
  {\bibinfo  {journal} {Phys. Rev. B}\ }\textbf {\bibinfo {volume} {106}},\
  \bibinfo {pages} {L060406} (\bibinfo {year} {2022})}\BibitemShut {NoStop}%
\bibitem [{\citenamefont {Chen}\ \emph {et~al.}(2021)\citenamefont {Chen},
  \citenamefont {Wang}, \citenamefont {Rebec}, \citenamefont {Jia},
  \citenamefont {Hashimoto}, \citenamefont {Lu}, \citenamefont {Moritz},
  \citenamefont {Moore}, \citenamefont {Devereaux},\ and\ \citenamefont
  {Shen}}]{chen}%
  \BibitemOpen
  \bibfield  {author} {\bibinfo {author} {\bibfnamefont {Z.}~\bibnamefont
  {Chen}}, \bibinfo {author} {\bibfnamefont {Y.}~\bibnamefont {Wang}}, \bibinfo
  {author} {\bibfnamefont {S.~N.}\ \bibnamefont {Rebec}}, \bibinfo {author}
  {\bibfnamefont {T.}~\bibnamefont {Jia}}, \bibinfo {author} {\bibfnamefont
  {M.}~\bibnamefont {Hashimoto}}, \bibinfo {author} {\bibfnamefont
  {D.}~\bibnamefont {Lu}}, \bibinfo {author} {\bibfnamefont {B.}~\bibnamefont
  {Moritz}}, \bibinfo {author} {\bibfnamefont {R.~G.}\ \bibnamefont {Moore}},
  \bibinfo {author} {\bibfnamefont {T.~P.}\ \bibnamefont {Devereaux}},\ and\
  \bibinfo {author} {\bibfnamefont {Z.-X.}\ \bibnamefont {Shen}},\ }\bibfield
  {title} {\bibinfo {title} {Anomalously strong near-neighbor attraction in
  doped {1D} cuprate chains},\ }\href {https://doi.org/10.1126/science.abf5174}
  {\bibfield  {journal} {\bibinfo  {journal} {Science}\ }\textbf {\bibinfo
  {volume} {373}},\ \bibinfo {pages} {1235} (\bibinfo {year}
  {2021})}\BibitemShut {NoStop}%
\bibitem [{\citenamefont {Li}\ \emph {et~al.}(2025)\citenamefont {Li},
  \citenamefont {Jost}, \citenamefont {Tang}, \citenamefont {Wang},
  \citenamefont {Zhong}, \citenamefont {Chen}, \citenamefont
  {Garcia-Fernandez}, \citenamefont {Pelliciari}, \citenamefont {Bisogni},
  \citenamefont {Moritz}, \citenamefont {Zhou}, \citenamefont {Wang},
  \citenamefont {Devereaux}, \citenamefont {Lee},\ and\ \citenamefont
  {Shen}}]{Li2025doping}%
  \BibitemOpen
  \bibfield  {author} {\bibinfo {author} {\bibfnamefont {J.}~\bibnamefont
  {Li}}, \bibinfo {author} {\bibfnamefont {D.}~\bibnamefont {Jost}}, \bibinfo
  {author} {\bibfnamefont {T.}~\bibnamefont {Tang}}, \bibinfo {author}
  {\bibfnamefont {R.}~\bibnamefont {Wang}}, \bibinfo {author} {\bibfnamefont
  {Y.}~\bibnamefont {Zhong}}, \bibinfo {author} {\bibfnamefont
  {Z.}~\bibnamefont {Chen}}, \bibinfo {author} {\bibfnamefont {M.}~\bibnamefont
  {Garcia-Fernandez}}, \bibinfo {author} {\bibfnamefont {J.}~\bibnamefont
  {Pelliciari}}, \bibinfo {author} {\bibfnamefont {V.}~\bibnamefont {Bisogni}},
  \bibinfo {author} {\bibfnamefont {B.}~\bibnamefont {Moritz}}, \bibinfo
  {author} {\bibfnamefont {K.}~\bibnamefont {Zhou}}, \bibinfo {author}
  {\bibfnamefont {Y.}~\bibnamefont {Wang}}, \bibinfo {author} {\bibfnamefont
  {T.~P.}\ \bibnamefont {Devereaux}}, \bibinfo {author} {\bibfnamefont {W.-S.}\
  \bibnamefont {Lee}},\ and\ \bibinfo {author} {\bibfnamefont {Z.-X.}\
  \bibnamefont {Shen}},\ }\bibfield  {title} {\bibinfo {title} {Doping
  dependence of 2-spinon excitations in the doped {1D} cuprate
  {${\mathrm{Ba}}_{2}{\mathrm{CuO}}_{3+\ensuremath{\delta}}$}},\ }\href
  {https://doi.org/10.1103/PhysRevLett.134.146501} {\bibfield  {journal}
  {\bibinfo  {journal} {Phys. Rev. Lett.}\ }\textbf {\bibinfo {volume} {134}},\
  \bibinfo {pages} {146501} (\bibinfo {year} {2025})}\BibitemShut {NoStop}%
\bibitem [{\citenamefont {Wang}\ \emph {et~al.}(2021)\citenamefont {Wang},
  \citenamefont {Chen}, \citenamefont {Shi}, \citenamefont {Moritz},
  \citenamefont {Shen},\ and\ \citenamefont {Devereaux}}]{wang}%
  \BibitemOpen
  \bibfield  {author} {\bibinfo {author} {\bibfnamefont {Y.}~\bibnamefont
  {Wang}}, \bibinfo {author} {\bibfnamefont {Z.}~\bibnamefont {Chen}}, \bibinfo
  {author} {\bibfnamefont {T.}~\bibnamefont {Shi}}, \bibinfo {author}
  {\bibfnamefont {B.}~\bibnamefont {Moritz}}, \bibinfo {author} {\bibfnamefont
  {Z.-X.}\ \bibnamefont {Shen}},\ and\ \bibinfo {author} {\bibfnamefont
  {T.~P.}\ \bibnamefont {Devereaux}},\ }\bibfield  {title} {\bibinfo {title}
  {Phonon-mediated long-range attractive interaction in one-dimensional
  cuprates},\ }\href {https://doi.org/10.1103/PhysRevLett.127.197003}
  {\bibfield  {journal} {\bibinfo  {journal} {Phys. Rev. Lett.}\ }\textbf
  {\bibinfo {volume} {127}},\ \bibinfo {pages} {197003} (\bibinfo {year}
  {2021})}\BibitemShut {NoStop}%
\bibitem [{\citenamefont {Tang}\ \emph {et~al.}(2023)\citenamefont {Tang},
  \citenamefont {Moritz}, \citenamefont {Peng}, \citenamefont {Shen},\ and\
  \citenamefont {Devereaux}}]{tangb}%
  \BibitemOpen
  \bibfield  {author} {\bibinfo {author} {\bibfnamefont {T.}~\bibnamefont
  {Tang}}, \bibinfo {author} {\bibfnamefont {B.}~\bibnamefont {Moritz}},
  \bibinfo {author} {\bibfnamefont {C.}~\bibnamefont {Peng}}, \bibinfo {author}
  {\bibfnamefont {Z.-X.}\ \bibnamefont {Shen}},\ and\ \bibinfo {author}
  {\bibfnamefont {T.~P.}\ \bibnamefont {Devereaux}},\ }\bibfield  {title}
  {\bibinfo {title} {Traces of electron-phonon coupling in one-dimensional
  cuprates},\ }\href {https://doi.org/10.1038/s41467-023-38408-6} {\bibfield
  {journal} {\bibinfo  {journal} {Nature Communications}\ }\textbf {\bibinfo
  {volume} {14}},\ \bibinfo {pages} {3129} (\bibinfo {year}
  {2023})}\BibitemShut {NoStop}%
\bibitem [{\citenamefont {Wang}\ \emph {et~al.}(2022)\citenamefont {Wang},
  \citenamefont {Wu}, \citenamefont {Jiang},\ and\ \citenamefont
  {Yao}}]{wang2b}%
  \BibitemOpen
  \bibfield  {author} {\bibinfo {author} {\bibfnamefont {H.-X.}\ \bibnamefont
  {Wang}}, \bibinfo {author} {\bibfnamefont {Y.-M.}\ \bibnamefont {Wu}},
  \bibinfo {author} {\bibfnamefont {Y.-F.}\ \bibnamefont {Jiang}},\ and\
  \bibinfo {author} {\bibfnamefont {H.}~\bibnamefont {Yao}},\ }\bibfield
  {title} {\bibinfo {title} {Spectral properties of {1D} extended {H}ubbard
  model from bosonization and time-dependent variational principle:
  applications to {1D} cuprate},\ }\href {https://arxiv.org/abs/2211.02031}
  {\bibfield  {journal} {\bibinfo  {journal} {arXiv:2211.02031}\ } (\bibinfo
  {year} {2022})}\BibitemShut {NoStop}%
\bibitem [{\citenamefont {Kumar}\ \emph {et~al.}(2018)\citenamefont {Kumar},
  \citenamefont {Nocera}, \citenamefont {Dagotto},\ and\ \citenamefont
  {Johnston}}]{Kumar2018multi}%
  \BibitemOpen
  \bibfield  {author} {\bibinfo {author} {\bibfnamefont {U.}~\bibnamefont
  {Kumar}}, \bibinfo {author} {\bibfnamefont {A.}~\bibnamefont {Nocera}},
  \bibinfo {author} {\bibfnamefont {E.}~\bibnamefont {Dagotto}},\ and\ \bibinfo
  {author} {\bibfnamefont {S.}~\bibnamefont {Johnston}},\ }\bibfield  {title}
  {\bibinfo {title} {Multi-spinon and antiholon excitations probed by resonant
  inelastic x-ray scattering on doped one-dimensional antiferromagnets},\
  }\href {https://doi.org/10.1088/1367-2630/aad00a} {\bibfield  {journal}
  {\bibinfo  {journal} {New Journal of Physics}\ }\textbf {\bibinfo {volume}
  {20}},\ \bibinfo {pages} {073019} (\bibinfo {year} {2018})}\BibitemShut
  {NoStop}%
\bibitem [{\citenamefont {Feiguin}\ \emph {et~al.}(2023)\citenamefont
  {Feiguin}, \citenamefont {Helman},\ and\ \citenamefont
  {Aligia}}]{Feiguin2023}%
  \BibitemOpen
  \bibfield  {author} {\bibinfo {author} {\bibfnamefont {A.~E.}\ \bibnamefont
  {Feiguin}}, \bibinfo {author} {\bibfnamefont {C.}~\bibnamefont {Helman}},\
  and\ \bibinfo {author} {\bibfnamefont {A.~A.}\ \bibnamefont {Aligia}},\
  }\bibfield  {title} {\bibinfo {title} {Effective one-band models for the
  one-dimensional cuprate
  {$\mathrm{Ba}_{2\ensuremath{-}x}{\mathrm{Sr}}_{x}\mathrm{CuO}_{3+\ensuremath{\delta}}$}},\
  }\href {https://doi.org/10.1103/PhysRevB.108.075125} {\bibfield  {journal}
  {\bibinfo  {journal} {Phys. Rev. B}\ }\textbf {\bibinfo {volume} {108}},\
  \bibinfo {pages} {075125} (\bibinfo {year} {2023})}\BibitemShut {NoStop}%
\bibitem [{\citenamefont {Zaanen}\ and\ \citenamefont
  {Ole\ifmmode~\acute{s}\else \'{s}\fi{}}(1988)}]{Zaanen1988}%
  \BibitemOpen
  \bibfield  {author} {\bibinfo {author} {\bibfnamefont {J.}~\bibnamefont
  {Zaanen}}\ and\ \bibinfo {author} {\bibfnamefont {A.~M.}\ \bibnamefont
  {Ole\ifmmode~\acute{s}\else \'{s}\fi{}}},\ }\bibfield  {title} {\bibinfo
  {title} {Canonical perturbation theory and the two-band model for
  high-${T}_{c}$ superconductors},\ }\href
  {https://doi.org/10.1103/PhysRevB.37.9423} {\bibfield  {journal} {\bibinfo
  {journal} {Phys. Rev. B}\ }\textbf {\bibinfo {volume} {37}},\ \bibinfo
  {pages} {9423} (\bibinfo {year} {1988})}\BibitemShut {NoStop}%
\bibitem [{\citenamefont {Varma}\ \emph {et~al.}(1993)\citenamefont {Varma},
  \citenamefont {Schmitt-Rink},\ and\ \citenamefont {Abrahams}}]{Varma1993}%
  \BibitemOpen
  \bibfield  {author} {\bibinfo {author} {\bibfnamefont {C.}~\bibnamefont
  {Varma}}, \bibinfo {author} {\bibfnamefont {S.}~\bibnamefont
  {Schmitt-Rink}},\ and\ \bibinfo {author} {\bibfnamefont {E.}~\bibnamefont
  {Abrahams}},\ }\bibfield  {title} {\bibinfo {title} {Charge transfer
  excitations and superconductivity in ``ionic'' metals},\ }\href
  {https://doi.org/https://doi.org/10.1016/0038-1098(93)90255-L} {\bibfield
  {journal} {\bibinfo  {journal} {Solid State Communications}\ }\textbf
  {\bibinfo {volume} {88}},\ \bibinfo {pages} {847} (\bibinfo {year}
  {1993})}\BibitemShut {NoStop}%
\bibitem [{\citenamefont {White}\ and\ \citenamefont
  {Feiguin}(2004)}]{white2004a}%
  \BibitemOpen
  \bibfield  {author} {\bibinfo {author} {\bibfnamefont {S.~R.}\ \bibnamefont
  {White}}\ and\ \bibinfo {author} {\bibfnamefont {A.~E.}\ \bibnamefont
  {Feiguin}},\ }\bibfield  {title} {\bibinfo {title} {Real-time evolution using
  the density matrix renormalization group},\ }\href
  {https://doi.org/10.1103/PhysRevLett.93.076401} {\bibfield  {journal}
  {\bibinfo  {journal} {Phys. Rev. Lett.}\ }\textbf {\bibinfo {volume} {93}},\
  \bibinfo {pages} {076401} (\bibinfo {year} {2004})}\BibitemShut {NoStop}%
\bibitem [{\citenamefont {Daley}\ \emph {et~al.}(2004)\citenamefont {Daley},
  \citenamefont {Kollath}, \citenamefont {Schollw\"ock},\ and\ \citenamefont
  {Vidal}}]{daley2004}%
  \BibitemOpen
  \bibfield  {author} {\bibinfo {author} {\bibfnamefont {A.~J.}\ \bibnamefont
  {Daley}}, \bibinfo {author} {\bibfnamefont {C.}~\bibnamefont {Kollath}},
  \bibinfo {author} {\bibfnamefont {U.}~\bibnamefont {Schollw\"ock}},\ and\
  \bibinfo {author} {\bibfnamefont {G.}~\bibnamefont {Vidal}},\ }\bibfield
  {title} {\bibinfo {title} {Time-dependent density-matrix
  renormalization-group using adaptive effective {H}ilbert spaces},\ }\href
  {https://doi.org/10.1088/1742-5468/2004/04/p04005} {\bibfield  {journal}
  {\bibinfo  {journal} {Journal of Statistical Mechanics: Theory and
  Experiment}\ }\textbf {\bibinfo {volume} {2004}},\ \bibinfo {pages} {P04005}
  (\bibinfo {year} {2004})}\BibitemShut {NoStop}%
\bibitem [{\citenamefont {Feiguin}(2011)}]{vietri}%
  \BibitemOpen
  \bibfield  {author} {\bibinfo {author} {\bibfnamefont {A.~E.}\ \bibnamefont
  {Feiguin}},\ }\bibfield  {title} {\bibinfo {title} {The density matrix
  renormalization group method and its time-dependent variants},\ }in\
  \href@noop {} {\emph {\bibinfo {booktitle} {XV Training Course in the Physics
  of Strongly Correlated Systems}}},\ Vol.\ \bibinfo {volume} {1419}\ (\bibinfo
   {publisher} {AIP Proceedings},\ \bibinfo {year} {2011})\ p.~\bibinfo {pages}
  {5}\BibitemShut {NoStop}%
\bibitem [{\citenamefont {Paeckel}\ \emph {et~al.}(2019)\citenamefont
  {Paeckel}, \citenamefont {K{\"o}hler}, \citenamefont {Swoboda}, \citenamefont
  {Manmana}, \citenamefont {Schollw{\"o}ck},\ and\ \citenamefont
  {Hubig}}]{Paeckel2019}%
  \BibitemOpen
  \bibfield  {author} {\bibinfo {author} {\bibfnamefont {S.}~\bibnamefont
  {Paeckel}}, \bibinfo {author} {\bibfnamefont {T.}~\bibnamefont {K{\"o}hler}},
  \bibinfo {author} {\bibfnamefont {A.}~\bibnamefont {Swoboda}}, \bibinfo
  {author} {\bibfnamefont {S.~R.}\ \bibnamefont {Manmana}}, \bibinfo {author}
  {\bibfnamefont {U.}~\bibnamefont {Schollw{\"o}ck}},\ and\ \bibinfo {author}
  {\bibfnamefont {C.}~\bibnamefont {Hubig}},\ }\bibfield  {title} {\bibinfo
  {title} {Time-evolution methods for matrix-product states},\ }\href
  {https://doi.org/https://doi.org/10.1016/j.aop.2019.167998} {\bibfield
  {journal} {\bibinfo  {journal} {Annals of Physics}\ }\textbf {\bibinfo
  {volume} {411}},\ \bibinfo {pages} {167998} (\bibinfo {year}
  {2019})}\BibitemShut {NoStop}%
\bibitem [{\citenamefont {Zawadzki}\ \emph {et~al.}(2023)\citenamefont
  {Zawadzki}, \citenamefont {Nocera},\ and\ \citenamefont
  {Feiguin}}]{Zawadzki2023timedependent}%
  \BibitemOpen
  \bibfield  {author} {\bibinfo {author} {\bibfnamefont {K.}~\bibnamefont
  {Zawadzki}}, \bibinfo {author} {\bibfnamefont {A.}~\bibnamefont {Nocera}},\
  and\ \bibinfo {author} {\bibfnamefont {A.~E.}\ \bibnamefont {Feiguin}},\
  }\bibfield  {title} {\bibinfo {title} {{A time-dependent momentum-resolved
  scattering approach to core-level spectroscopies}},\ }\href
  {https://doi.org/10.21468/SciPostPhys.15.4.166} {\bibfield  {journal}
  {\bibinfo  {journal} {SciPost Phys.}\ }\textbf {\bibinfo {volume} {15}},\
  \bibinfo {pages} {166} (\bibinfo {year} {2023})}\BibitemShut {NoStop}%
\bibitem [{\citenamefont {Oppenheim}\ \emph {et~al.}(1999)\citenamefont
  {Oppenheim}, \citenamefont {Schafer},\ and\ \citenamefont
  {Buck}}]{oppenheim99}%
  \BibitemOpen
  \bibfield  {author} {\bibinfo {author} {\bibfnamefont {A.~V.}\ \bibnamefont
  {Oppenheim}}, \bibinfo {author} {\bibfnamefont {R.~W.}\ \bibnamefont
  {Schafer}},\ and\ \bibinfo {author} {\bibfnamefont {J.~R.}\ \bibnamefont
  {Buck}},\ }\href@noop {} {\emph {\bibinfo {title} {Discrete-Time Signal
  Processing}}},\ \bibinfo {edition} {2nd}\ ed.\ (\bibinfo  {publisher}
  {Prentice-hall Englewood Cliffs},\ \bibinfo {year} {1999})\BibitemShut
  {NoStop}%
\bibitem [{\citenamefont {Ogata}\ and\ \citenamefont
  {Shiba}(1990)}]{Ogata1990}%
  \BibitemOpen
  \bibfield  {author} {\bibinfo {author} {\bibfnamefont {M.}~\bibnamefont
  {Ogata}}\ and\ \bibinfo {author} {\bibfnamefont {H.}~\bibnamefont {Shiba}},\
  }\bibfield  {title} {\bibinfo {title} {Bethe-ansatz wave function, momentum
  distribution, and spin correlation in the one-dimensional strongly correlated
  {H}ubbard model},\ }\href {https://doi.org/10.1103/PhysRevB.41.2326}
  {\bibfield  {journal} {\bibinfo  {journal} {Phys. Rev. B}\ }\textbf {\bibinfo
  {volume} {41}},\ \bibinfo {pages} {2326} (\bibinfo {year}
  {1990})}\BibitemShut {NoStop}%
\bibitem [{\citenamefont {Penc}\ \emph {et~al.}(1997)\citenamefont {Penc},
  \citenamefont {Hallberg}, \citenamefont {Mila},\ and\ \citenamefont
  {Shiba}}]{Penc1997b}%
  \BibitemOpen
  \bibfield  {author} {\bibinfo {author} {\bibfnamefont {K.}~\bibnamefont
  {Penc}}, \bibinfo {author} {\bibfnamefont {K.}~\bibnamefont {Hallberg}},
  \bibinfo {author} {\bibfnamefont {F.}~\bibnamefont {Mila}},\ and\ \bibinfo
  {author} {\bibfnamefont {H.}~\bibnamefont {Shiba}},\ }\bibfield  {title}
  {\bibinfo {title} {Spectral functions of the one-dimensional {H}ubbard model
  in the {$U \rightarrow +\infty$} limit: How to use the factorized wave
  function},\ }\href {https://doi.org/10.1103/PhysRevB.55.15475} {\bibfield
  {journal} {\bibinfo  {journal} {Phys. Rev. B}\ }\textbf {\bibinfo {volume}
  {55}},\ \bibinfo {pages} {15475} (\bibinfo {year} {1997})}\BibitemShut
  {NoStop}%
\bibitem [{\citenamefont {Benthien}\ and\ \citenamefont
  {Jeckelmann}(2007)}]{Benthien2007}%
  \BibitemOpen
  \bibfield  {author} {\bibinfo {author} {\bibfnamefont {H.}~\bibnamefont
  {Benthien}}\ and\ \bibinfo {author} {\bibfnamefont {E.}~\bibnamefont
  {Jeckelmann}},\ }\bibfield  {title} {\bibinfo {title} {Spin and charge
  dynamics of the one-dimensional extended {H}ubbard model},\ }\href
  {https://doi.org/10.1103/PhysRevB.75.205128} {\bibfield  {journal} {\bibinfo
  {journal} {Phys. Rev. B}\ }\textbf {\bibinfo {volume} {75}},\ \bibinfo
  {pages} {205128} (\bibinfo {year} {2007})}\BibitemShut {NoStop}%
\bibitem [{\citenamefont {de~Groot}\ and\ \citenamefont
  {Kotani}(2008)}]{deGroot2008}%
  \BibitemOpen
  \bibfield  {author} {\bibinfo {author} {\bibfnamefont {F.}~\bibnamefont
  {de~Groot}}\ and\ \bibinfo {author} {\bibfnamefont {A.}~\bibnamefont
  {Kotani}},\ }\href {https://doi.org/10.1201/9781420008425} {\emph {\bibinfo
  {title} {Core Level Spectroscopy of Solids}}}\ (\bibinfo  {publisher} {CRC
  Press},\ \bibinfo {address} {Boca Raton, FL},\ \bibinfo {year}
  {2008})\BibitemShut {NoStop}%
\bibitem [{\citenamefont {Chantler}\ \emph {et~al.}(2024)\citenamefont
  {Chantler}, \citenamefont {Bunker}, \citenamefont {D'Angelo},\ and\
  \citenamefont {Diaz-Moreno}}]{Chantler2024xasprimer}%
  \BibitemOpen
  \bibfield  {author} {\bibinfo {author} {\bibfnamefont {C.~T.}\ \bibnamefont
  {Chantler}}, \bibinfo {author} {\bibfnamefont {G.}~\bibnamefont {Bunker}},
  \bibinfo {author} {\bibfnamefont {P.}~\bibnamefont {D'Angelo}},\ and\
  \bibinfo {author} {\bibfnamefont {S.}~\bibnamefont {Diaz-Moreno}},\
  }\bibfield  {title} {\bibinfo {title} {X-ray absorption spectroscopy},\
  }\href {https://doi.org/10.1038/s43586-024-00366-8} {\bibfield  {journal}
  {\bibinfo  {journal} {Nature Reviews Methods Primers}\ }\textbf {\bibinfo
  {volume} {4}},\ \bibinfo {pages} {89} (\bibinfo {year} {2024})}\BibitemShut
  {NoStop}%
\bibitem [{\citenamefont {{Nocera, Alberto}}\ and\ \citenamefont {{Feiguin,
  Adrian}}(2023)}]{Nocera2023}%
  \BibitemOpen
  \bibfield  {author} {\bibinfo {author} {\bibnamefont {{Nocera, Alberto}}}\
  and\ \bibinfo {author} {\bibnamefont {{Feiguin, Adrian}}},\ }\bibfield
  {title} {\bibinfo {title} {Auger spectroscopy beyond the ultra-short
  core-hole relaxation time approximation},\ }\href
  {https://doi.org/10.1140/epjp/s13360-023-04717-4} {\bibfield  {journal}
  {\bibinfo  {journal} {Eur. Phys. J. Plus}\ }\textbf {\bibinfo {volume}
  {138}},\ \bibinfo {pages} {1106} (\bibinfo {year} {2023})}\BibitemShut
  {NoStop}%
\bibitem [{\citenamefont {Bourges}\ \emph {et~al.}(1997)\citenamefont
  {Bourges}, \citenamefont {Casalta}, \citenamefont {Ivanov},\ and\
  \citenamefont {Petitgrand}}]{Bourges1997superexchange}%
  \BibitemOpen
  \bibfield  {author} {\bibinfo {author} {\bibfnamefont {P.}~\bibnamefont
  {Bourges}}, \bibinfo {author} {\bibfnamefont {H.}~\bibnamefont {Casalta}},
  \bibinfo {author} {\bibfnamefont {A.~S.}\ \bibnamefont {Ivanov}},\ and\
  \bibinfo {author} {\bibfnamefont {D.}~\bibnamefont {Petitgrand}},\ }\bibfield
   {title} {\bibinfo {title} {Superexchange coupling and spin susceptibility
  spectral weight in undoped monolayer cuprates},\ }\href
  {https://doi.org/10.1103/PhysRevLett.79.4906} {\bibfield  {journal} {\bibinfo
   {journal} {Phys. Rev. Lett.}\ }\textbf {\bibinfo {volume} {79}},\ \bibinfo
  {pages} {4906} (\bibinfo {year} {1997})}\BibitemShut {NoStop}%
\bibitem [{\citenamefont {Coldea}\ \emph {et~al.}(2001)\citenamefont {Coldea},
  \citenamefont {Hayden}, \citenamefont {Aeppli}, \citenamefont {Perring},
  \citenamefont {Frost}, \citenamefont {Mason}, \citenamefont {Cheong},\ and\
  \citenamefont {Fisk}}]{Coldea2001spin}%
  \BibitemOpen
  \bibfield  {author} {\bibinfo {author} {\bibfnamefont {R.}~\bibnamefont
  {Coldea}}, \bibinfo {author} {\bibfnamefont {S.~M.}\ \bibnamefont {Hayden}},
  \bibinfo {author} {\bibfnamefont {G.}~\bibnamefont {Aeppli}}, \bibinfo
  {author} {\bibfnamefont {T.~G.}\ \bibnamefont {Perring}}, \bibinfo {author}
  {\bibfnamefont {C.~D.}\ \bibnamefont {Frost}}, \bibinfo {author}
  {\bibfnamefont {T.~E.}\ \bibnamefont {Mason}}, \bibinfo {author}
  {\bibfnamefont {S.-W.}\ \bibnamefont {Cheong}},\ and\ \bibinfo {author}
  {\bibfnamefont {Z.}~\bibnamefont {Fisk}},\ }\bibfield  {title} {\bibinfo
  {title} {Spin waves and electronic interactions in
  {${\mathrm{La}}_{2}{\mathrm{CuO}}_{4}$}},\ }\href
  {https://doi.org/10.1103/PhysRevLett.86.5377} {\bibfield  {journal} {\bibinfo
   {journal} {Phys. Rev. Lett.}\ }\textbf {\bibinfo {volume} {86}},\ \bibinfo
  {pages} {5377} (\bibinfo {year} {2001})}\BibitemShut {NoStop}%
\bibitem [{\citenamefont {Lorenzana}\ \emph {et~al.}(2005)\citenamefont
  {Lorenzana}, \citenamefont {Seibold},\ and\ \citenamefont
  {Coldea}}]{Lorenzana2005sum}%
  \BibitemOpen
  \bibfield  {author} {\bibinfo {author} {\bibfnamefont {J.}~\bibnamefont
  {Lorenzana}}, \bibinfo {author} {\bibfnamefont {G.}~\bibnamefont {Seibold}},\
  and\ \bibinfo {author} {\bibfnamefont {R.}~\bibnamefont {Coldea}},\
  }\bibfield  {title} {\bibinfo {title} {Sum rules and missing spectral weight
  in magnetic neutron scattering in the cuprates},\ }\href
  {https://doi.org/10.1103/PhysRevB.72.224511} {\bibfield  {journal} {\bibinfo
  {journal} {Phys. Rev. B}\ }\textbf {\bibinfo {volume} {72}},\ \bibinfo
  {pages} {224511} (\bibinfo {year} {2005})}\BibitemShut {NoStop}%
\bibitem [{\citenamefont {Walters}\ \emph
  {et~al.}(2009{\natexlab{b}})\citenamefont {Walters}, \citenamefont {Perring},
  \citenamefont {Caux}, \citenamefont {Savici}, \citenamefont {Gu},
  \citenamefont {Lee}, \citenamefont {Ku},\ and\ \citenamefont
  {Zaliznyak}}]{Walters2009effect}%
  \BibitemOpen
  \bibfield  {author} {\bibinfo {author} {\bibfnamefont {A.~C.}\ \bibnamefont
  {Walters}}, \bibinfo {author} {\bibfnamefont {T.~G.}\ \bibnamefont
  {Perring}}, \bibinfo {author} {\bibfnamefont {J.-S.}\ \bibnamefont {Caux}},
  \bibinfo {author} {\bibfnamefont {A.~T.}\ \bibnamefont {Savici}}, \bibinfo
  {author} {\bibfnamefont {G.~D.}\ \bibnamefont {Gu}}, \bibinfo {author}
  {\bibfnamefont {C.-C.}\ \bibnamefont {Lee}}, \bibinfo {author} {\bibfnamefont
  {W.}~\bibnamefont {Ku}},\ and\ \bibinfo {author} {\bibfnamefont {I.~A.}\
  \bibnamefont {Zaliznyak}},\ }\bibfield  {title} {\bibinfo {title} {Effect of
  covalent bonding on magnetism and the missing neutron intensity in copper
  oxide compounds},\ }\href {https://doi.org/10.1038/nphys1405} {\bibfield
  {journal} {\bibinfo  {journal} {Nature Physics}\ }\textbf {\bibinfo {volume}
  {5}},\ \bibinfo {pages} {867} (\bibinfo {year}
  {2009}{\natexlab{b}})}\BibitemShut {NoStop}%
\bibitem [{\citenamefont {Eskes}\ \emph {et~al.}(1991)\citenamefont {Eskes},
  \citenamefont {Meinders},\ and\ \citenamefont
  {Sawatzky}}]{Eskes1991anomalous}%
  \BibitemOpen
  \bibfield  {author} {\bibinfo {author} {\bibfnamefont {H.}~\bibnamefont
  {Eskes}}, \bibinfo {author} {\bibfnamefont {M.~B.~J.}\ \bibnamefont
  {Meinders}},\ and\ \bibinfo {author} {\bibfnamefont {G.~A.}\ \bibnamefont
  {Sawatzky}},\ }\bibfield  {title} {\bibinfo {title} {Anomalous transfer of
  spectral weight in doped strongly correlated systems},\ }\href
  {https://doi.org/10.1103/PhysRevLett.67.1035} {\bibfield  {journal} {\bibinfo
   {journal} {Phys. Rev. Lett.}\ }\textbf {\bibinfo {volume} {67}},\ \bibinfo
  {pages} {1035} (\bibinfo {year} {1991})}\BibitemShut {NoStop}%
\bibitem [{\citenamefont {Jia}\ \emph {et~al.}(2016)\citenamefont {Jia},
  \citenamefont {Wohlfeld}, \citenamefont {Wang}, \citenamefont {Moritz},\ and\
  \citenamefont {Devereaux}}]{Jia2016using}%
  \BibitemOpen
  \bibfield  {author} {\bibinfo {author} {\bibfnamefont {C.}~\bibnamefont
  {Jia}}, \bibinfo {author} {\bibfnamefont {K.}~\bibnamefont {Wohlfeld}},
  \bibinfo {author} {\bibfnamefont {Y.}~\bibnamefont {Wang}}, \bibinfo {author}
  {\bibfnamefont {B.}~\bibnamefont {Moritz}},\ and\ \bibinfo {author}
  {\bibfnamefont {T.~P.}\ \bibnamefont {Devereaux}},\ }\bibfield  {title}
  {\bibinfo {title} {Using {RIXS} to uncover elementary charge and spin
  excitations},\ }\href {https://doi.org/10.1103/PhysRevX.6.021020} {\bibfield
  {journal} {\bibinfo  {journal} {Phys. Rev. X}\ }\textbf {\bibinfo {volume}
  {6}},\ \bibinfo {pages} {021020} (\bibinfo {year} {2016})}\BibitemShut
  {NoStop}%
\bibitem [{\citenamefont {Thomas}\ \emph {et~al.}(2025)\citenamefont {Thomas},
  \citenamefont {Banerjee}, \citenamefont {Nocera},\ and\ \citenamefont
  {Johnston}}]{Thomas2025theory}%
  \BibitemOpen
  \bibfield  {author} {\bibinfo {author} {\bibfnamefont {J.}~\bibnamefont
  {Thomas}}, \bibinfo {author} {\bibfnamefont {D.}~\bibnamefont {Banerjee}},
  \bibinfo {author} {\bibfnamefont {A.}~\bibnamefont {Nocera}},\ and\ \bibinfo
  {author} {\bibfnamefont {S.}~\bibnamefont {Johnston}},\ }\bibfield  {title}
  {\bibinfo {title} {Theory of electron-phonon interactions in extended
  correlated systems probed by resonant inelastic x-ray scattering},\ }\href
  {https://doi.org/10.1103/PhysRevX.15.021030} {\bibfield  {journal} {\bibinfo
  {journal} {Phys. Rev. X}\ }\textbf {\bibinfo {volume} {15}},\ \bibinfo
  {pages} {021030} (\bibinfo {year} {2025})}\BibitemShut {NoStop}%
\bibitem [{\citenamefont {Laurell}\ \emph {et~al.}(2022)\citenamefont
  {Laurell}, \citenamefont {Scheie}, \citenamefont {Tennant}, \citenamefont
  {Okamoto}, \citenamefont {Alvarez},\ and\ \citenamefont
  {Dagotto}}]{Laurell2022magnetic}%
  \BibitemOpen
  \bibfield  {author} {\bibinfo {author} {\bibfnamefont {P.}~\bibnamefont
  {Laurell}}, \bibinfo {author} {\bibfnamefont {A.}~\bibnamefont {Scheie}},
  \bibinfo {author} {\bibfnamefont {D.~A.}\ \bibnamefont {Tennant}}, \bibinfo
  {author} {\bibfnamefont {S.}~\bibnamefont {Okamoto}}, \bibinfo {author}
  {\bibfnamefont {G.}~\bibnamefont {Alvarez}},\ and\ \bibinfo {author}
  {\bibfnamefont {E.}~\bibnamefont {Dagotto}},\ }\bibfield  {title} {\bibinfo
  {title} {Magnetic excitations, nonclassicality, and quantum wake spin
  dynamics in the {H}ubbard chain},\ }\href
  {https://doi.org/10.1103/PhysRevB.106.085110} {\bibfield  {journal} {\bibinfo
   {journal} {Phys. Rev. B}\ }\textbf {\bibinfo {volume} {106}},\ \bibinfo
  {pages} {085110} (\bibinfo {year} {2022})}\BibitemShut {NoStop}%
\bibitem [{\citenamefont {Scheie}\ \emph
  {et~al.}(2025{\natexlab{a}})\citenamefont {Scheie}, \citenamefont {Laurell},
  \citenamefont {Simeth}, \citenamefont {Dagotto},\ and\ \citenamefont
  {Tennant}}]{Scheie2025tutorial}%
  \BibitemOpen
  \bibfield  {author} {\bibinfo {author} {\bibfnamefont {A.}~\bibnamefont
  {Scheie}}, \bibinfo {author} {\bibfnamefont {P.}~\bibnamefont {Laurell}},
  \bibinfo {author} {\bibfnamefont {W.}~\bibnamefont {Simeth}}, \bibinfo
  {author} {\bibfnamefont {E.}~\bibnamefont {Dagotto}},\ and\ \bibinfo {author}
  {\bibfnamefont {D.~A.}\ \bibnamefont {Tennant}},\ }\bibfield  {title}
  {\bibinfo {title} {Tutorial: Extracting entanglement signatures from neutron
  spectroscopy},\ }\href
  {https://doi.org/https://doi.org/10.1016/j.mtquan.2024.100020} {\bibfield
  {journal} {\bibinfo  {journal} {Materials Today Quantum}\ }\textbf {\bibinfo
  {volume} {5}},\ \bibinfo {pages} {100020} (\bibinfo {year}
  {2025}{\natexlab{a}})}\BibitemShut {NoStop}%
\bibitem [{\citenamefont {Padma}\ \emph {et~al.}(2025)\citenamefont {Padma},
  \citenamefont {Thomas}, \citenamefont {TenHuisen}, \citenamefont {He},
  \citenamefont {Guan}, \citenamefont {Li}, \citenamefont {Lee}, \citenamefont
  {Wang}, \citenamefont {Lee}, \citenamefont {Mao}, \citenamefont {Jang},
  \citenamefont {Bisogni}, \citenamefont {Pelliciari}, \citenamefont {Dean},
  \citenamefont {Johnston},\ and\ \citenamefont {Mitrano}}]{Padma2025beyond}%
  \BibitemOpen
  \bibfield  {author} {\bibinfo {author} {\bibfnamefont {H.}~\bibnamefont
  {Padma}}, \bibinfo {author} {\bibfnamefont {J.}~\bibnamefont {Thomas}},
  \bibinfo {author} {\bibfnamefont {S.~F.~R.}\ \bibnamefont {TenHuisen}},
  \bibinfo {author} {\bibfnamefont {W.}~\bibnamefont {He}}, \bibinfo {author}
  {\bibfnamefont {Z.}~\bibnamefont {Guan}}, \bibinfo {author} {\bibfnamefont
  {J.}~\bibnamefont {Li}}, \bibinfo {author} {\bibfnamefont {B.}~\bibnamefont
  {Lee}}, \bibinfo {author} {\bibfnamefont {Y.}~\bibnamefont {Wang}}, \bibinfo
  {author} {\bibfnamefont {S.~H.}\ \bibnamefont {Lee}}, \bibinfo {author}
  {\bibfnamefont {Z.}~\bibnamefont {Mao}}, \bibinfo {author} {\bibfnamefont
  {H.}~\bibnamefont {Jang}}, \bibinfo {author} {\bibfnamefont {V.}~\bibnamefont
  {Bisogni}}, \bibinfo {author} {\bibfnamefont {J.}~\bibnamefont {Pelliciari}},
  \bibinfo {author} {\bibfnamefont {M.~P.~M.}\ \bibnamefont {Dean}}, \bibinfo
  {author} {\bibfnamefont {S.}~\bibnamefont {Johnston}},\ and\ \bibinfo
  {author} {\bibfnamefont {M.}~\bibnamefont {Mitrano}},\ }\bibfield  {title}
  {\bibinfo {title} {Beyond-{H}ubbard pairing in a cuprate ladder},\ }\href
  {https://doi.org/10.1103/PhysRevX.15.021049} {\bibfield  {journal} {\bibinfo
  {journal} {Phys. Rev. X}\ }\textbf {\bibinfo {volume} {15}},\ \bibinfo
  {pages} {021049} (\bibinfo {year} {2025})}\BibitemShut {NoStop}%
\bibitem [{\citenamefont {Scheie}\ \emph
  {et~al.}(2025{\natexlab{b}})\citenamefont {Scheie}, \citenamefont {Laurell},
  \citenamefont {Thomas}, \citenamefont {Sharma}, \citenamefont {Kolesnikov},
  \citenamefont {Granroth}, \citenamefont {Zhang}, \citenamefont {Lake},
  \citenamefont {Jr.}, \citenamefont {Bewley}, \citenamefont {Eccleston},
  \citenamefont {Akimitsu}, \citenamefont {Dagotto}, \citenamefont {Batista},
  \citenamefont {Alvarez}, \citenamefont {Johnston},\ and\ \citenamefont
  {Tennant}}]{scheie2025cooper}%
  \BibitemOpen
  \bibfield  {author} {\bibinfo {author} {\bibfnamefont {A.}~\bibnamefont
  {Scheie}}, \bibinfo {author} {\bibfnamefont {P.}~\bibnamefont {Laurell}},
  \bibinfo {author} {\bibfnamefont {J.}~\bibnamefont {Thomas}}, \bibinfo
  {author} {\bibfnamefont {V.}~\bibnamefont {Sharma}}, \bibinfo {author}
  {\bibfnamefont {A.~I.}\ \bibnamefont {Kolesnikov}}, \bibinfo {author}
  {\bibfnamefont {G.~E.}\ \bibnamefont {Granroth}}, \bibinfo {author}
  {\bibfnamefont {Q.}~\bibnamefont {Zhang}}, \bibinfo {author} {\bibfnamefont
  {B.}~\bibnamefont {Lake}}, \bibinfo {author} {\bibfnamefont {M.~M.}\
  \bibnamefont {Jr.}}, \bibinfo {author} {\bibfnamefont {R.~I.}\ \bibnamefont
  {Bewley}}, \bibinfo {author} {\bibfnamefont {R.~S.}\ \bibnamefont
  {Eccleston}}, \bibinfo {author} {\bibfnamefont {J.}~\bibnamefont {Akimitsu}},
  \bibinfo {author} {\bibfnamefont {E.}~\bibnamefont {Dagotto}}, \bibinfo
  {author} {\bibfnamefont {C.~D.}\ \bibnamefont {Batista}}, \bibinfo {author}
  {\bibfnamefont {G.}~\bibnamefont {Alvarez}}, \bibinfo {author} {\bibfnamefont
  {S.}~\bibnamefont {Johnston}},\ and\ \bibinfo {author} {\bibfnamefont
  {D.~A.}\ \bibnamefont {Tennant}},\ }\bibfield  {title} {\bibinfo {title}
  {Cooper-pair localization in the magnetic dynamics of a cuprate ladder},\
  }\href {https://arxiv.org/abs/2501.10296} {\bibfield  {journal} {\bibinfo
  {journal} {arXiv:2501.10296}\ } (\bibinfo {year}
  {2025}{\natexlab{b}})}\BibitemShut {NoStop}%
\end{thebibliography}%
    
\end{document}